\documentclass[aps,prd,twocolumn,superscriptaddress,nofootinbib,showpacs]{revtex4-2}
\usepackage{amsmath}
\usepackage{amssymb}
\usepackage{graphicx}
\usepackage{xcolor,color}
\usepackage{url}
\usepackage{bm}
\usepackage{mathrsfs}
\usepackage[utf8]{inputenc}
\usepackage{hyperref}
\usepackage{enumerate}
\usepackage{amsthm}
\usepackage{bbm}
\usepackage[normalem]{ulem}
\usepackage{upgreek}
\usepackage{tensor}
\usepackage{siunitx}
\usepackage{orcidlink}
\usepackage[caption=false]{subfig}
\usepackage{float}
\usepackage{colortbl}

\definecolor{colour1}{HTML}{0571b0} %--- Blue
\definecolor{colour2}{HTML}{92c5de} %--- Cyan
\definecolor{colour3}{HTML}{f4a582} %--- Orange
\definecolor{colour4}{HTML}{ca0020} %--- Maroon
\definecolor{colour5}{HTML}{fe4a49} %--- Red
\definecolor{colour6}{HTML}{2d3092} %--- Purple

\definecolor{colour99}{HTML}{0000FF} %---
\hypersetup{colorlinks=true, linkcolor=colour99, citecolor=colour99,
filecolor=colour99, urlcolor=colour99}

\theoremstyle{definition}

\usepackage{color,xcolor}
\usepackage{ulem}
\usepackage{float}
\usepackage{subfig}
\usepackage{cleveref}
\usepackage{multirow}
\usepackage{booktabs}
\usepackage{booktabs}
\usepackage{threeparttable}

\usepackage[export]{adjustbox}
\usepackage{times}
\usepackage{commath}
\usepackage{lipsum}

\crefname{equation}{Eq.}{Eq.}
\Crefname{equation}{Eqs.}{Eqs.}

\crefname{section}{Sec.}{Sec.}
\Crefname{section}{Secs.}{Secs.}

\crefname{table}{Table}{Table}
\Crefname{table}{Tables}{Tables}
\crefname{figure}{Figure}{Figure}
\Crefname{figure}{Figures.}{Figures}

\crefname{appendix}{Appendix}{Appendix}

\usepackage{threeparttable}
\graphicspath{{Figures/}}
\begin{document}
%\pagewiselinenumbers% 按页重新编号
%\switchlinenumbers

\title{Complete total-transmission modes of Kerr black holes}

\author{Changkai Chen\,\orcidlink{0000-0002-4023-0682}}
\affiliation{School of Physics and Astronomy, Beijing Normal University, Beijing, 100875, China}
%\affiliation{School of Mathematics, Yunnan Normal University, Kunming 650500, China}

\author{Xiaohua Zhang\,\orcidlink{0000-0002-0914-5455}}\email[Contact author: ]{zhangxiaohua@ynnu.edu.cn}
\affiliation{School of Mathematics, Yunnan Normal University, Kunming 650500, China}

\author{Zhoujian Cao\,\orcidlink{0000-0002-1932-7295} } \email[Contact author: ]{zjcao@bnu.edu.cn}
\affiliation{School of Physics and Astronomy, Beijing Normal University, Beijing, 100875, China}
\affiliation{Institute for Frontiers in Astronomy and Astrophysics, Beijing Normal University, Beijing, 102206, China}

\affiliation{School of Fundamental Physics and Mathematical Sciences, Hangzhou Institute for Advanced Study, UCAS, Hangzhou, 310024, China}

\author{Jiliang Jing\,\orcidlink{0000-0002-2803-7900} }%\email[Contact author: ]{jljing@hunnu.edu.cn}
\affiliation{Department of Physics, Key Laboratory of Low Dimensional Quantum Structures and Quantum Control of Ministry of Education, Synergetic Innovation Center for Quantum Effects and Applications, Hunan Normal University, Changsha, 410081, Hunan, China}
\author{Sheng Long\,\orcidlink{0009-0009-0163-2724}}
\affiliation{School of Fundamental Physics and Mathematical Sciences, Hangzhou Institute for Advanced Study, UCAS, Hangzhou, 310024, China}
\date{\today}

\date{\today}

\begin{abstract}
We construct the complete spectrum of gravitational total-transmission modes (TTMs) of Kerr black holes and find four globally continuous families, $n_\infty=1,2,3,4$, each with two complex-conjugate frequency branches. Compared with the three-family classification of Cook and Lu, our global continuation resolves the mirror-related sectors of their $n=2$ family into separate families without introducing additional symmetry-unrelated roots. This organization distinguishes complex conjugation at fixed $m$ from the mirror symmetry connecting the $m$ and $-m$ spectra. The $n_\infty=3$ family approaches the Schwarzschild algebraically special frequency, whereas the $n_\infty=1,2,$ and $4$ families diverge as $\omega\propto a^{-4/3}$ along lower-half-plane directions $-150^\circ$, $-90^\circ$, and $-30^\circ$, respectively. High-precision data up to $\ell=32$ confirm the Cook--Lu small-spin asymptotics and reveal how the divergent branches are embedded in the global four-family spectrum, with a family-dependent relation between the spherical-limit angular labels and the large-$|a\omega|$ ordering. For axisymmetric perturbations, the $n_\infty=2$ and $n_\infty=3$ branches coalesce at genuine exceptional points, where the scattering spectral function develops a second-order zero and the TTM excitation factors are strongly enhanced, offering a different interpretation of the structure previously described by Cook and Lu as an overtone-multiplet splitting. Finally, we identify a previously unreported anomalous proximity between the $n_\infty=3$ TTM family and an unconventional Kerr quasinormal-mode sequence, with frequency separations reaching order $10^{-8}$ in the representative case studied.
\end{abstract}

\maketitle
%%%%%%%%%%%%%%%%%%%%%%%%%%%%%%%%%%%%%%%%%%%%%%%%%%%%%%%%%%%%%%%%
%\newpage
%\tableofcontents
%\newpage

\section{Introduction}\label{sec:introduction}
Black hole perturbation theory provides one of the principal frameworks for studying dynamics in strong gravitational fields, gravitational-wave (GW) propagation, and black hole spectroscopy~\cite{Sasaki:2003xr,Berti:2009kk}. On the Kerr background, scalar, electromagnetic, and gravitational perturbations are governed by the Teukolsky equation~\cite{Teukolsky:1973ha,Teukolsky:1974yv}.
After separation of the temporal, azimuthal, radial, and angular dependences, the problem reduces to a coupled angular eigenvalue problem and a radial scattering problem.
The radial Teukolsky equation is non-Hermitian and supports several inequivalent spectral problems, depending on the boundary conditions imposed at the event horizon and at spatial infinity~\cite{Berti:2025hly}.

The most familiar spectrum~\cite{Berti:2009kk,Berti:2025hly,BertiWeb,Cook:2014cta,Chen:2025sbz,ChenQNM} is that of quasinormal modes (QNMs).
QNMs satisfy  purely ingoing boundary conditions at the event horizon and purely outgoing boundary conditions at spatial infinity.
Their complex frequencies determine the characteristic oscillation and damping timescales of the binary black hole ringdown.
QNMs therefore play a central role in black hole spectroscopy, GW modeling, and tests of general relativity~\cite{Dreyer:2003bv, Berti:2005ys, Cardoso:2016ryw, Berti:2018vdi, Isi:2019aib, Isi:2021iql, Capano:2021etf,wanghe}.
The Kerr QNM spectrum has been studied using a variety of analytical and numerical techniques, including continued fractions, spectral decompositions, direct integration, and confluent Heun representations~\cite{Berti:2009kk,Berti:2025hly,Cook:2014cta,Chen:2025sbz}.

The radial Teukolsky equation also admits a distinct class of discrete solutions. TTMs are frequencies for which a reflection amplitude in the radial scattering problem vanishes~\cite{Wald:1973wwa,Chandrasekhar:1984mgh}.
A perturbation incident from one side of the effective potential is then transmitted to the other side without a reflected component. There are two types of TTMs. Left total-transmission modes, denoted by
$\mathrm{TTM}_\mathrm{L}$, are outgoing at both the event horizon and spatial infinity.
Right total-transmission modes, denoted by $\mathrm{TTM}_\mathrm{R}$, are ingoing at both boundaries.
These correspond to different radial boundary-value problems, although their frequency spectra coincide for vacuum Kerr perturbations.
The corresponding radial solutions and their analytic representations are nevertheless distinct~\cite{Chen:2025sbz}.

The algebraically special mode provides the historical origin of Kerr TTMs~\cite{Wald:1973wwa,Chandrasekhar:1984mgh}.
For a Schwarzschild black hole, the algebraically special frequency is purely imaginary and can be written as
\cite{Couch:1973zc}
\begin{equation}
 \omega_{\ell}^{\mathrm{AS}}
 =
 \pm i\,
 \frac{(\ell-1)\ell(\ell+1)(\ell+2)}{12},
 \label{eq:ASmode}
\end{equation}
where $\ell$ is the angular harmonic index.
In Kerr spacetime, algebraically special solutions are associated with the vanishing of the
Starobinsky--Teukolsky constant (STC)~\cite{Teukolsky:1973ha,Teukolsky:1974yv,Wald:1973wwa,Chandrasekhar:1984mgh}.
This algebraic condition provides a useful criterion for identifying candidate TTM frequencies.
Its explicit form and its relation to the radial Teukolsky equation are given in Sec.~\ref{sec:method}.

Early numerical studies of Kerr algebraically special perturbations showed that the Schwarzschild algebraically special frequency does not generate a single simple family as the Kerr spin is increased
\cite{Chandrasekhar:1984mgh,Couch:1973zc}.
Depending on $\ell$ and $m$, a frequency sequence can acquire a nonzero real part or remain on the imaginary axis.
High-accuracy studies of Kerr TTMs subsequently revealed a richer
structure~\cite{Cook:2014cta,Cook:2016fge,Cook:2018ses,Cook:2022kbb,cook_2025_16801267,greg_cook_2026_19070504}.
In particular, algebraically special modes of Schwarzschild black holes can satisfy both QNM and $\mathrm{TTM}_\mathrm{L}$ boundary conditions~\cite{Chen:2025sbz}. These results also showed that the imaginary axis requires special care in numerical calculations: mode sequences can develop nonsmooth behavior, and their identification depends on the continuation of both the radial solution and the angular eigenvalue branch.

Beyond the Kerr TTM sequences that approach the Schwarzschild algebraically special frequency~\eqref{eq:ASmode} as $a\to0$, an important advance was the discovery of an additional $m=0$ branch of Kerr TTMs~\cite{Cook:2018ses}.
This branch has a singular small-$a$ behavior: along the Kerr sequence, its frequency satisfies $|\omega|\to\infty$ as $a\to0$, rather than approaching a finite Schwarzschild frequency. This divergent behavior is consistent with the STC. At exactly $a=0$, all terms containing $a\omega$ vanish and the condition reduces to the Schwarzschild problem with a finite frequency, whereas the divergent Kerr sequences arise in the singular limit in which $a$ becomes arbitrarily small while $|a\omega|$ remains large.

Cook and Lu subsequently constructed extensive numerical sequences of gravitational TTMs for $2\leq\ell\leq8$ and classified the known solutions into three families~\cite{Cook:2022kbb}, labeled by $n=0$, $n=1$, and $n=2$. The $n=0$ family consists of the sequences that approach the finite Schwarzschild frequency~\eqref{eq:ASmode}.
By contrast, the $n=1$ and $n=2$ families exhibit a singular small-$a$ limit: for nonzero $a$ sufficiently close to zero, their frequencies diverge with leading magnitude proportional to $a^{-4/3}$, rather than approaching the finite Schwarzschild frequency~\eqref{eq:ASmode}. These results demonstrated that the Kerr TTM spectrum is not exhausted by the finite Schwarzschild frequency~\eqref{eq:ASmode}.

Their numerical classification also led to an apparently nonuniform description of the symmetry between the frequency branches~\cite{Cook:2022kbb}.
It is useful first to recall the conventional Kerr QNM notation, in which $\omega^{+}_{\ell m n}$ and $\omega^{-}_{\ell m n}$ denote the positive and negative frequency branches associated with the same mode labels $(\ell,m,n)$, and satisfy
\begin{equation}
\omega^{-}_{\ell m n}=-\left(\omega^{+}_{\ell,-m,n}\right)^{*},
\label{eq:mirror-symmetry}
\end{equation}
so that the negative frequency branch is also the mirror counterpart of the positive frequency branch. In this setting, the second frequency branch and the mirror mode refer to the same solution branch.

Cook and Lu adopted the same formal relation~\eqref{eq:mirror-symmetry} for TTMs~\cite{Cook:2022kbb}. They used $\omega^{+}_{\ell m n}$ for the branch conventionally chosen to have a positive real part whenever possible, while $\omega^{-}_{\ell m n}$ denoted the corresponding mirror branch. The transformation~\eqref{eq:mirror-symmetry} reflects the frequency through the imaginary axis while simultaneously interchanging the spectra with $m$ and $-m$.
Although this notation is natural for QNMs, where the negative-frequency branch and the mirror branch coincide, its use for TTMs requires additional care because the complete TTM root structure contains an independent complex-conjugation relation at fixed $(\ell,m)$.

The STC possesses a complex-conjugation symmetry at fixed $(\ell,m)$: if $\omega$ is a root, then $\omega^{*}$ is also a root. Thus, the TTM roots at fixed $(\ell,m)$ occur in complex-conjugate pairs. This relation is distinct from \cref{eq:mirror-symmetry}, which connects the spectra at $m$ and $-m$ and reflects the corresponding frequency through the imaginary axis.
Therefore, unlike in the conventional QNM notation, a conjugate pair at fixed $(\ell,m)$ and a mirror pair connecting $m$ and $-m$ need not represent the same pair of TTM roots.

Within this convention, Cook and Lu reported that the $n=0$ and $n=1$ families satisfy the additional relation
\begin{equation}
\omega^{-}_{\ell m n}=\omega^{+}_{\ell m n},\qquad n\in\{0,1\},
\label{eq:degenerate-mirror}
\end{equation}
and consequently described their mirror modes as degenerate~\cite{Cook:2022kbb}.
In contrast, the $n=2$ family was described as possessing a distinct and nondegenerate set of mirror modes.
This results in an apparently nonuniform description of the overtone families, in which the $n=0$ and $n=1$ families are assigned a degenerate mirror structure, whereas the $n=2$ family is not.

In this work, we build on the confluent Heun (HeunC) formalism used in our previous work~\cite{Chen:2025sbz} to construct the gravitational TTM spectrum of Kerr black holes.
A global root search is performed at the extreme-Kerr limit, and the resulting branches are continuously traced toward smaller spin.
Enlarging the search region yields no additional roots.
The complete TTM spectrum is thereby organized into four continuously tracked families, $n_\infty\in\{1,2,3,4\}$.
This four-family classification differs from the Cook--Lu classification~\cite{Cook:2022kbb} in the global assignment of TTM roots among the families, rather than through the appearance of additional symmetry-unrelated roots.
The symmetry structure, angular labeling, Schwarzschild-limit behavior, small-spin asymptotics, exceptional-point structure, and relation to the unconventional Kerr QNM sequence are then examined within this global classification.

The remainder of this paper is organized as follows.
In \cref{sec:method}, we formulate the angular and radial Teukolsky equations
within the HeunC framework, derive the TTM conditions and
the associated excitation factors, and establish the exact
frequency symmetries of the complete TTM spectrum.
In \cref{sec:Result}, we first classify the extreme-Kerr roots into four
families and clarify their relation to the Cook--Lu classification.
We then trace the complete Kerr TTM spectrum across the spin range,
analyze the exceptional points and excitation-factor enhancement in
the axisymmetric sector, derive the small-$a$ asymptotic behavior of
the four families, and examine the anomalous spectral proximity
between TTMs and QNMs.
Section~\ref{sec:Conclusion} summarizes our main results.
Throughout this paper, we set $M=1$.

\section{Methodology}\label{sec:method}

In this section, we adapt the framework used in our previous work~\cite{Chen:2025sbz} to compute the complete TTM spectrum of Kerr black holes accurately and efficiently.
The precise correspondence between the Teukolsky master equation and the HeunC equation is first established.

%--------------------------------------------------------------------
\subsection{Teukolsky and HeunC equations}\label{subsec:teukolsky_heun}

For a perturbation field of spin weight $s\in\{0,\pm1,\pm2\}$ on the Kerr
background, the homogeneous Teukolsky equation~\cite{Teukolsky:1973ha}
is separated via the ansatz
\begin{equation}\label{eq:separationEq-Teuko}
  \psi(t,r,\theta,\phi)
  = \mathrm{e}^{-i\omega t}\mathrm{e}^{im\phi}S_{\ell m}(\Theta)R_{\ell m}(r),
  \quad \Theta \equiv \cos\theta ,
\end{equation}
where $\omega$ is the eigenfrequency and $m$ the azimuthal
number. This yields the angular Teukolsky equation (ATE)
\begin{equation}\label{eq:GFoATE}
  \left[
    \frac{\mathrm{d}}{\mathrm{d}\Theta}
    \!\left(\nabla\frac{\mathrm{d}}{\mathrm{d}\Theta}\right)
    + U(\Theta)
  \right]S_{\ell m} = 0 ,
\end{equation}
with $\nabla = (1-\Theta)(1+\Theta)$ and
\begin{equation}\label{eq:U_ATE_Kerr}
  U(\Theta) = a^2\omega^2\Theta^2
            - \frac{(m+s\Theta)^2}{1-\Theta^2}
            - 2a\omega s\Theta + s + {}_sA_{\ell m} ,
\end{equation}
and the radial Teukolsky equation (RTE)
\begin{equation}\label{eq:GFoRTE}
  \left[
    \Delta^{-s+1}\frac{\mathrm{d}}{\mathrm{d}r}
    \!\left(\Delta^{s+1}\frac{\mathrm{d}}{\mathrm{d}r}\right)
    + V(r)
  \right]R_{\ell m} = 0 ,
\end{equation}
with
\begin{equation}\label{eq:V_RTE_Kerr}
  V(r) = K^2 - isK\Delta' + (2isK' - {\lambda})\Delta .
\end{equation}
Here $\Delta = (r-r_-)(r-r_+)$, $r_\pm = M\pm\sqrt{M^2-a^2}$,
$K = (r^2+a^2)\omega - am$, and
\begin{equation}\label{eq:lambda_hat}
  {\lambda} = {}_sA_{\ell m}(a\omega) + a^2\omega^2 - 2am\omega .
\end{equation}
The angular eigenvalue satisfies
\begin{equation}\label{eq:A_lm_a0}
  {}_sA_{\ell m}(a=0) = \ell(\ell+1) - s(s+1) .
\end{equation}
The ATE \eqref{eq:GFoATE} determines ${}_sA_{\ell m}$ and hence the separation constant $\lambda$ through \cref{eq:lambda_hat}, while $\lambda$ enters the RTE~\eqref{eq:GFoRTE} and couples the angular and radial problems.

Each of \Cref{eq:GFoATE,eq:GFoRTE} possesses two regular singularities—at
$\Theta=\pm1$ and $r=r_\pm$, respectively—together with an irregular
singularity at infinity, and can be systematically reduced to the standard
form of the HeunC equation. This reduction proceeds via two
transformations. First, a M\"obius transformation of the independent variable,
\begin{equation}\label{eq:mtran0}
  x = \frac{\hat{a}y + \hat{b}}{\hat{c}y + \hat{d}} ,
\end{equation}
with $y=\Theta$ for the ATE \eqref{eq:GFoATE} and $y=r$ for the RTE \eqref{eq:GFoRTE}, where the constants
$\{\hat{a},\hat{b},\hat{c},\hat{d}\}$ map the physical singularities onto the singular points $x=0,1,\infty$ of the HeunC equation.
Second, an $S$-homotopic transformation is employed to factor out the singular behavior of the dependent variable,
\begin{align}
  S_{\ell m}(\Theta) &= S_\theta(x)\,\mathbf{H}(x), \label{eq:Sm-Sol}\\
  R_{\ell m}(r)      &= S_r(x)\,\mathbf{H}(x),   \label{eq:Rm-Sol}
\end{align}
where $\mathbf{H}(x)={\rm HeunC}(\alpha,\beta,\gamma,\delta,\eta;x)$ denotes a HeunC function. The angular and radial
prefactors are
\begin{align}
  S_\theta(\alpha, \beta,\gamma;x) &= (-x)^{\tfrac{1}{2}\beta}(1-x)^{\tfrac{1}{2}\gamma}
                \mathrm{e}^{\tfrac{1}{2}\alpha x}, \label{eq:S-homotopic-theta-d2}\\
  S_r(\alpha, \beta,\gamma;x)      &= (-x)^{\tfrac{1}{2}(\beta-s)}(1-x)^{\tfrac{1}{2}(\gamma-s)}
                \mathrm{e}^{\tfrac{1}{2}\alpha x}, \label{eq:S-homotopic-r}
\end{align}
differing only by the spin weight $s$ carried by the radial field.

\subsection{TTM conditions}
\label{sec:BC_RTE}
The asymptotic solutions~\cite{Teukolsky:1973ha,Teukolsky:1974yv} of the RTE~\eqref{eq:GFoRTE} can be expressed as
\begin{equation}\label{eq:RTE-asympt}
  R_{\ell m}(r) \to
  \begin{cases}
      Z_{\mathrm{in}}^{r_-} R_{\mathrm{in}}^{r_-} + Z_{\mathrm{up}}^{r_-} R_{\mathrm{up}}^{r_-},
    & r \to r_-, \\
    Z_{\mathrm{in}}^{r_+} R_{\mathrm{in}}^{r_+} + Z_{\mathrm{up}}^{r_+} R_{\mathrm{up}}^{r_+},
    & r \to r_+, \\
    Z_{\mathrm{in}}^{\infty} R_{\mathrm{in}}^{\infty}
    + Z_{\mathrm{up}}^{\infty} R_{\mathrm{up}}^{\infty},
    & r \to \infty ,
  \end{cases}
\end{equation}
and
\begin{align}
R_{\mathrm{in}}^{r_-} &= \Delta^{-s} \mathrm{e}^{-i P_- r^*},\quad R_{\mathrm{up}}^{r_-} = \mathrm{e}^{i P_- r^*};\\
R_{\mathrm{in}}^{r_+} &= \Delta^{-s} \mathrm{e}^{-i P_+ r^*},\quad R_{\mathrm{up}}^{r_+} = \mathrm{e}^{i P_+ r^*};\\
R_{\mathrm{in}}^{\infty} &= r^{-1} \mathrm{e}^{-i \omega r^*},\quad \,\, R_{\mathrm{up}}^{\infty} = r^{-1-2s} \mathrm{e}^{i \omega r^*};
\end{align}
where $P_{\pm} = \omega - m \Omega_{\pm}$ is the frequency measured in the frame co-rotating with the horizon, and $\Omega_{\pm} = \frac{a}{2M r_{\pm}}$ is the angular velocity of the inner/outer horizon.
Here, $Z_{\mathrm{in}}^{r_{\pm}}$ and $Z_{\mathrm{up}}^{r_{\pm}}$ are the complex amplitudes of the ingoing and outgoing waves near the inner/outer horizon, and $Z_{\mathrm{in}}^{\infty}$ and $Z_{\mathrm{up}}^{\infty}$ are the corresponding amplitudes near infinity.

Two types of complex-frequency spectra are distinguished by their boundary conditions~\cite{Berti:2009kk}:
\begin{description}
  \item[(1)] The standard QNMs are defined by purely ingoing waves at the outer horizon and purely outgoing waves at spatial infinity~\cite{Cook:2014cta}:
\begin{equation}\label{eq:QNM_Condition}
{R_{\ell m}} \to \left\{ {\begin{array}{*{20}{l}}
{Z_{{\rm{in}}}^{{r_ + }}R_{{\rm{in}}}^{{r_ + }}}&{r \to {r_ + },}\\
{Z_{{\rm{up}}}^\infty R_{{\rm{up}}}^\infty ,}&{r \to  + \infty .}
\end{array}} \right.
\end{equation}
The necessary and sufficient condition for the QNM spectrum is
\begin{equation}\label{eq:QNM_BC_exact}
Z_{{\rm{up}}}^{{r_ + }}=Z_{{\rm{in}}}^\infty  = 0.
\end{equation}

  \item[(2)] From the scattering-amplitude viewpoint, the TTM is defined by requiring a single wave type throughout the entire exterior domain $(r_+,\infty)$. It splits into two branches according to the wave propagation direction:
\begin{subequations}\label{eq:TTM_LR_BC}
\begin{align}
\text{TTM}_{\mathrm{L}}:\quad
& {R_{\ell m}} \to \left\{ {\begin{array}{*{20}{l}}
{Z_{{\rm{up}}}^{{r_ + }}R_{{\rm{up}}}^{{r_ + }}}&{r \to {r_ + },}\\
{Z_{{\rm{up}}}^\infty R_{{\rm{up}}}^\infty ,}&{r \to  + \infty .}
\end{array}} \right.
\label{eq:TTML_BC_scatt} \\
\text{TTM}_{\mathrm{R}}:\quad
& {R_{\ell m}} \to \left\{ {\begin{array}{*{20}{l}}
{Z_{{\rm{in}}}^{{r_ + }}R_{{\rm{in}}}^{{r_ + }}}&{r \to {r_ + },}\\
{Z_{{\rm{in}}}^\infty R_{{\rm{in}}}^\infty ,}&{r \to  + \infty .}
\end{array}} \right.
\label{eq:TTMR_BC_scatt}
\end{align}
\end{subequations}
Equations~\eqref{eq:TTM_LR_BC} constrain only the two asymptotic endpoints, requiring $Z_{{\rm{in}}}^{{r_ + }}=Z_{{\rm{in}}}^\infty  = 0$ for $\text{TTM}_{\mathrm{L}}$, and $Z_{{\rm{up}}}^{{r_ + }}=Z_{{\rm{up}}}^\infty  = 0$ for $\text{TTM}_{\mathrm{R}}$.
\end{description}
The physical TTM condition, however, demands a single wave type throughout the entire exterior domain $(r_+,\infty)$. Cook proved that the necessary and sufficient condition for this is the existence of a confluent Heun polynomial solution~\cite{Cook:2014cta}. The existence of such a solution, in turn, requires that the same single wave type be imposed at the inner horizon $r_-$ as well. The complete TTM boundary conditions therefore read
\begin{subequations}\label{eq:TTM_full_BC}
\begin{align}
\text{TTM}_{\mathrm{L}}:\quad
& {R_{\ell m}} \to \left\{ {\begin{array}{*{20}{l}}
{Z_{{\rm{up}}}^{{r_-}}R_{{\rm{up}}}^{{r_-}}}&{r \to {r_-},}\\
{Z_{{\rm{up}}}^{{r_ + }}R_{{\rm{up}}}^{{r_ + }}}&{r \to {r_ + },}\\
{Z_{{\rm{up}}}^\infty R_{{\rm{up}}}^\infty ,}&{r \to  + \infty .}
\end{array}} \right.
\\
\text{TTM}_{\mathrm{R}}:\quad
& {R_{\ell m}} \to \left\{ {\begin{array}{*{20}{l}}
{Z_{{\rm{in}}}^{{r_-}}R_{{\rm{in}}}^{{r_-}}}&{r \to {r_-},}\\
{Z_{{\rm{in}}}^{{r_ + }}R_{{\rm{in}}}^{{r_ + }}}&{r \to {r_ + },}\\
{Z_{{\rm{in}}}^\infty R_{{\rm{in}}}^\infty ,}&{r \to  + \infty .}
\end{array}} \right.
\end{align}
\end{subequations}
which enforce a single wave type across $(r_+,\infty)$ and are mathematically equivalent to the vanishing of the Starobinsky--Teukolsky constant (STC)~\cite{Wald:1973wwa,Chandrasekhar:1984mgh,Teukolsky:1974yv,Cook:2018ses,Cook:2022kbb}. For gravitational perturbations, the STC is given by
\begin{align}\label{eq:Starobinsky_const}
  {|{\bf{C}}_{\rm ST}|}^2=\lambdabar^2(\lambdabar&+2)^2+ 8\lambdabar{a}\omega\left(6({a}\omega+m) -5\lambdabar({a}\omega-m)\right)\nonumber\\
      &+ 144\omega^2\left(1+{a}^2({a}\omega-m)^2\right)=0,
\end{align}
where $\lambdabar$ is related to the separation constant $\lambda$ according to
\begin{subequations}\label{eq:lambdabar_def}
  \begin{align}
    &\lambdabar= \lambda,  \quad\quad & s=-2,  \label{eq:lambdabar_neg2}\\
    &\lambdabar=\lambda+2s, \quad &s=+2. \label{eq:lambdabar_pos2}
  \end{align}
\end{subequations}
Adopting the time dependence $\psi \sim \mathrm{e}^{-i\omega t}$, the $\mathrm{TTM}_{\mathrm{L}}$ solutions of $\psi_4$ carry spin weight $s=-2$,
while the $\mathrm{TTM}_{\mathrm{R}}$ solutions of $\psi_0$ carry spin weight $s=+2$.

In our previous work~\cite{Chen:2025sbz}, numerical TTM results obtained directly from the boundary conditions~\eqref{eq:TTM_LR_BC} were found to agree with those computed from the STC~\eqref{eq:Starobinsky_const}. For branches connected to a known TTM seed, such as the Schwarzschild algebraically special frequency~\eqref{eq:ASmode}, continuation in small steps of the rotation parameter $a$ can reliably track the solution while preserving the TTM boundary conditions. Similar continuation strategies have also been adopted in recent numerical studies of TTMs in other black-hole spacetimes~\cite{Zhou:2025xdo,Wu:2026hvf,Dong:2026zxy}. Without a reliable seed, however, a blind root search in the complex-frequency plane can produce spurious solutions or make the identification of genuine TTMs ambiguous. To determine the complete Kerr TTM spectrum in this work, we therefore use the STC~\eqref{eq:Starobinsky_const} as the primary spectral condition.
\subsection{Unified excitation factors}
\label{sec:excitation-factor}

The excitation factors of QNMs and TTMs can be formulated within a unified scattering framework.
The index ${\rm X}\in\{{\rm Q},{\rm L},{\rm R}\}$ distinguishes standard QNMs, ${\rm TTM}_{\rm L}$, and ${\rm TTM}_{\rm R}$, respectively.
For each problem, we introduce a dimensionless spectral function $\mathcal{F}_{\rm X}(\omega)$ defined by the ratio between the vanishing boundary amplitude and a nonvanishing reference amplitude, along with a normalized transmission amplitude $\mathcal{T}_{\rm X}(\omega)$:

\begin{subequations}
  \begin{align}
\mathcal{F}_{\rm Q}(\omega)
&=\frac{Z_{\rm in}^{\infty}}{Z_{\rm up}^{\infty}},
&
\mathcal{T}_{\rm Q}(\omega)
&=\frac{Z_{\rm in}^{r_+}}{Z_{\rm up}^{\infty}},
\label{eq:QNM-spectral-ratio}
\\
\mathcal{F}_{\rm R}(\omega)
&=\frac{Z_{\rm up}^{\infty}}{Z_{\rm in}^{\infty}},
&
\mathcal{T}_{\rm R}(\omega)
&=\frac{Z_{\rm in}^{r_+}}{Z_{\rm in}^{\infty}},
\label{eq:TTMR-spectral-ratio}
\\
\mathcal{F}_{\rm L}(\omega)
&=\frac{Z_{\rm in}^{r_+}}{Z_{\rm up}^{r_+}},
&
\mathcal{T}_{\rm L}(\omega)
&=\frac{Z_{\rm up}^{\infty}}{Z_{\rm up}^{r_+}}.\label{eq:TTML-spectral-ratio}
\end{align}
\end{subequations}
The eigenfrequencies $\omega_{\rm X}$ are universally determined by $\mathcal{F}_{\rm X}(\omega_{\rm X})=0$.
Formulating $\mathcal{F}_{\rm X}$ and $\mathcal{T}_{\rm X}$ as amplitude ratios ensures their invariance under any overall normalization of the radial solution.

We define the normalized spectral response function as
\begin{equation}
\mathcal{G}_{\rm X}(\omega)
=
\frac{\mathcal{T}_{\rm X}(\omega)}
{2k_{\rm X}(\omega)\mathcal{F}_{\rm X}(\omega)},
\label{eq:unified-spectral-response}
\end{equation}
where the characteristic wave numbers are
\begin{equation}
k_{\rm Q}=\omega,
\qquad
k_{\rm R}=\omega,
\qquad
k_{\rm L}=P_+=\omega-m\Omega_+.
\label{eq:characteristic-wavenumbers}
\end{equation}
For a simple root of $\mathcal{F}_{\rm X}$, the excitation factor is defined as the residue of $\mathcal{G}_{\rm X}(\omega)$ at $\omega = \omega_{\rm X}$,
\begin{equation}
\mathscr{B}_{\rm X}
\equiv
\underset{\omega=\omega_{\rm X}}{\operatorname{Res}}\,
\mathcal{G}_{\rm X}(\omega)
=
\left.
\frac{\mathcal{T}_{\rm X}(\omega)}
{2k_{\rm X}(\omega)\mathcal{F}'_{\rm X}(\omega)}
\right|_{\omega=\omega_{\rm X}}.
\label{eq:unified-excitation-residue}
\end{equation}
Equivalently, defining $\alpha_{\rm X} = i\,{\rm d}\mathcal{F}_{\rm X}/{\rm d}\omega$, we obtain
\begin{equation}\label{eq:unified-excitation-factor}
\mathscr{B}_{\rm X}
=
-\left.
\frac{\mathcal{T}_{\rm X}(\omega)}
{2ik_{\rm X}(\omega)\alpha_{\rm X}(\omega)}
\right|_{\omega=\omega_{\rm X}}.
\end{equation}
Here, $d\mathcal{F}_{\rm X}/d\omega$ denotes the total derivative at fixed $(a,\ell,m,s)$, accounting for the implicit frequency dependence of the angular eigenvalue ${}_sA_{\ell m}(a\omega)$.

For the axisymmetric modes ($m=0$) considered here, $P_+=\omega$, so that all three excitation factors share the same wave-number normalization $k_{\rm X}=\omega$. Nevertheless, $\mathscr{B}_{\rm Q}$, $\mathscr{B}_{\rm L}$, and $\mathscr{B}_{\rm R}$ remain distinct because they correspond to inequivalent radial boundary-value problems.

%%--------------------------------------------------------------------
\subsection{Properties of the STC}
\label{subsec:STC_properties}

Two structural properties of the STC are particularly useful for the numerical determination of TTMs: the emergence of roots at complex infinity in the limit $a\to0$, and the exact symmetry relations satisfied by the TTM frequencies. These properties provide useful prior information for identifying TTM roots in the complex-frequency plane.

\subsubsection{Complex-infinity roots}
\label{subsubsec:STC_infinity}
For finite frequencies, the Schwarzschild limit of the STC is obtained by taking $a=0$ while keeping $\omega$ fixed. Using \Cref{eq:A_lm_a0,eq:lambdabar_def}, the quantity $\lambdabar$ approaches the same Schwarzschild value
$\lambdabar_0=\ell(\ell+1)-2$
for both gravitational representations $s=\pm2$. The STC therefore reduces to
\begin{equation}
\left.
\left|{\bf C}_{\mathrm{ST}}\right|^2
\right|_{a=0}
=
\lambdabar_0^2(\lambdabar_0+2)^2
+144\omega^2
=0,
\label{eq:STC_Schwarzschild}
\end{equation}
which gives the algebraically special frequencies~\eqref{eq:ASmode}.

This finite-frequency limit does not exhaust the solutions of the Kerr STC~\eqref{eq:Starobinsky_const}. In particular, the limit $a\to0$ is singular for solutions whose frequencies diverge as the spin decreases. For such solutions, one cannot set $a=0$ at fixed $\omega$, because the product $a\omega$ need not vanish. Instead, the terms containing $a\omega$ remain relevant, and the frequency must be determined together with the large-$|a\omega|$ asymptotic behavior of the angular separation constant.

The $n=1$ and $n=2$ TTM families in the Cook--Lu classification provide examples of this singular behavior~\cite{Cook:2022kbb}. For nonzero $a$ sufficiently close to zero, their frequencies diverge with a fractional-power dependence on $a$ rather than approaching a finite Schwarzschild frequency. Their asymptotic expansions therefore cannot be obtained by expanding the STC at fixed $\omega$; they must instead be constructed together with the corresponding asymptotic branch of the angular separation constant. The same distinction is essential when searching for additional TTM families.

Consequently, numerical continuation of the TTM spectrum over the full range $0\leq a\leq1$ must treat the finite Schwarzschild limit and the singular behavior as $a\to0$ separately.
In the Cook--Lu classification~\cite{Cook:2022kbb}, the algebraically special frequencies in \cref{eq:STC_Schwarzschild} identify the $n=0$ family, whereas the singular families require the large-$|a\omega|$ behavior of the angular separation constant.

\subsubsection{Symmetry relations of the TTM spectrum}
\label{subsubsec:STC_symmetry}

We now establish two exact symmetry operations of the complete TTM spectrum and derive their special consequence for $m=0$. These relations follow directly from the analytic structure of the STC and the angular eigenvalue problem. The mirror transformation itself is the standard symmetry used by Cook and Lu~\cite{Cook:2022kbb}, but retaining it together with the independent complex-conjugation symmetry at fixed $m$ leads to a different classification of the complete TTM spectrum. These relations are derived below and verified numerically using both our HeunC method~\cite{ChenQNM,Chen:2025sbz} and an independent implementation of Cook's method~\cite{greg_cook_2026_19070504}; the numerical results are presented in \cref{sec:Result}.

\begin{enumerate}

\item Symmetry about the real axis at fixed $m$.

The angular eigenvalue satisfies the identity
\begin{equation}
\label{eq:angular_conjugation}
{}_sA_{\ell m}^{*}(\omega)
=
{}_sA_{\ell m}(\omega^{*}).
\end{equation}
Since $a$ and $m$ are real, \cref{eq:lambdabar_def} implies
\begin{equation}
\label{eq:lambda_conjugation}
\lambdabar(\omega^{*})
=
[\lambdabar(\omega)]^{*}.
\end{equation}
Applying complex conjugation to each term in \cref{eq:Starobinsky_const} therefore gives
\begin{equation}
\label{eq:STC_conjugation}
\left|{\bf C}_{\rm ST}\right|^{2}(\omega^{*})
=
\left[
\left|{\bf C}_{\rm ST}\right|^{2}(\omega)
\right]^{*}.
\end{equation}
Consequently, if $\omega$ is a root of the STC for fixed $(\ell,m)$, then $\omega^{*}$ is also a root for the same $(\ell,m)$. The complete TTM spectrum is therefore symmetric under reflection about the real axis.

To distinguish this conjugate pair from the mirror pair discussed below, we denote the two frequency branches by $\omega_{\mathrm{I}}$ and $\omega_{\mathrm{II}}$, continuing the notation introduced in our previous work~\cite{Chen:2025sbz}, with $\operatorname{Im}(\omega_{\mathrm{I}})<0$ and $\operatorname{Im}(\omega_{\mathrm{II}})>0$, respectively. In the numerical spectrum considered here, no TTM branch is found to cross the real axis, so this assignment provides an unambiguous labeling of the two conjugate roots. We therefore avoid the superscripts $\pm$, which may obscure the distinction between the two frequency branches at fixed $(\ell,m)$ and the independent mirror symmetry connecting the spectra at $m$ and $-m$. Our notation is defined by
\begin{equation}
\label{eq:sym_real_axis}
\omega_{\mathrm{II}}(\ell,m)
=
\omega_{\mathrm{I}}^{*}(\ell,m).
\end{equation}
Thus, the two frequency branches at fixed $(\ell,m)$ form a complex-conjugate pair.
For each family $n_\infty$, the two members $\omega_{\mathrm{I}}$ and $\omega_{\mathrm{II}}$ constitute one complex-conjugate pair. The four families therefore account for eight root branches in the full complex-frequency plane. The upper-half-plane roots are included in this algebraic count, although they correspond to temporally growing perturbations under the convention $\mathrm{e}^{-i\omega t}$ and are not part of the physical damped spectrum usually displayed.

\item Symmetry under $m\rightarrow -m$ about the imaginary axis.

In addition to \cref{eq:angular_conjugation}, the angular eigenvalue satisfies~\cite{Cook:2022kbb}
\begin{equation}
\label{eq:angular_m_symmetry}
{}_sA_{\ell,-m}(\omega)
=
{}_sA_{\ell m}(-\omega).
\end{equation}
Combining \Cref{eq:angular_conjugation,eq:angular_m_symmetry} gives
\begin{equation}
\label{eq:angular_m_conjugation}
{}_sA_{\ell,-m}(-\omega^{*})
=
\left[
{}_sA_{\ell m}(\omega)
\right]^{*}.
\end{equation}
Since $a$ and $m$ are real, the simultaneous transformation
\begin{equation}
\label{eq:mirror_transformation}
m\rightarrow -m,
\qquad
\omega\rightarrow -\omega^{*}
\end{equation}
also maps $\lambdabar$ into its complex conjugate through \cref{eq:lambdabar_def}. Substitution of \cref{eq:mirror_transformation} into \cref{eq:Starobinsky_const} therefore produces the complex conjugate of the original STC. Hence, if $\omega$ is a TTM root for $m$, then $-\omega^{*}$ is a TTM root for $-m$.

For either of the two frequency branches introduced above, we denote the corresponding mirror root by
\begin{equation}
\label{eq:sym_m_conjugation}
\omega_{\mathrm{I/II}}^{\mathrm{mir}}(\ell,m)
=
-\omega_{\mathrm{I/II}}^{*}(\ell,-m).
\end{equation}
Here the superscript $\mathrm{mir}$ denotes the mirror image under the simultaneous transformation $m\rightarrow-m$ and $\omega\rightarrow-\omega^{*}$, while the labels $\mathrm{I/II}$ continue to distinguish the two frequency branches related by complex conjugation. Since $\omega\rightarrow-\omega^{*}$ changes the sign of the real part while leaving the imaginary part unchanged, this transformation corresponds geometrically to reflection about the imaginary axis. Equation~\eqref{eq:sym_m_conjugation} therefore represents the usual mirror symmetry of the Teukolsky system. This mirror symmetry is distinct from the complex-conjugation symmetry in \cref{eq:sym_real_axis}: the former connects the spectra at $m$ and $-m$, whereas the latter relates the two frequency branches at fixed $m$.

\item Special fourfold symmetry for $m=0$.

For $m=0$, the transformation $m\rightarrow-m$ leaves $m$ unchanged. The two symmetry operations established above therefore both act within the same $m=0$ spectrum for a given $\ell$. Starting from a root $\omega_{\mathrm{I}}(\ell,0)$, reflection about the real axis gives
\begin{equation}
\omega_{\mathrm{II}}(\ell,0)
=
\omega_{\mathrm{I}}^{*}(\ell,0),
\end{equation}
while reflection about the imaginary axis gives
\begin{equation}
\omega_{\mathrm{I}}^{\mathrm{mir}}(\ell,0)
=
-\omega_{\mathrm{I}}^{*}(\ell,0).
\end{equation}
Applying both transformations gives the fourth root
\begin{equation}
\omega_{\mathrm{II}}^{\mathrm{mir}}(\ell,0)
=
-\omega_{\mathrm{I}}(\ell,0).
\end{equation}
Thus, a generic TTM root in the $m=0$ spectrum generates the four-element set
\begin{equation}
\left\{
\omega_{\rm I},
\omega_{\rm I}^{*},
-\omega_{\rm I}^{*},
-\omega_{\rm I}
\right\}.
\label{eq:m0FourfoldSet}
\end{equation}
Writing
\begin{equation}
\omega_{\rm I}=x+iy,
\end{equation}
the corresponding frequencies are
\begin{equation}
x+iy,
\quad
x-iy,
\quad
-x+iy,
\quad
-x-iy.
\label{eq:m0FourRoots}
\end{equation}
The complete $m=0$ TTM spectrum is therefore invariant under reflection about the real axis, reflection about the imaginary axis, and inversion through the origin. For a generic root with $x\neq0$ and $y\neq0$, all four frequencies in \cref{eq:m0FourRoots} are distinct. 
If the root lies on either coordinate axis, some members of the set coincide and the number of distinct frequencies is correspondingly reduced.
This fourfold orbit acts within the already identified eight-root set and does not generate additional roots. At coalescence points, the root count is understood with algebraic multiplicity.
\end{enumerate}
%%--------------------------------------------------------------------
\subsection{Angular eigenvalue and regularity condition}
\label{subsec:heunc_ATE}

Using the basis functions defined in \cref{eq:Sm-Sol}, the general solution of the ATE~\eqref{eq:GFoATE} is constructed as a linear combination of two linearly independent HeunC solutions:
\begin{align}\label{eq:GSol_ATE}
{{\mathbb{S}}_{\ell m}} &= {\mathfrak{D}_1}S_\theta(\alpha,\beta,\gamma;x){\rm{HeunC}}(\alpha,\beta,\gamma,\delta,\eta;x) \nonumber\\
&+{\mathfrak{D}_2}S_\theta(\alpha,-\beta,\gamma;x){\rm{HeunC}}(\alpha,-\beta,\gamma,\delta,\eta;x),
\end{align}
where $\mathfrak{D}_1$ and $\mathfrak{D}_2$ are arbitrary constants.

The ATE possesses regular singularities at the poles $\Theta=\pm1$. To construct solutions regular at each pole, we introduce the local coordinates
\begin{align}
x_{+1} &= \tfrac{1}{2}(1-\Theta) \quad \text{for } \Theta=+1 \text{ (north pole)}, \label{eq:x_plus}\\
x_{-1} &= \tfrac{1}{2}(1+\Theta) \quad \text{for } \Theta=-1 \text{ (south pole)}. \label{eq:x_minus}
\end{align}
Depending on the chosen coordinate, the HeunC parameters $(\alpha,\beta,\gamma,\delta,\eta)$ take distinct forms. For Kerr black holes, these parameter sets are
\begin{subequations}\label{eq:ATE_HCprameters_Kerr}
\begin{align}
&\alpha_{-1}=4a\omega,\quad\quad\quad\quad\; \alpha_{+1}=-4a\omega,\\
&\beta_{-1}=|m-s|,\quad\quad\quad\;\; \beta_{+1}=|m+s|,\\
&\gamma_{-1}=|m+s|,\quad\quad\quad \gamma_{+1}=|m-s|,\\
&\delta_{-1}=4sa\omega,\quad\quad\quad\quad \delta_{+1}=-4sa\omega,\\
&\eta_{-1}=\tfrac{1}{2}m^2-\tfrac{1}{2}s^2-s-2ma\omega-2sa\omega-{\lambda},\\
&\eta_{+1}=\tfrac{1}{2}m^2-\tfrac{1}{2}s^2-s-2ma\omega+2sa\omega-{\lambda}.
\end{align}
\end{subequations}
Substituting the HeunC parameters~\eqref{eq:ATE_HCprameters_Kerr} into the general solution~\eqref{eq:GSol_ATE} gives the corresponding local solutions around the two poles. We denote by $S_{\ell m}(x_{+1})$ the solution regular at the north pole and by $S_{\ell m}(x_{-1})$ the solution regular at the south pole, obtained using the parameter sets labeled by $+1$ and $-1$, respectively. Both are functions of the same angular variable $\Theta$ through the local coordinates $x_{+1}$ and $x_{-1}$ defined in \Cref{eq:x_plus,eq:x_minus}.

For a globally regular angular solution, these two locally regular solutions must represent the same solution in their common domain and therefore be linearly dependent. This condition is equivalent to the vanishing of their Wronskian~\cite{Chen:2025sbz}:
\begin{equation}\label{eq:Wronskian_def}
W_\theta=S_{\ell m}(x_{-1})\,\frac{\mathrm{d}}{\mathrm{d}\Theta}S_{\ell m}(x_{+1})-S_{\ell m}(x_{+1})\,\frac{\mathrm{d}}{\mathrm{d}\Theta}S_{\ell m}(x_{-1}).
\end{equation}

We evaluate the Wronskian at the equatorial plane $\Theta=0$, where $x_{+1}=x_{-1}=1/2$. The common S-homotopic prefactors cancel from the condition $W_\theta=0$. Using $dx_{+1}/d\Theta=-1/2$ and $dx_{-1}/d\Theta=1/2$ then gives
\begin{widetext}
\begin{equation}\label{eq:ATE_Wronskian}
\begin{split}
W_\theta &\propto \mathrm{HeunC}(\alpha_{-1},\beta_{-1},\gamma_{-1},\delta_{-1},\eta_{-1};\tfrac{1}{2})\,\mathrm{HeunC}'(\alpha_{+1},\beta_{+1},\gamma_{+1},\delta_{+1},\eta_{+1};\tfrac{1}{2})\\
&\quad+\mathrm{HeunC}(\alpha_{+1},\beta_{+1},\gamma_{+1},\delta_{+1},\eta_{+1};\tfrac{1}{2})\,\mathrm{HeunC}'(\alpha_{-1},\beta_{-1},\gamma_{-1},\delta_{-1},\eta_{-1};\tfrac{1}{2}),
\end{split}
\end{equation}
\end{widetext}
where $\mathrm{HeunC}'(\cdots;x)$ denotes the derivative with respect to $x$.
The condition $W_\theta=0$ determines the allowed values of the separation constant $\lambda$, from which the corresponding angular eigenvalue ${}_sA_{\ell m}(a\omega)$ follows through \cref{eq:lambda_hat}. The numerical procedure used to extract these angular eigenvalues is described in detail in our previous work~\cite{Chen:2025sbz}.

\begin{figure*}[htbp]
\centering
\includegraphics[width=7in]{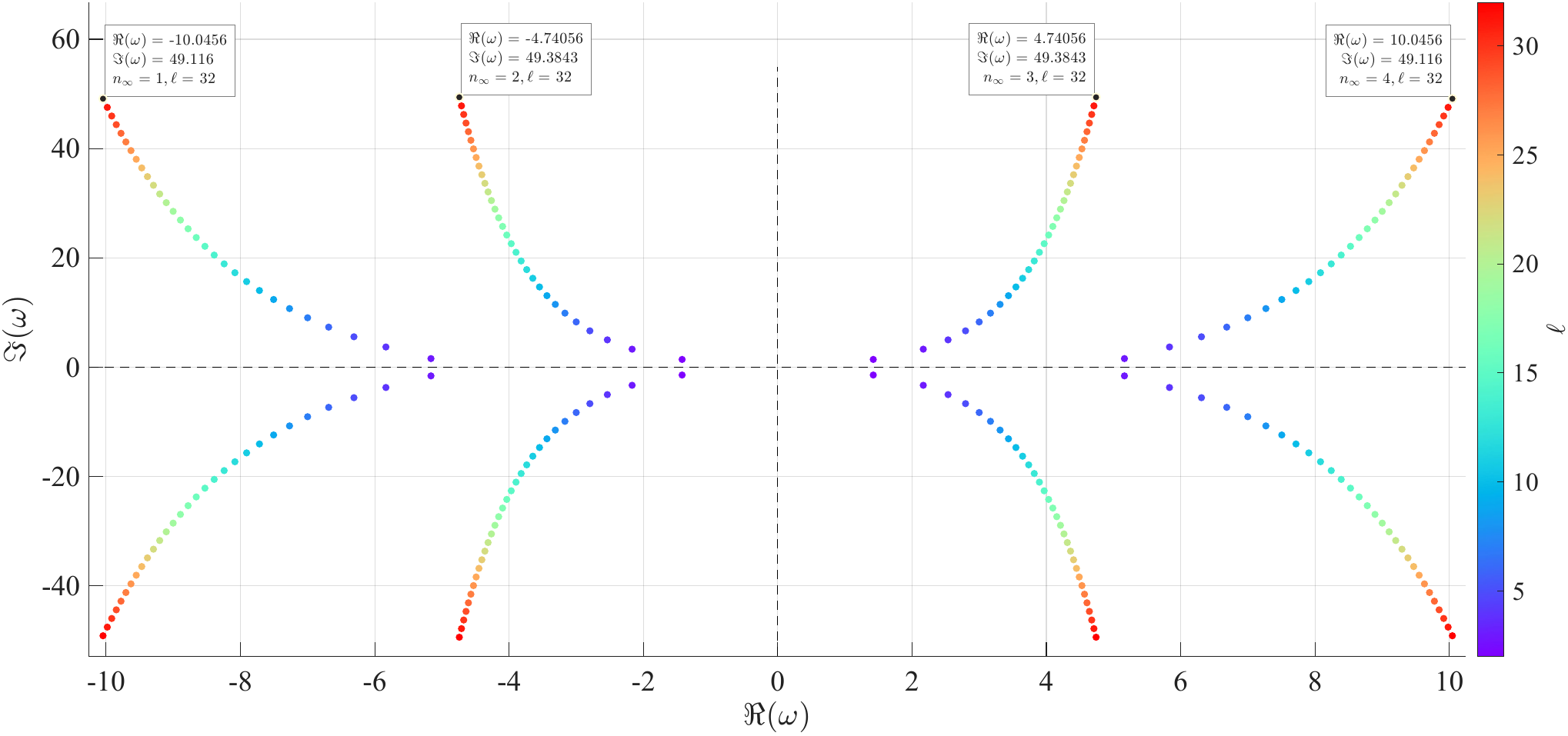}
\caption{Extreme-Kerr TTM frequencies for $m=0$ and $2\leq\ell\leq32$.}
\label{fig:EKerrm0}
\end{figure*}
\begin{figure*}[htbp]
\centering
\subfloat[$n_{\infty}=1$]{\includegraphics[width=3.4in]{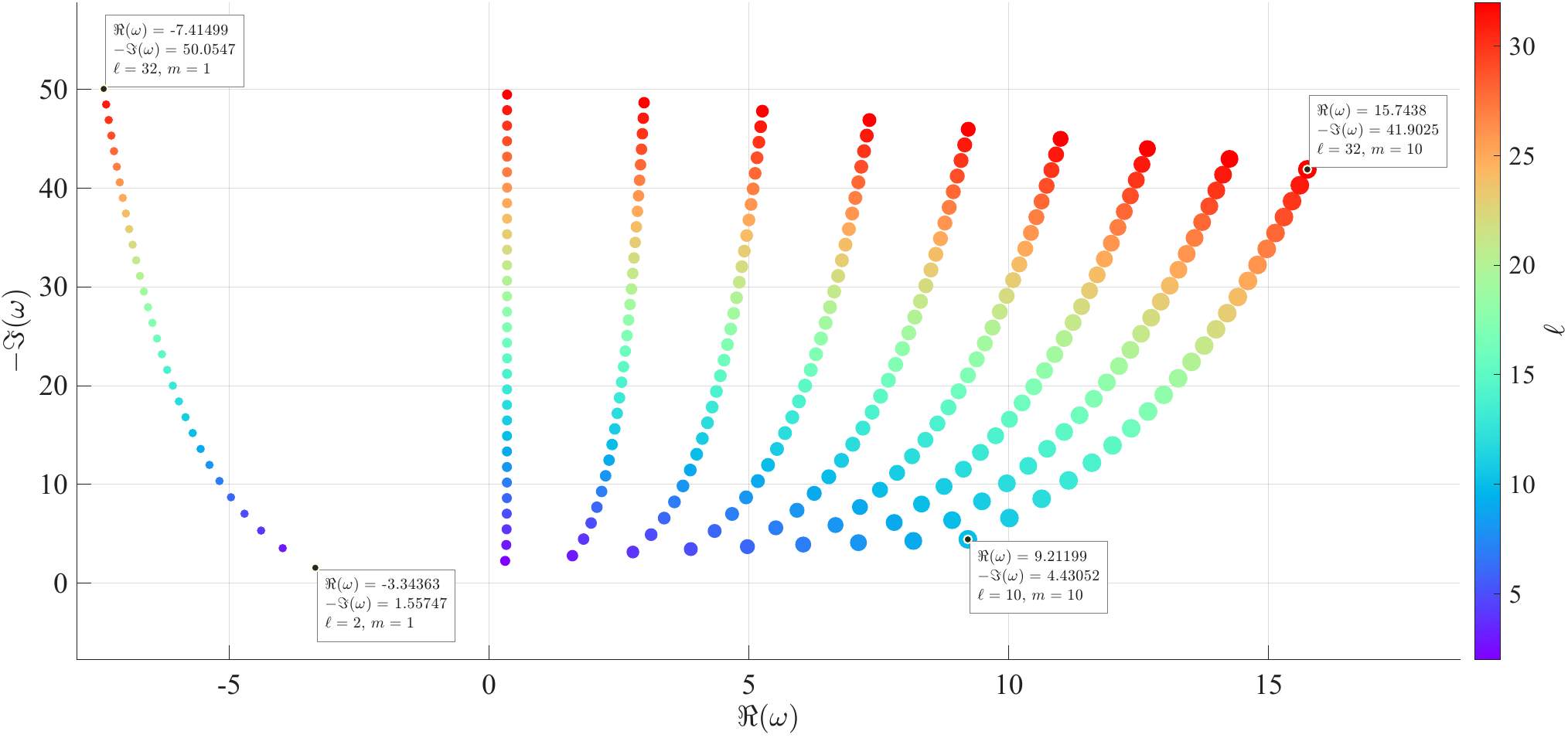}}
\subfloat[$n_{\infty}=2$]{\includegraphics[width=3.4in]{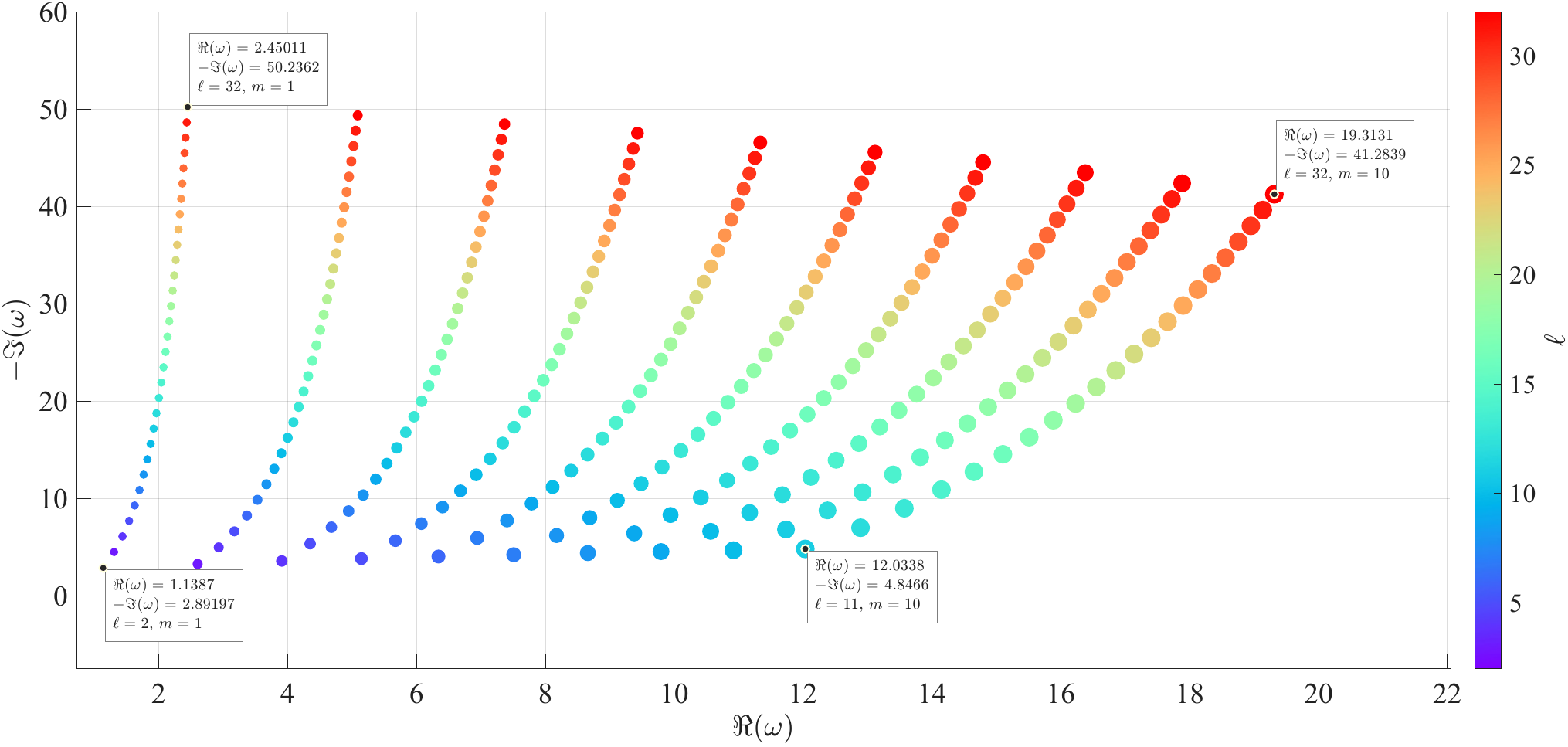}}\\
\subfloat[$n_{\infty}=3$]{\includegraphics[width=3.4in]{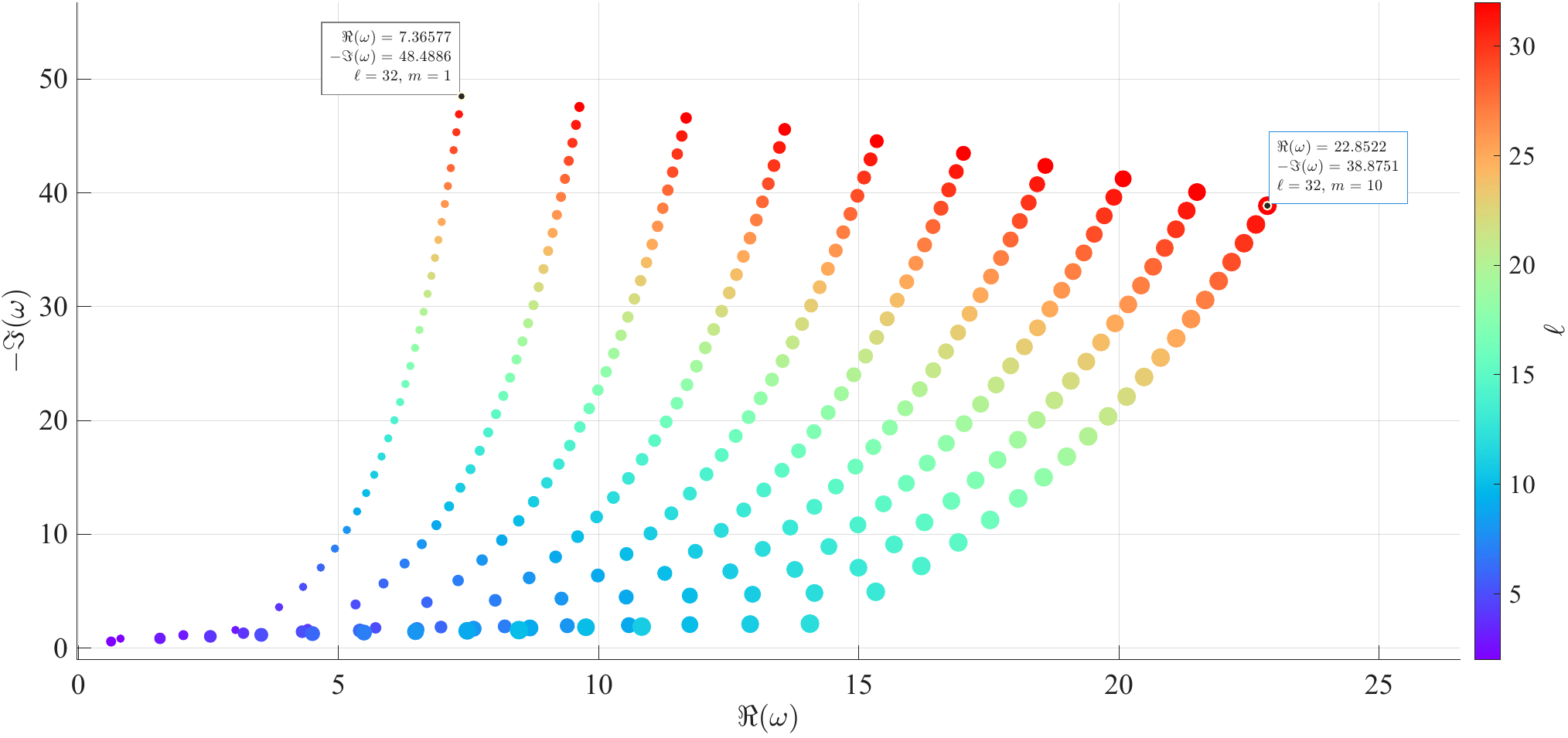}}
\subfloat[$n_{\infty}=4$]{\includegraphics[width=3.4in]{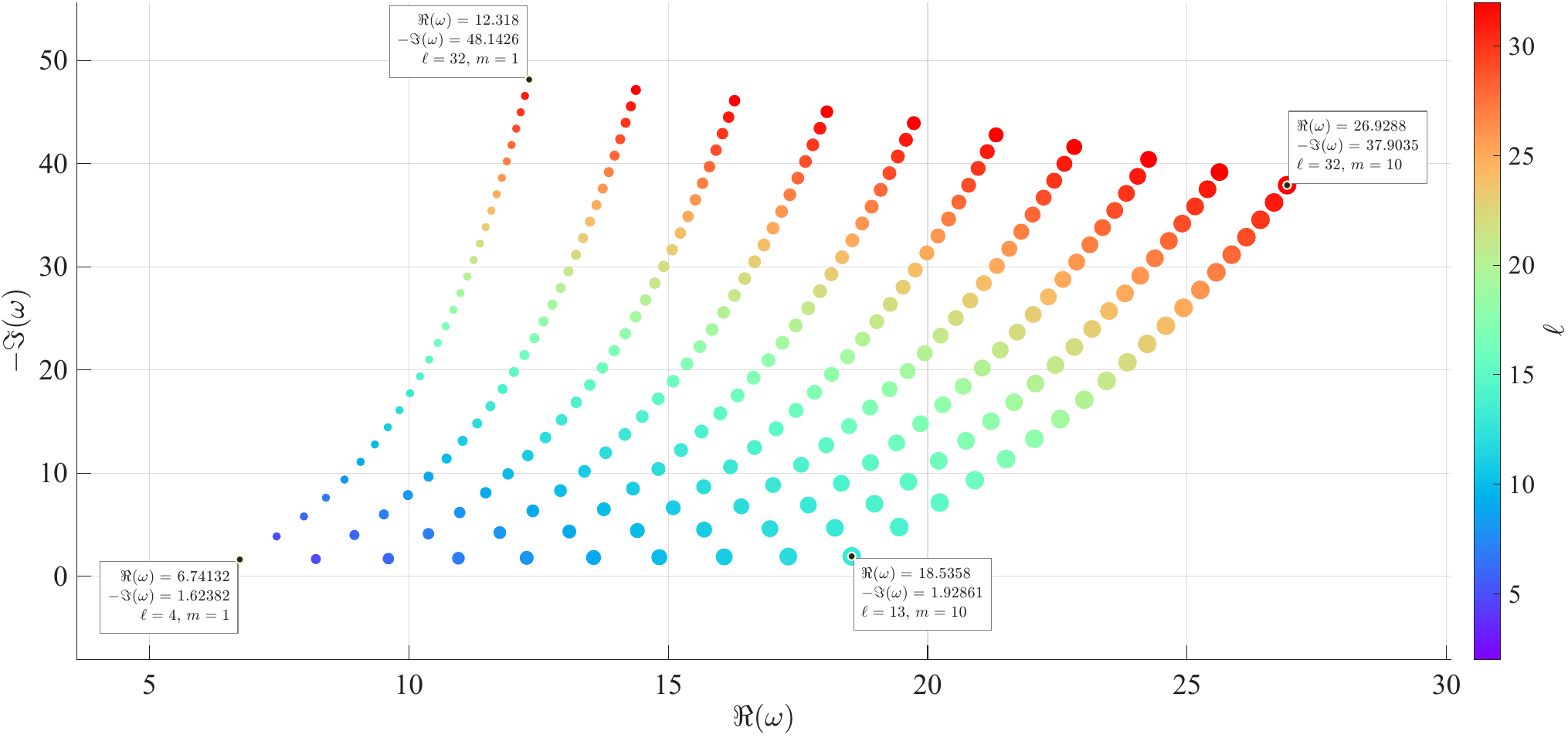}}
\caption{Extreme-Kerr TTM frequencies in the $\omega_{\rm I}$ branch for $1\leq m\leq10$ and $2\leq\ell\leq32$, grouped by $n_\infty$. The vertical axis shows the negative imaginary part of the frequency. The marker size encodes the value of $m$, and the color encodes the value of $\ell$.}
\label{fig:EKerrm1_m10}
\end{figure*}
\begin{figure}[htbp]
\centering
\includegraphics[width=3.4in]{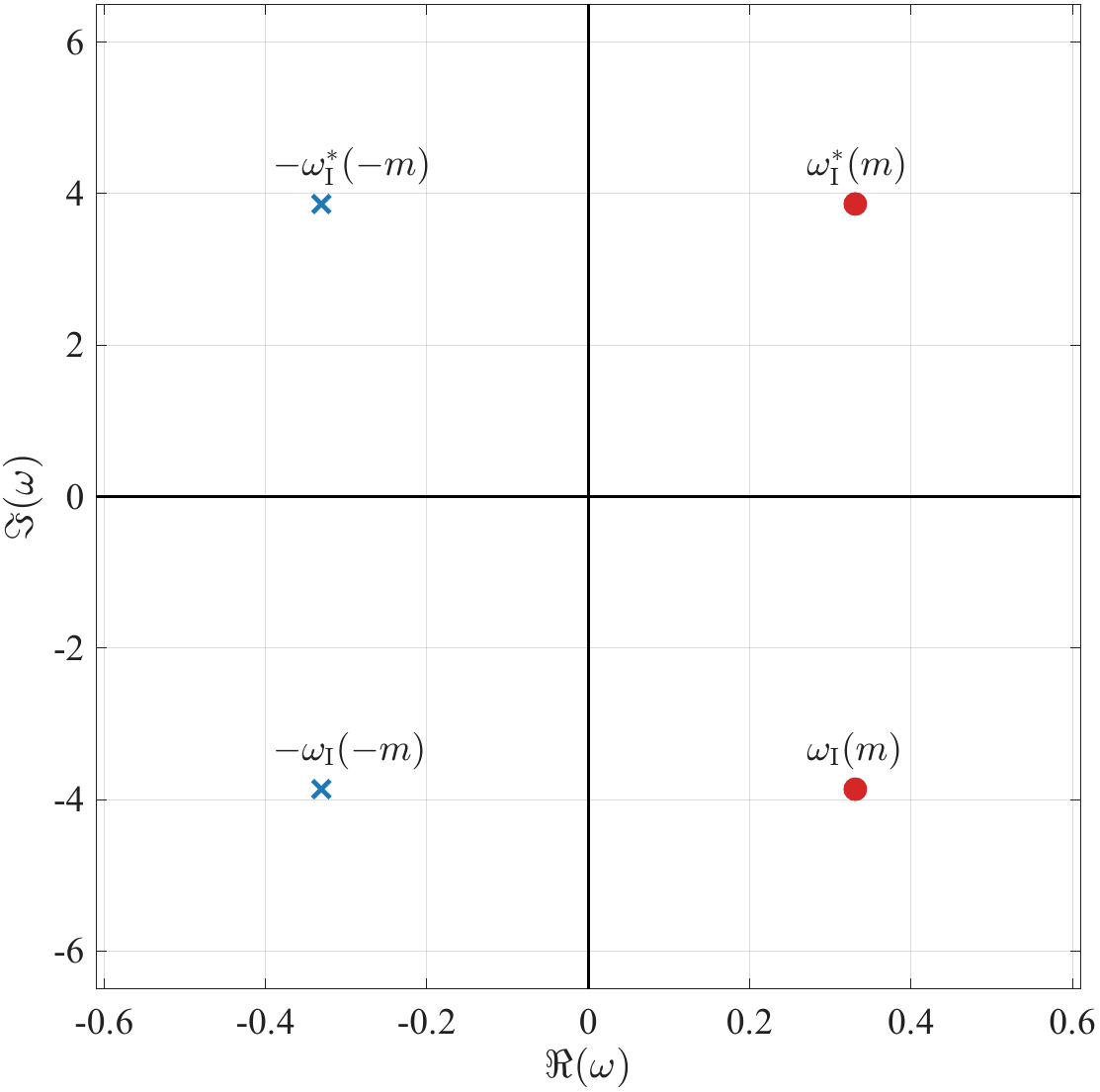}
\caption{Four symmetry-related extreme-Kerr TTM roots for $n_\infty=1$ and $\ell=m=2$ at $a=1$.}
\label{fig:l2m2_Ekerr}
\end{figure}

\begin{table*}[htbp]
\centering
\caption{Residuals used to test the symmetry relations for the lowest available $\ell$ in each of the four $n_\infty$ families with $m=2$. Each entry gives the residuals of the STC~\eqref{eq:Starobinsky_const} and the angular regularity condition~\eqref{eq:ATE_Wronskian}, respectively.}
\label{tab:first_ell}
\begin{tabular}{ccccc}
\toprule
$n_\infty$ & {1} & {2} & {3} & {4} \\
\midrule
$\ell$ & {2} & {3} & {2} & {5} \\
\midrule
${\omega _{\rm{I}}}(m)$ & $\{2.1\times 10^{-92}, 1.3\times 10^{-78}\}$ & $\{9.0\times 10^{-80}, 4.8\times 10^{-83}\}$ & $\{6.8\times 10^{-78}, 1.2\times 10^{-80}\}$ & $\{1.4\times 10^{-81}, 8.8\times 10^{-86}\}$ \\
$\omega _{\rm{I}}^*(m)$ & $\{2.1\times 10^{-92}, 1.3\times 10^{-78}\}$ & $\{9.0\times 10^{-80}, 4.8\times 10^{-83}\}$ & $\{6.8\times 10^{-78}, 1.2\times 10^{-80}\}$ & $\{1.4\times 10^{-81}, 8.8\times 10^{-86}\}$ \\
$-\omega _{\rm{I}}^*(m)$ & $\{1.7\times 10^{4}, 5.2\times 10^{-79}\}$ & $\{6.4\times 10^{5}, 1.5\times 10^{-79}\}$ & $\{3.2\times 10^{3}, 3.0\times 10^{-82}\}$ & $\{9.5\times 10^{7}, 1.3\times 10^{-75}\}$ \\
$-\omega _{\rm{I}}^*(-m)$ & $\{1.1\times 10^{-74}, 2.1\times 10^{-79}\}$ & $\{1.4\times 10^{-73}, 3.6\times 10^{-77}\}$ & $\{6.9\times 10^{-78}, 3.4\times 10^{-83}\}$ & $\{9.8\times 10^{-70}, 6.3\times 10^{-75}\}$ \\
$-\omega _{\rm{I}}(-m)$ & $\{1.1\times 10^{-74}, 2.1\times 10^{-79}\}$ & $\{1.4\times 10^{-73}, 3.6\times 10^{-77}\}$ & $\{6.9\times 10^{-78}, 3.4\times 10^{-83}\}$ & $\{9.8\times 10^{-70}, 6.3\times 10^{-75}\}$ \\
\bottomrule
\end{tabular}
\end{table*}

\section{Numerical Results}\label{sec:Result}
The complete Kerr TTM spectrum is computed using a high-precision extension of the numerical framework developed in our previous work~\cite{Chen:2025sbz,ChenQNM}. The calculations are carried out with precision substantially beyond standard floating-point accuracy, and the corresponding data sets and numerical programs are available from the online repository~\cite{ChenTTM}. The numerical accuracy is assessed primarily through the residuals of the STC~\eqref{eq:Starobinsky_const} and the angular regularity condition~\eqref{eq:ATE_Wronskian}, which are typically smaller than $10^{-50}$ for the solutions reported below. As an independent cross-check, selected solutions have also been verified using the scattering-amplitude method developed in our previous work~\cite{Chen:2025sbz}, confirming the vanishing of the corresponding reflection amplitude; the relevant formulas are summarized in Appendix~\ref{sec:TTM-Scatter-Amp}.

A complete search for TTMs cannot in general be initialized from the Schwarzschild limit in the same manner as a conventional QNM continuation. At $a=0$, the available finite frequency seed is the Schwarzschild frequency~\eqref{eq:ASmode}, whereas the additional TTM families discussed in \cref{subsubsec:STC_infinity} approach complex infinity as $a\to0$.
A continuation starting only from $a=0$ would therefore recover the branches connected to the finite Schwarzschild solution but can miss the singular families. To avoid this limitation, the search is initiated from the extreme-Kerr limit $a=1$, where a broad root search is first performed, and the resulting branches are subsequently continued toward smaller values of $a$.

\subsection{Extreme-Kerr TTM}

At $a=1$, a dense root search is performed over the region $-100\leq\Re(\omega)\leq100$ and $-50\leq\Im(\omega)\leq50$. The search is repeated over larger regions of the complex-frequency plane, yielding no additional roots.
For $m=0$, \cref{fig:EKerrm0} shows eight root sequences in the full complex-frequency plane, forming four complex-conjugate pairs. For $1\leq m\leq10$, \cref{fig:EKerrm1_m10} shows the four lower-half-plane sequences for each fixed $m$; their upper-half-plane conjugates follow from the symmetry established in \cref{subsubsec:STC_symmetry}. Together, these numerical results and the conjugation symmetry establish the same eight-sequence structure for each fixed $m$ in the investigated range, while the negative-$m$ spectra follow from the mirror transformation.
A family label $n_\infty\in\{1,2,3,4\}$ is therefore introduced to classify the four independent pairs.
At the extreme-Kerr limit, they are ordered from left to right according to increasing $\Re(\omega)$, with the two conjugate members of each pair assigned the same value of $n_\infty$.
The label $n_\infty$ is intended to classify the complete TTM spectrum rather than to extend directly the conventional QNM overtone index. Cook and Lu~\cite{Cook:2022kbb} retained the QNM-inspired notation $n$ and classified the previously known TTM solutions into the $n=0$, $n=1$, and $n=2$ families according to their continuation and small-$a$ behavior. Unlike the QNM spectrum, however, the TTM families do not share a common finite frequency limit at $a=0$, so the Schwarzschild limit does not provide a common reference point for ordering all TTM branches. By contrast, the extreme-Kerr spectrum provides a common reference configuration for the roots identified here, and the label $n_\infty$ allows each family to be tracked continuously as $a$ decreases.

\subsubsection{Symmetry structure of the extreme-Kerr spectrum}
Figure~\ref{fig:EKerrm0} shows the $m=0$ roots for $2\leq\ell\leq32$. The numerical spectrum exhibits the fourfold symmetry derived in \cref{subsubsec:STC_symmetry}: the roots form complex-conjugate pairs about the real axis and mirror pairs about the imaginary axis. Under the mirror symmetry, the $n_\infty=1$ and $n_\infty=4$ families are mapped into each other, while $n_\infty=2$ and $n_\infty=3$ form the second mirror pair. The complete $m=0$ spectrum is therefore symmetric about both coordinate axes and under inversion through the origin.
The same symmetry structure persists for $m\neq0$, with the important distinction that reflection about the imaginary axis simultaneously interchanges the $m$ and $-m$ sectors.

Figure~\ref{fig:l2m2_Ekerr} illustrates this structure for $n_\infty=1$ and $\ell=m=2$. Starting from $\omega_{\rm I}(m)$, complex conjugation gives the second root $\omega_{\rm I}^{*}(m)=\omega_{\rm II}(m)$ at the same $m$, while the corresponding mirror branch is $-\omega_{\rm I}^{*}(-m)$.
Its complex conjugate produces the fourth symmetry-related root $-\omega_{\rm I}(-m)$.
The four frequencies therefore form the rectangular pattern expected from the two independent symmetry operations derived in \cref{subsubsec:STC_symmetry}.
\cref{tab:first_ell} provides a quantitative test of these symmetry relations for $m=2$. For each $n_\infty$, the residuals of \Cref{eq:Starobinsky_const,eq:ATE_Wronskian} are reported for the lowest available $\ell$.
The residuals of $\omega_{\rm I}(m)$ and $\omega_{\rm I}^{*}(m)$ confirm the complex-conjugation symmetry at fixed $m$. The comparison between $-\omega_{\rm I}^{*}(m)$ and $-\omega_{\rm I}^{*}(-m)$ is especially informative: reflection about the imaginary axis at fixed $m$ does not in general produce another TTM, whereas the simultaneous transformation $m\rightarrow-m$ and $\omega\rightarrow-\omega^{*}$ does. The residuals therefore provide a direct numerical distinction between the complex-conjugation symmetry at fixed $m$ and the mirror symmetry connecting the $m$ and $-m$ spectra.

In \cref{fig:EKerrm1_m10}, only the lower-half-plane branch $\omega_{\rm I}$ with $\mathrm{Im}(\omega)<0$ is plotted for clarity; the conjugate $\omega_{\rm II}$ spectrum follows from reflection across the real axis, and the negative-$m$ modes are obtained via the mirror transformation. In this and subsequent figures, $-\mathrm{Im}(\omega)$ is plotted on the vertical axis so that frequencies with $\mathrm{Im}(\omega)<0$ appear above the horizontal axis, consistent with the standard presentation of Kerr QNM spectra.

\subsubsection{Polar-index $\ell$ assignment}

For the $n_\infty=3$ family, the polar index $\ell$ is unambiguous because the TTM sequence approaches the finite Schwarzschild algebraically special frequency as $a\to0$, and hence $c=a\omega\to0$. The associated angular eigensolution therefore connects directly to the spherical limit, where $\ell=\ell_{\min}+L$ with $\ell_{\min}=\max(|m|,|s|)$ and $L=0,1,2,\ldots$.
For the remaining $n_\infty=1$, $n_\infty=2$, and $n_\infty=4$ families, the physical TTM sequences have singular small-$a$ limits and do not approach $c=0$. Cook and Wang~\cite{Cook:2026gpm} emphasized that the polar index is a label of an angular eigensolution rather than a parameter of the angular equation, and that the ordering of complex angular eigenvalues can change along a continuous path in the complex $c$ plane. Motivated by this distinction, we assign $\ell$ by continuously tracing each numerically identified angular eigensolution from its target value $c_0=a_0\omega_0$ to the spherical limit $c=0$. The same branch is identified through overlaps of neighboring angular eigenvectors rather than by instantaneous eigenvalue ordering, and the spherical branch reached at $c=0$ fixes the polar index $\ell$ of the TTM.
This spherical-limit assignment is distinct from the asymptotic labeling previously used for singular TTM sequences~\cite{Cook:2022kbb}. As shown below, the two labels are related but need not coincide, and their difference depends on the TTM family.

The densely resolved and consistently labeled extreme-Kerr spectrum also provides a useful numerical basis for extending the TTM calculation to larger values of $m$ and $\ell$. A blind search over a large region of the complex-frequency plane becomes increasingly inefficient at high angular indices and, more importantly, an isolated root obtained in such a search does not by itself determine its polar index $\ell$ or its family assignment $n_\infty$. The continuation data obtained here avoid this ambiguity by providing neighboring, already identified TTM solutions from which initial guesses for new values of $m$ and $\ell$ can be constructed by local extrapolation or fitting and subsequently refined by the full TTM root solver. This extrapolation-guided strategy substantially reduces the need for repeated large-domain root searches and preserves the spectral identification of the resulting solutions during the iterative extension of the data set. Existing numerical data from Cook and Lu~\cite{Cook:2022kbb} provide an important reference for this purpose, but cover a substantially smaller range in $\ell$ and employ a different angular-mode labeling for the singular TTM branches.
The present high-$\ell$ data set, together with the spherical-limit assignment described above, therefore provides a complementary starting point for systematic extensions toward larger $m$ and $\ell$.

\subsubsection{Relation to the Cook--Lu classification}
The classification adopted here, illustrated in \Cref{fig:EKerrm0,fig:EKerrm1_m10}, differs from the classification of Cook and Lu~\cite{Cook:2022kbb}. Figure~\ref{fig:EKerrm0} makes explicit the mirror pairing of the $n_\infty$ families, with $n_\infty=1$ mapped to $n_\infty=4$ and $n_\infty=2$ mapped to $n_\infty=3$, while \cref{fig:EKerrm1_m10} displays the $\omega_{\rm I}$ branch for positive $m$, from which the conjugate branch and the negative-$m$ spectrum are generated by the two symmetry operations.
Cook and Lu~\cite{Cook:2022kbb} instead organize the spectrum in terms of the mirror branches $\omega^\pm$, related by $\omega^-_{\ell mn}=-(\omega^+_{\ell,-m,n})^*$. For their $n=0$ and $n=1$ families, they found $\omega^-_{\ell mn}=\omega^+_{\ell mn}$ and therefore retained only the $m\geq0$ sector of the $\omega^+$ solutions, treating the omitted $m<0$ sector as redundant mirror data. By contrast, this degeneracy is absent for their $n=2$ family, for which all values of $m$ were retained.
For the complete TTM spectrum, the fixed-$m$ conjugate branches and the mirror transformation are kept distinct.
With this classification, part of the negative-$m$ sector assigned to the $n=2$ family in Ref.~\cite{Cook:2022kbb} is instead identified with the $m<0$ continuation of the $n_\infty=1$ family.

The resulting difference therefore lies primarily in the global assignment of roots to the TTM families rather than in the underlying root set.
The enlarged searches recover the same eight-sequence root set, organized here into four conjugate pairs.
This numerical completeness is distinct from the comparison of family labels: Cook and Lu group the damped roots into three families, with their $n=2$ family containing distinct mirror-related sectors~\cite{Cook:2022kbb}.
Our convention assigns these sectors to separate global continuations, $n_\infty=1$ and $4$, while the imaginary-direction divergent and finite continuations are associated with $n_\infty=2$ and $3$, respectively.
Thus, the four-family classification is not obtained by adding previously omitted conjugate roots to the Cook--Lu damped spectrum, but represents a different global organization of the root branches, where the conjugate partners are retained explicitly to describe the full complex-frequency root structure. No additional symmetry-unrelated TTM roots are implied by this reclassification.

\begin{figure*}[htbp]
	\centering
	\subfloat[$n_{\infty}=2$]{\includegraphics[width=3.5in]{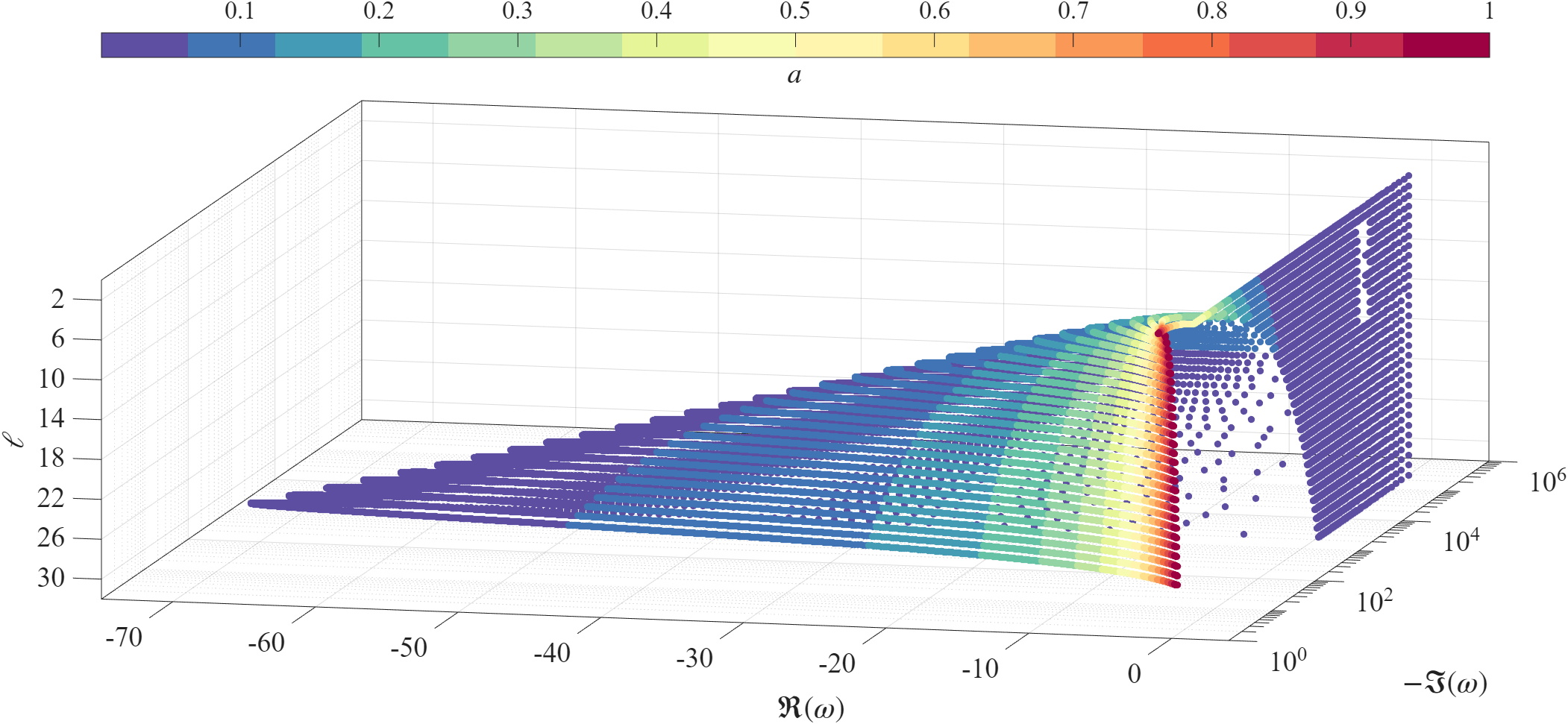}}\,\,
	\subfloat[$n_{\infty}=3$]{\includegraphics[width=3.36in]{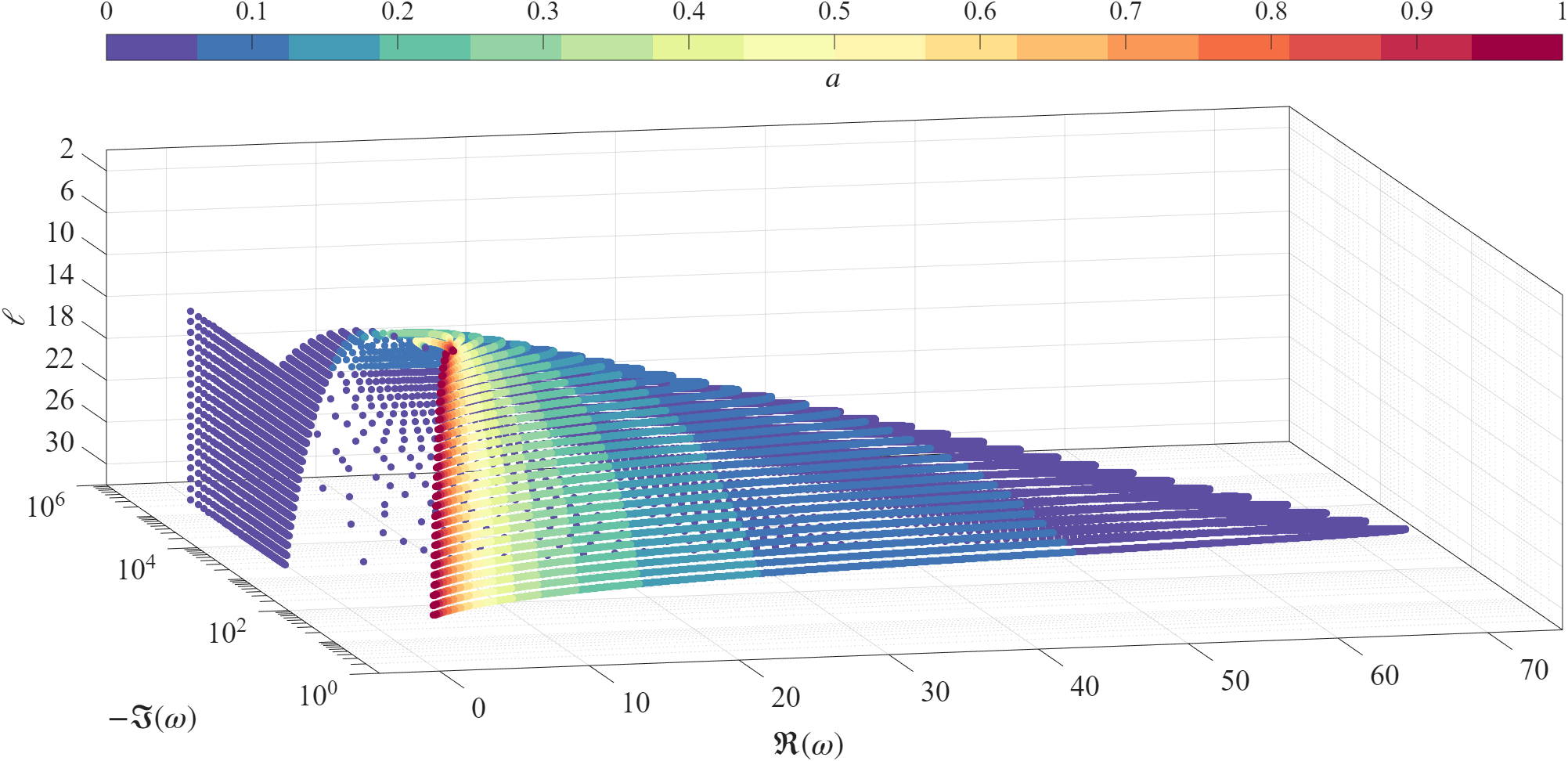}}\,\,
	\subfloat[$n_{\infty}=1$]{\includegraphics[width=3.4in]{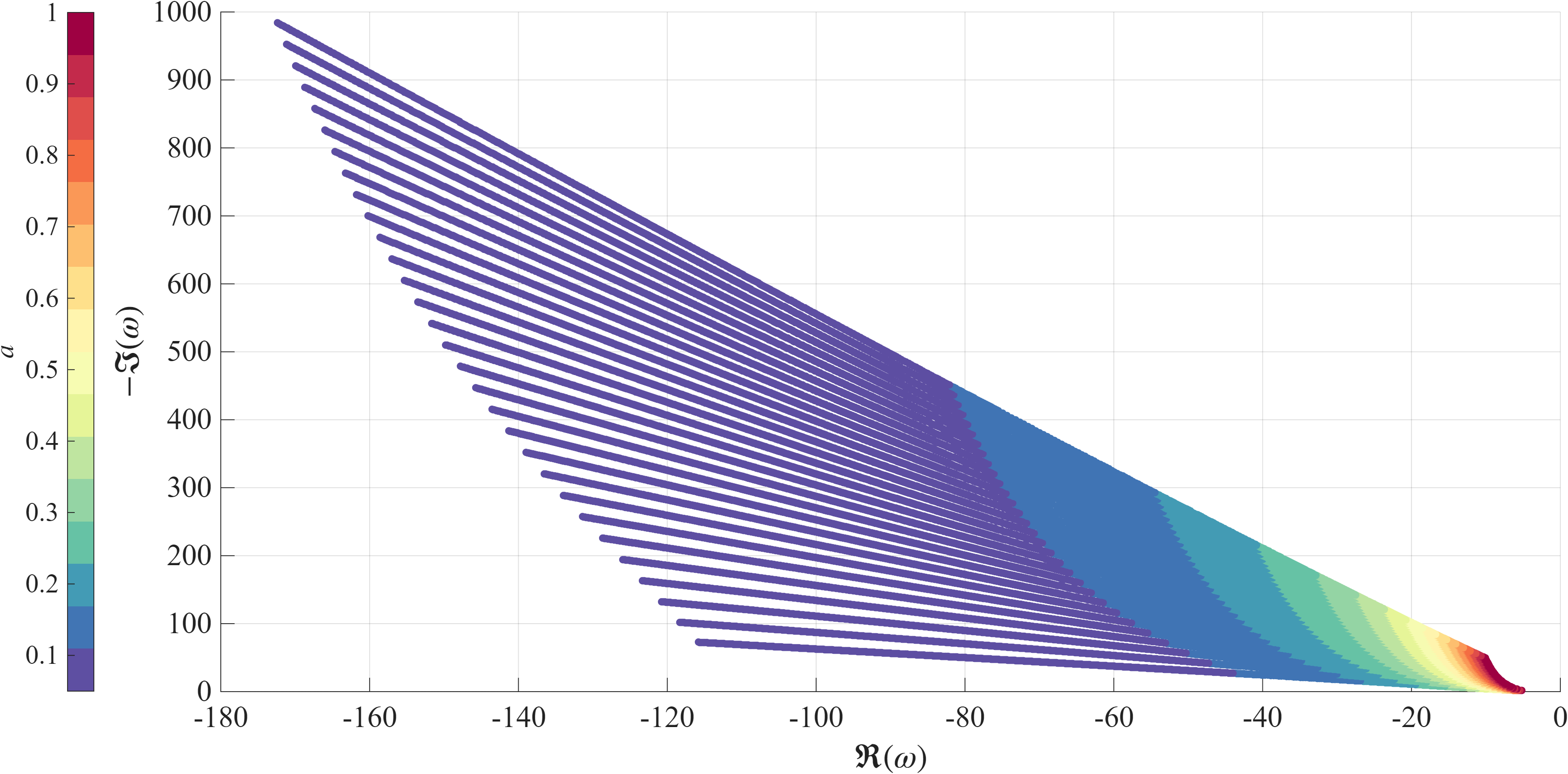}}\,\,
	\subfloat[$n_{\infty}=4$]{\includegraphics[width=3.4in]{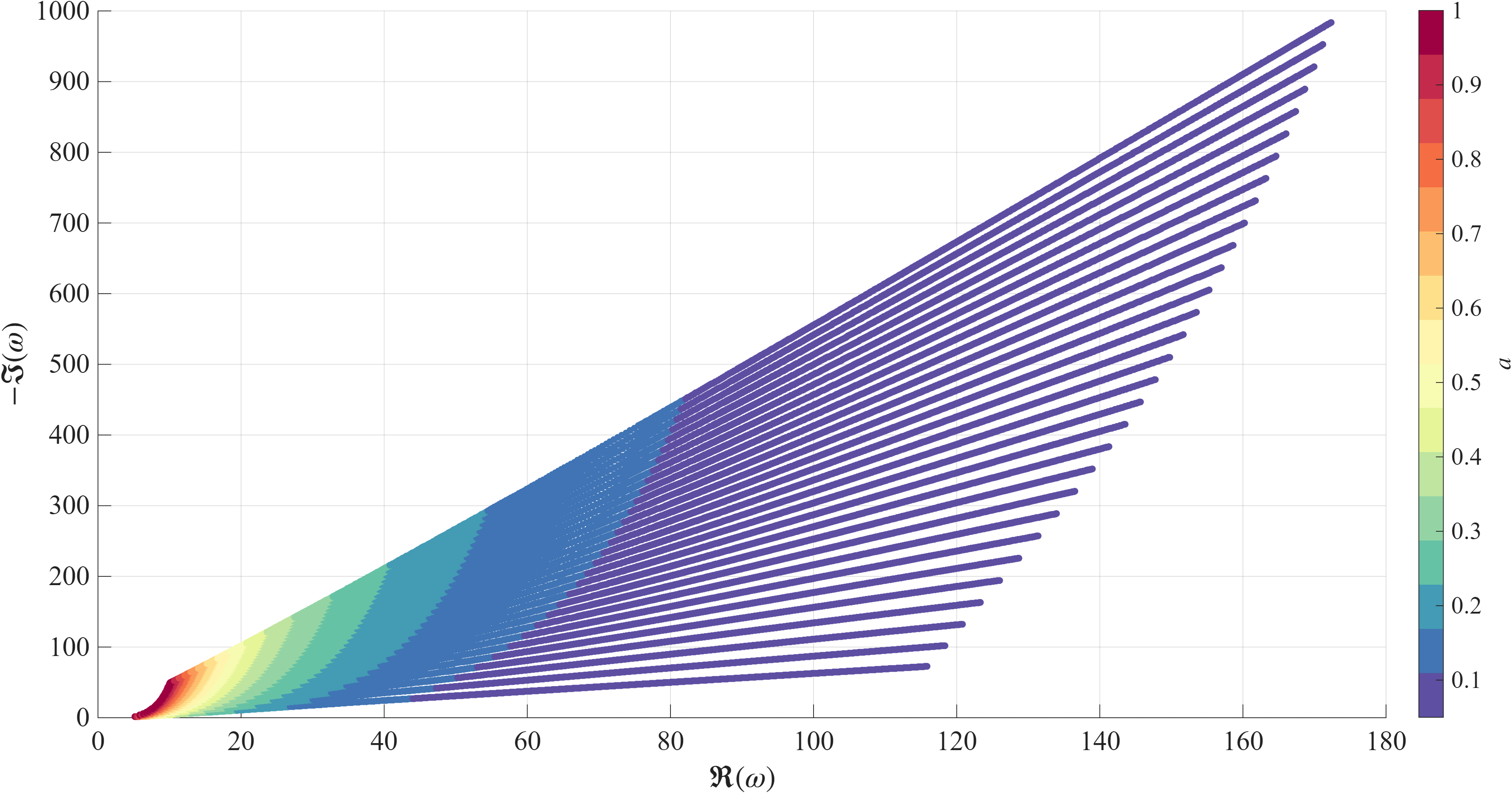}}\,\,
	\caption{Kerr TTM sequences in the $m=0$ sector for $2\leq\ell\leq32$, grouped by $n_{\infty}=1,2,3,4$. The $n_{\infty}=3$ sequences are continued to $a=0$, whereas the $n_{\infty}=1,2,$ and $4$ sequences are shown down to $a=0.05$.}
	\label{fig:TTM_M0}
\end{figure*}

\begin{figure*}[htbp]
	\centering
	\subfloat[$n_{\infty}=2,\ell=2$]{\includegraphics[width=3.6in]{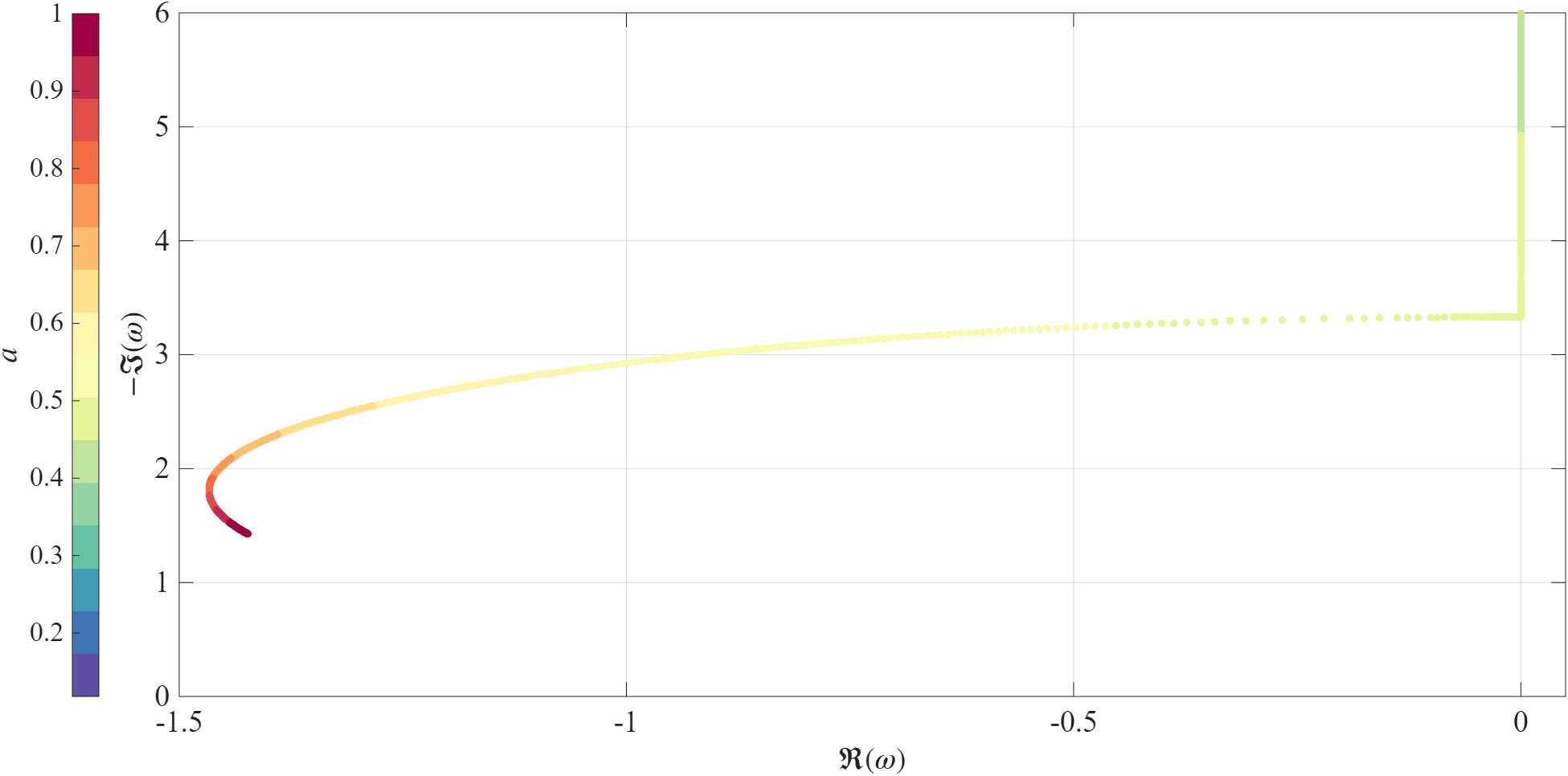}}
	\subfloat[$n_{\infty}=3,\ell=2$]{\includegraphics[width=3.33in]{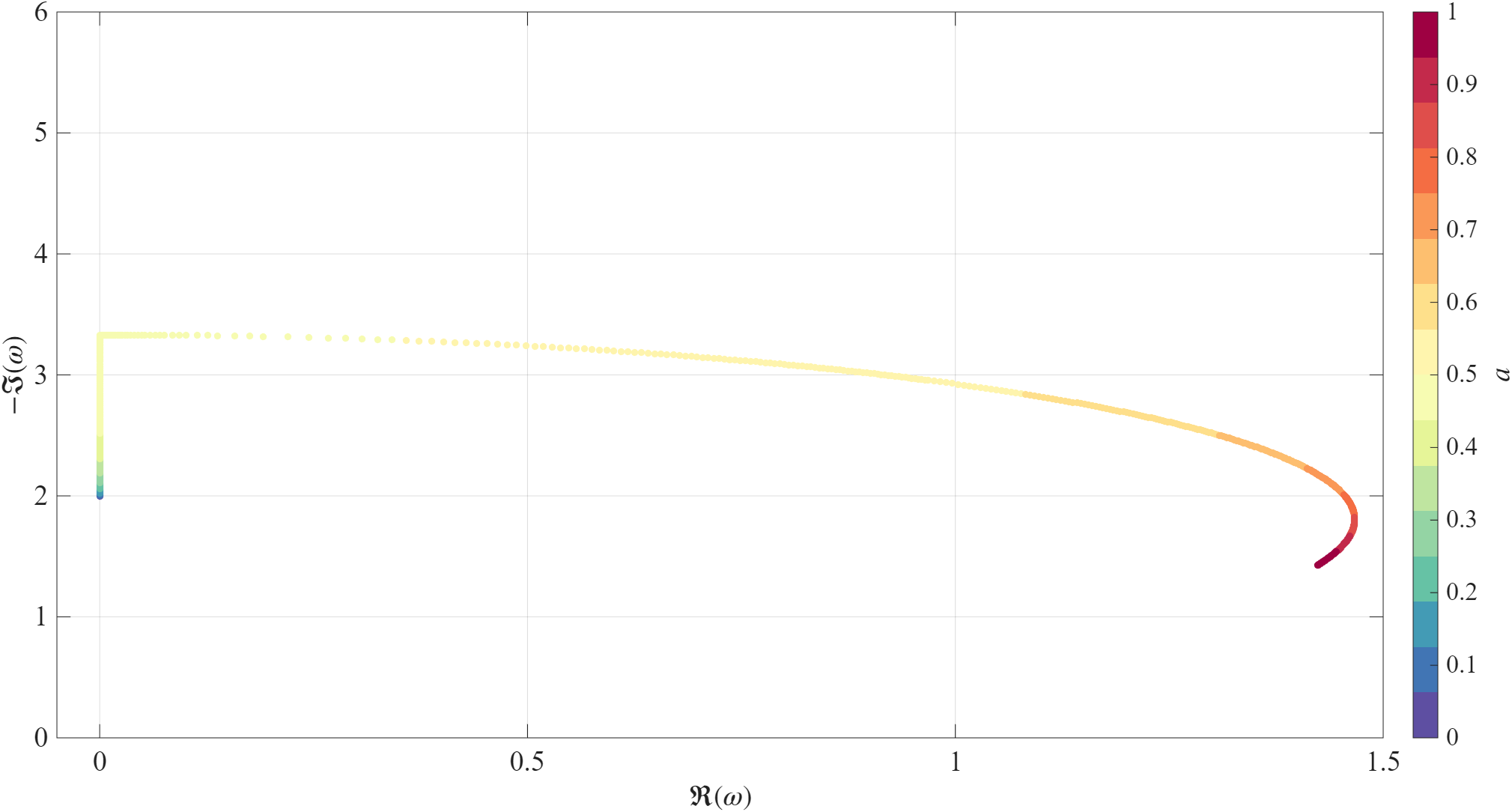}}\,\,
	\subfloat[$n_{\infty}=2,\ell=4$]{\includegraphics[width=3.6in]{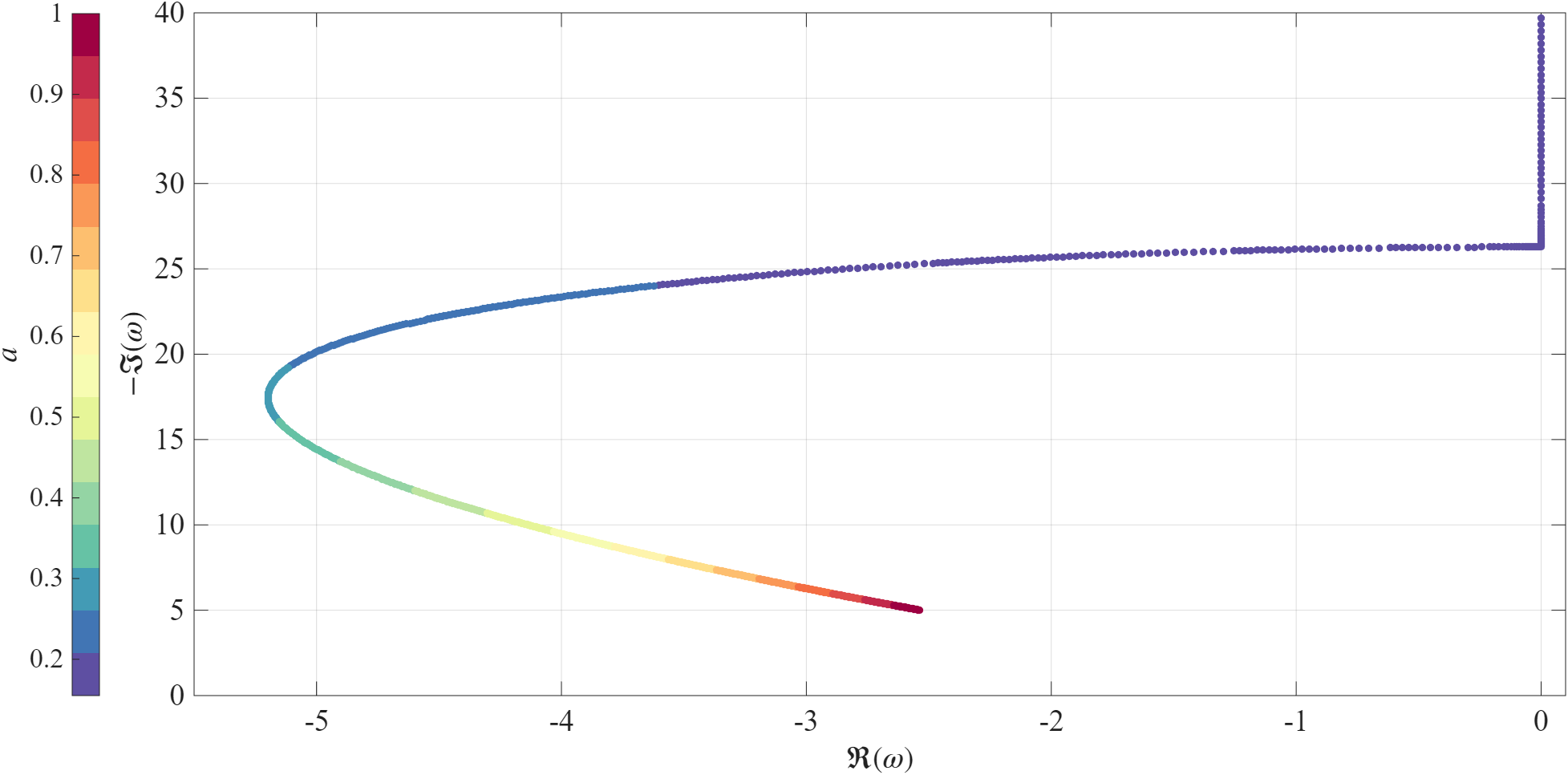}}
	\subfloat[$n_{\infty}=3,\ell=4$]{\includegraphics[width=3.33in]{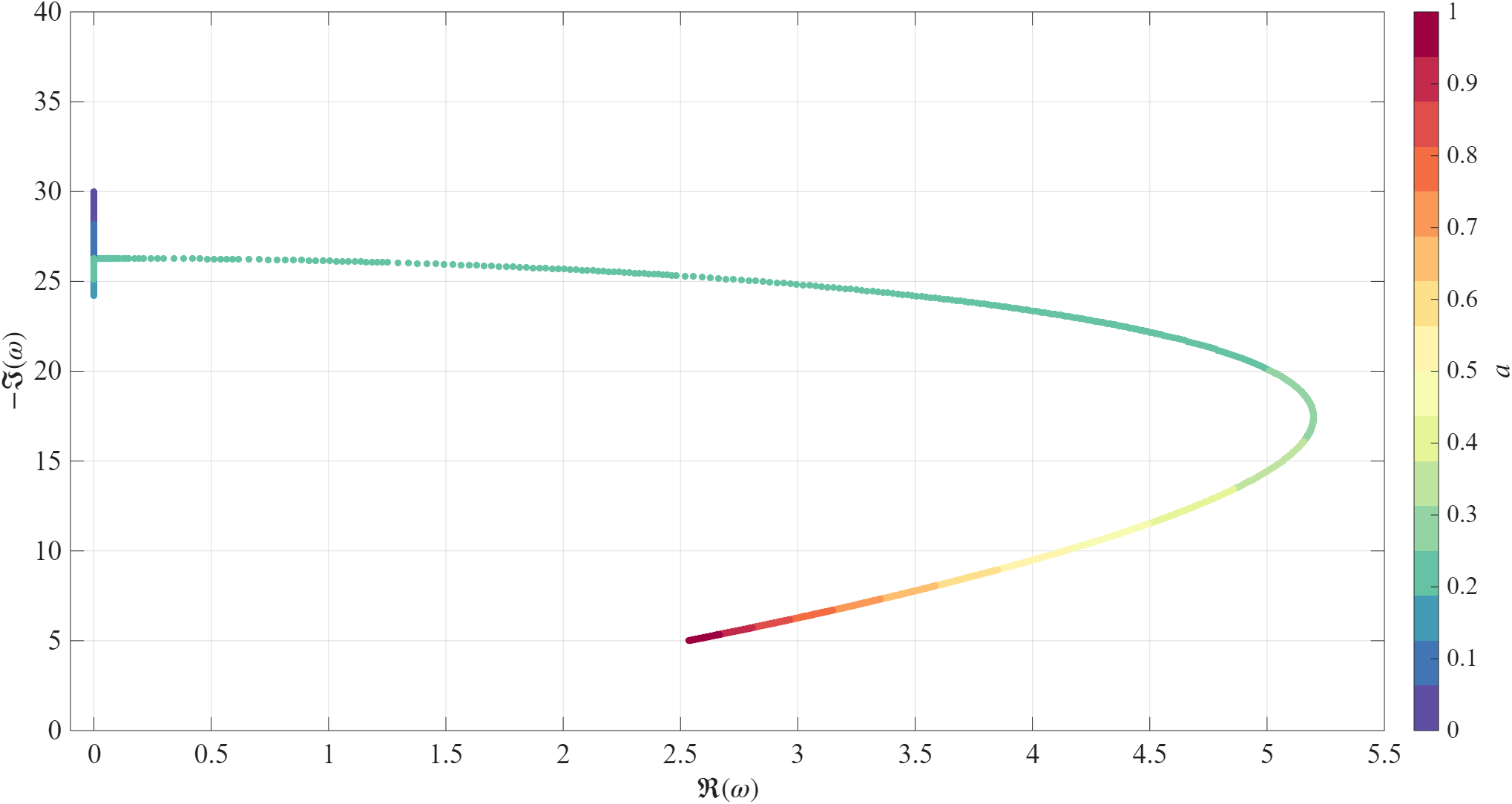}}\,\,
	\subfloat[$n_{\infty}=2,\ell=8$]{\includegraphics[width=3.6in]{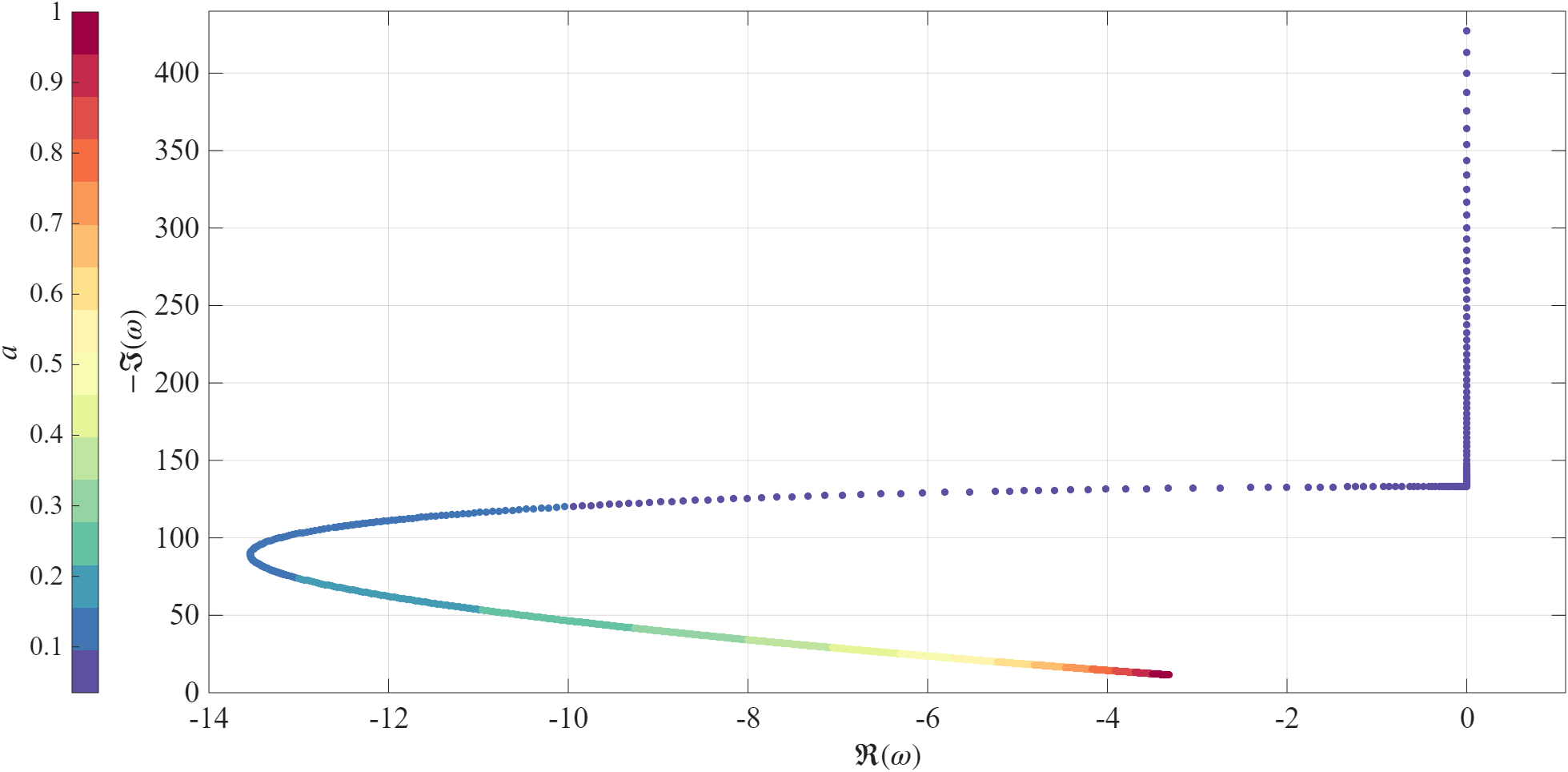}}
	\subfloat[$n_{\infty}=3,\ell=8$]{\includegraphics[width=3.33in]{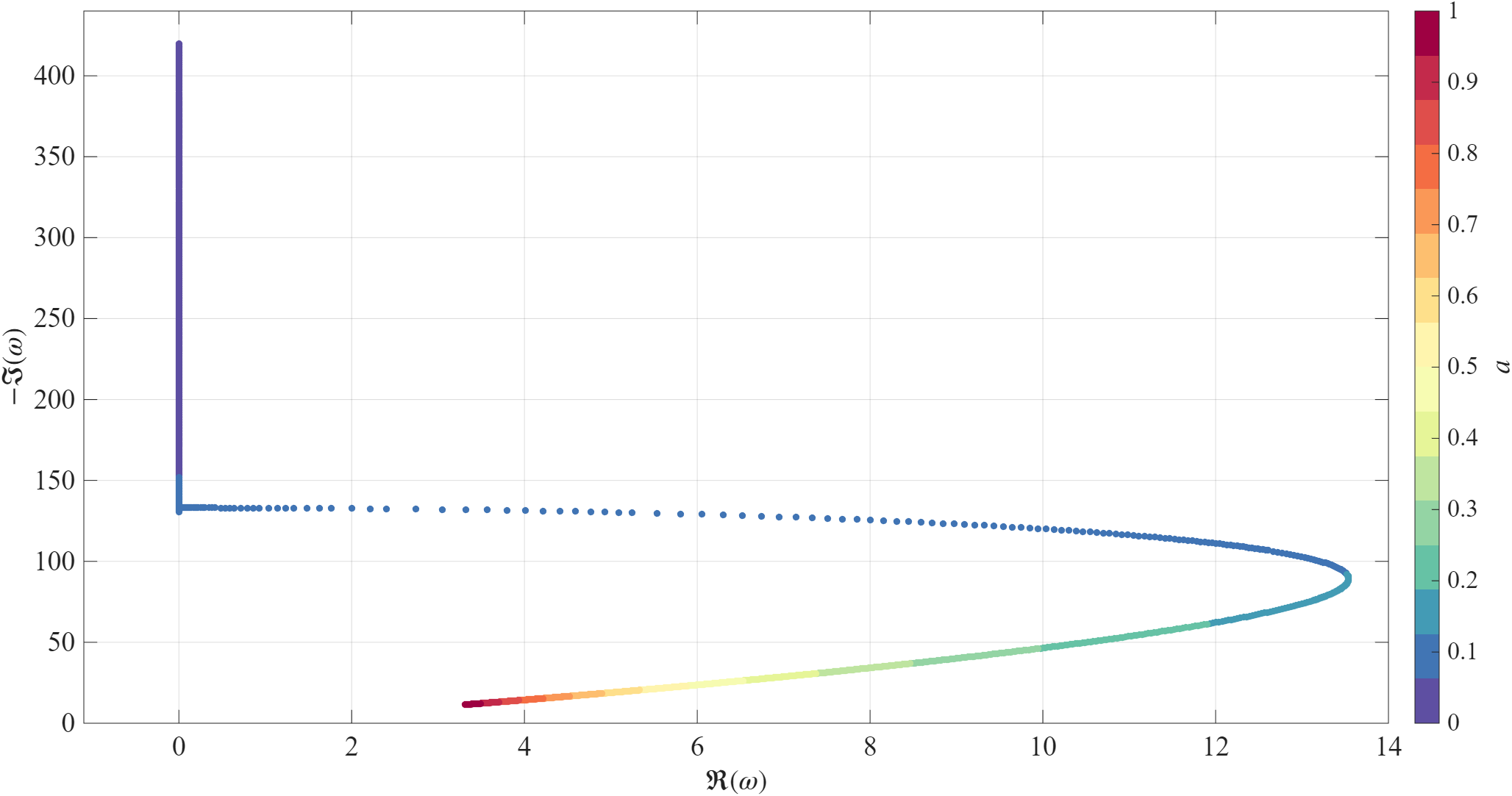}}\,\,
	\caption{Kerr TTM frequency trajectories in the $m=0$ sector for the $n_{\infty}=2$ and $n_{\infty}=3$ families with $\ell=2,4,8$.}
	\label{fig:TTM_M0_L2_4_8}
\end{figure*}

\begin{figure*}[htbp]
	\centering
	\subfloat[$n_{\infty}=3,\ell=4$]{\includegraphics[width=3.4in]{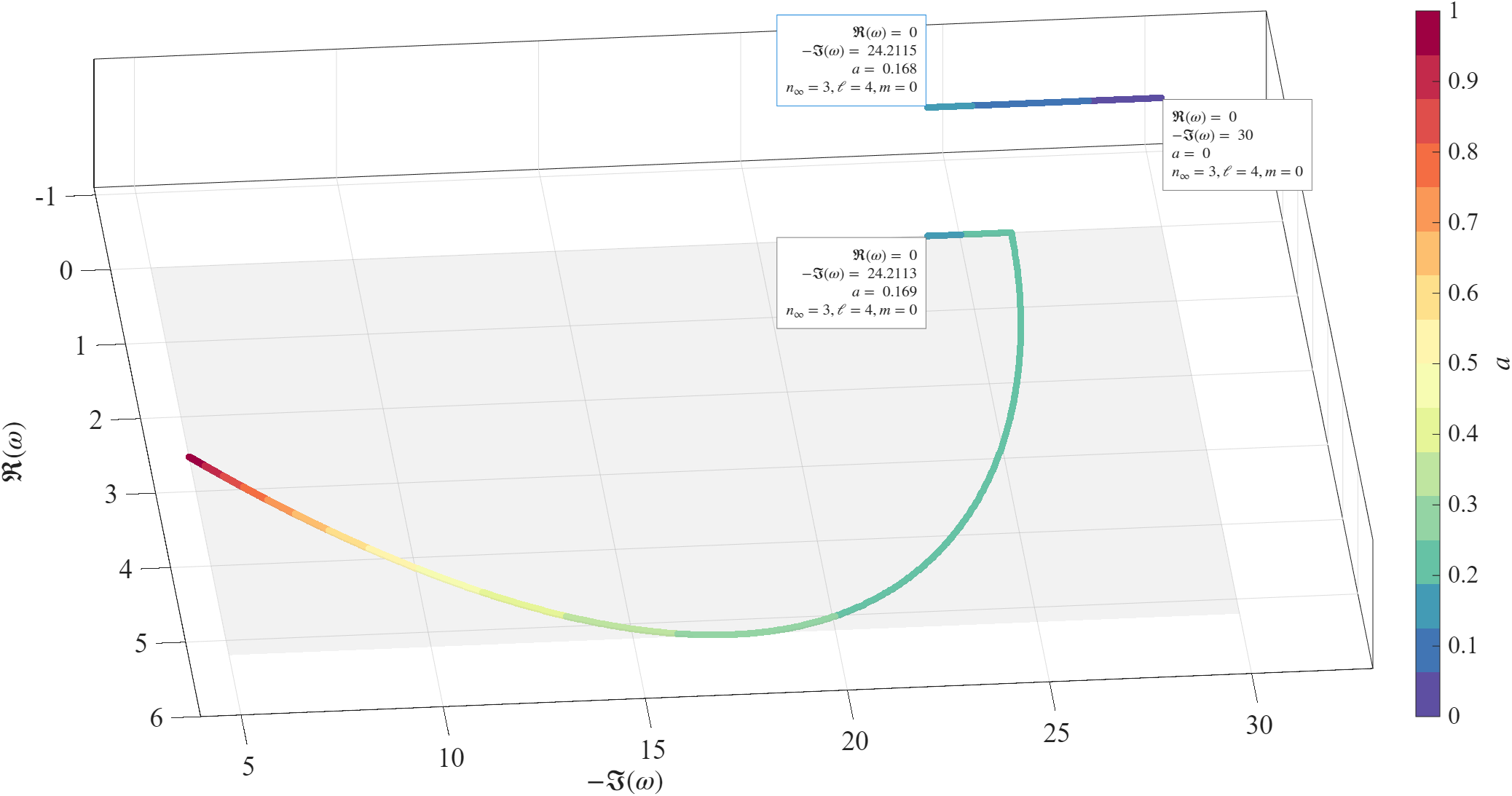}}\,\,
	\subfloat[$n_{\infty}=3,\ell=8$]{\includegraphics[width=3.4in]{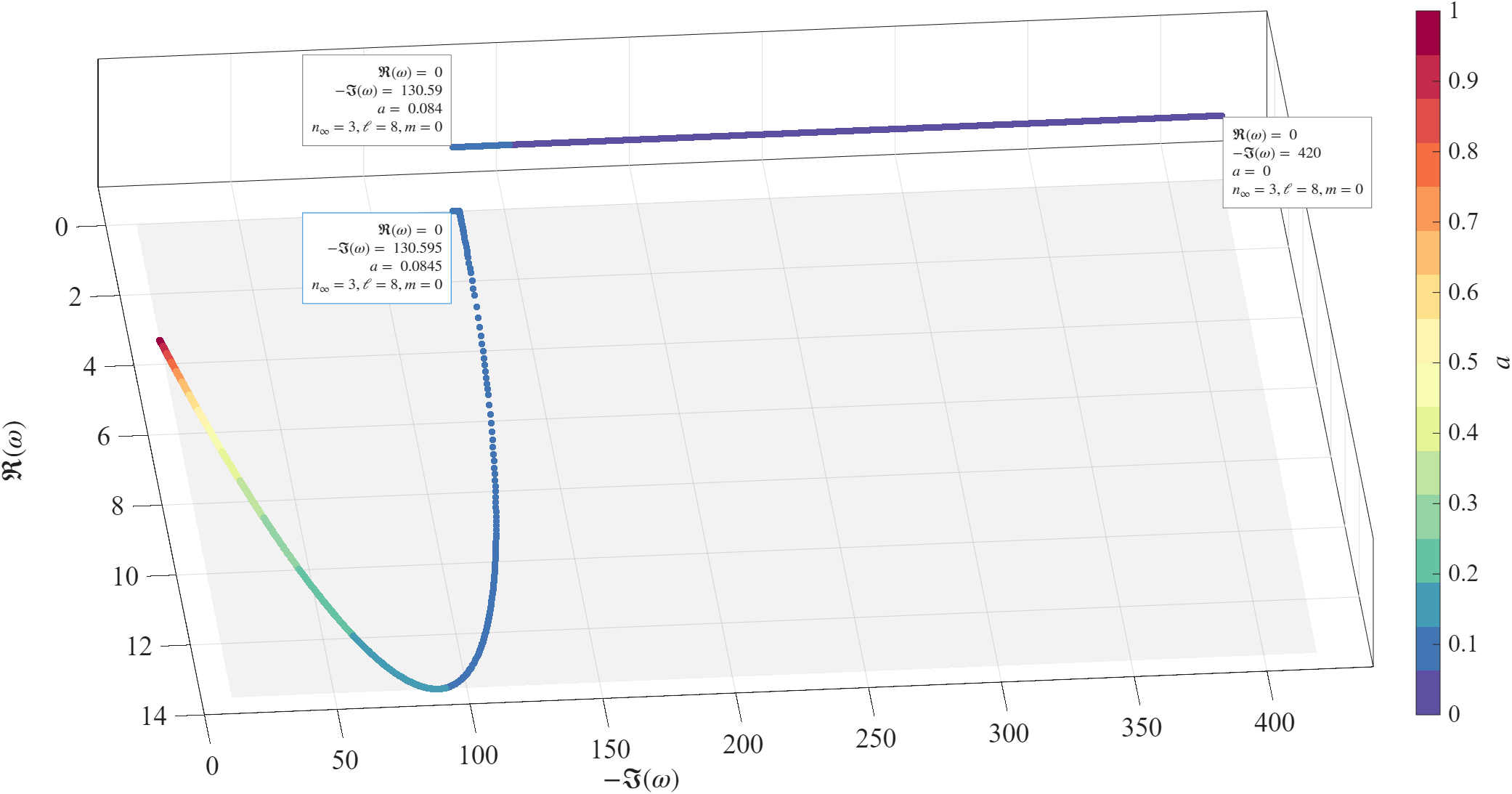}}\,\,
	\caption{Three-dimensional views of the $n_{\infty}=3$ Kerr TTM trajectories in the $m=0$ sector for $\ell=4$ and $\ell=8$, illustrating their evolution through the exceptional point and toward the Schwarzschild limit.}
	\label{fig:TTM_M0_4_8}
\end{figure*}

\begin{figure*}[htbp]
	\centering
	\subfloat[Magnitude of the excitation factor\label{subfig:TTM_ExcFac_L2_mag}]{\includegraphics[width=2.2in]{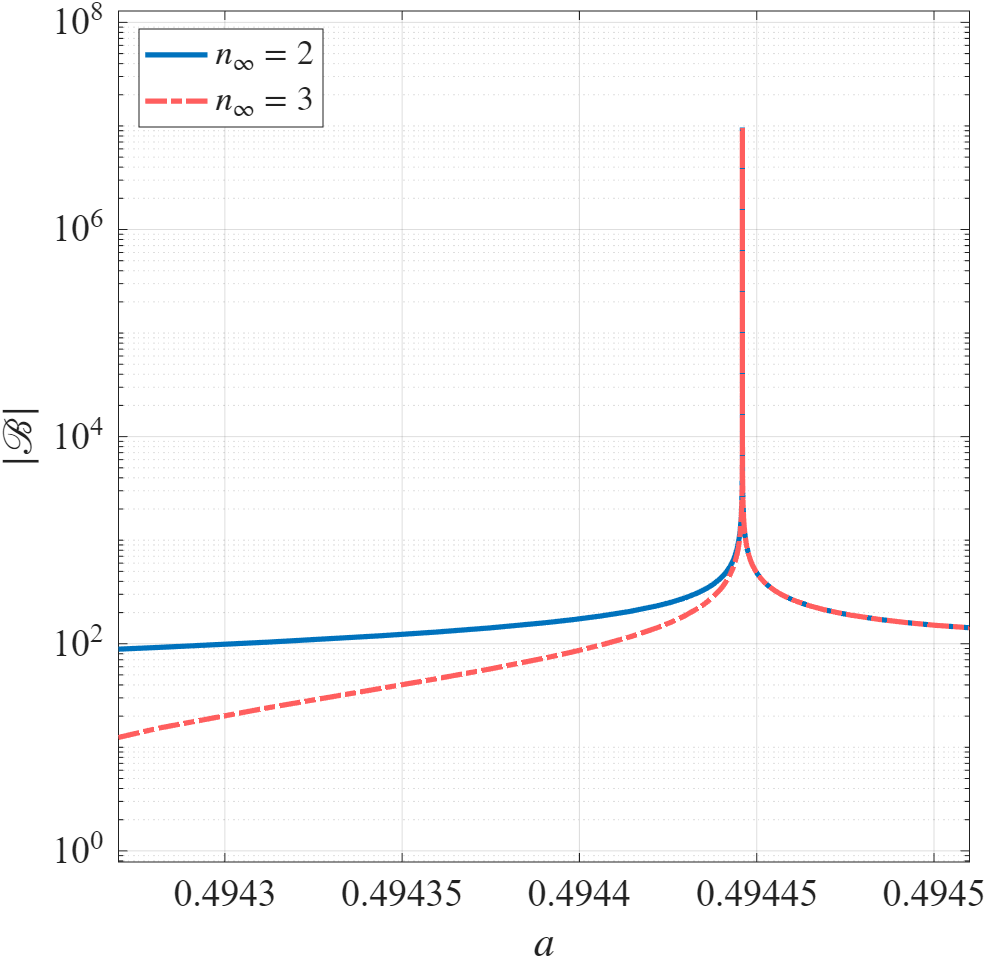}}\,\,
	\subfloat[Global evolution of the excitation factor\label{subfig:TTM_ExcFac_L2_global}]{\includegraphics[width=2.45in]{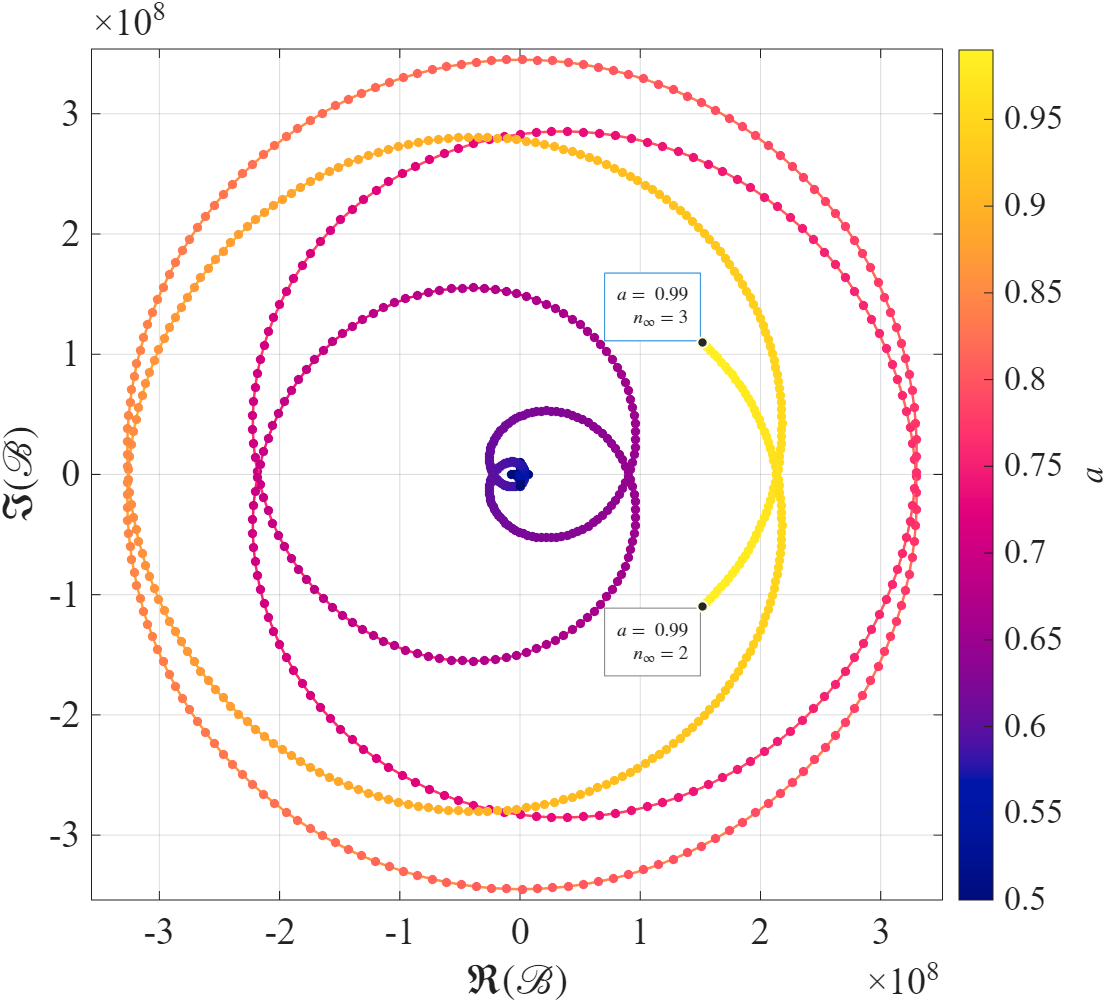}}\,\,
	\subfloat[Detailed view near the exceptional point\label{subfig:TTM_ExcFac_L2_detail}]{\includegraphics[width=2.2in]{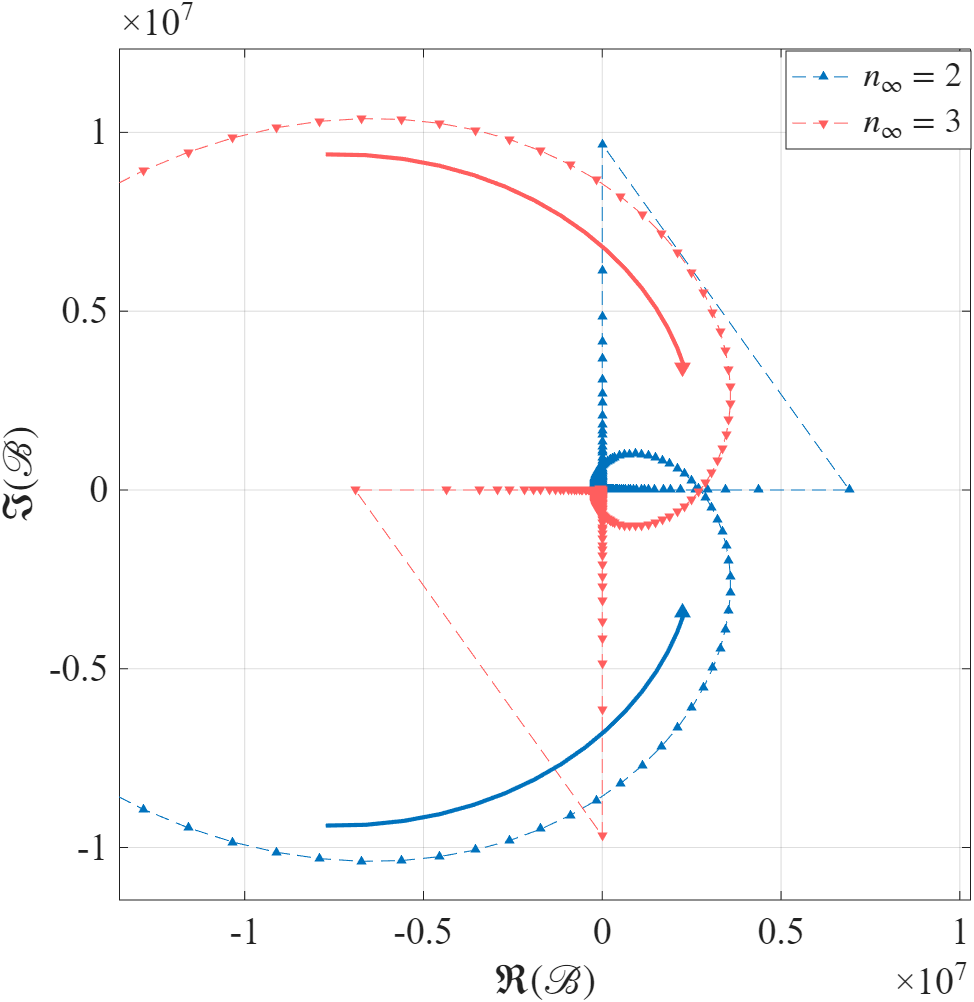}}
	\caption{TTM excitation factors for the $n_{\infty}=2$ and $n_{\infty}=3$ branches with $m=0$ and $\ell=2$ over $a\in[0.394,0.99]$. The strong enhancement occurs near the exceptional point where the two TTM branches coalesce.}
	\label{fig:TTM_M0_ExcFac_L2}
\end{figure*}

\begin{figure*}[htbp]
	\centering
	\subfloat[Magnitude of the excitation factor\label{subfig:TTM_ExcFac_L3_mag}]{\includegraphics[width=2.2in]{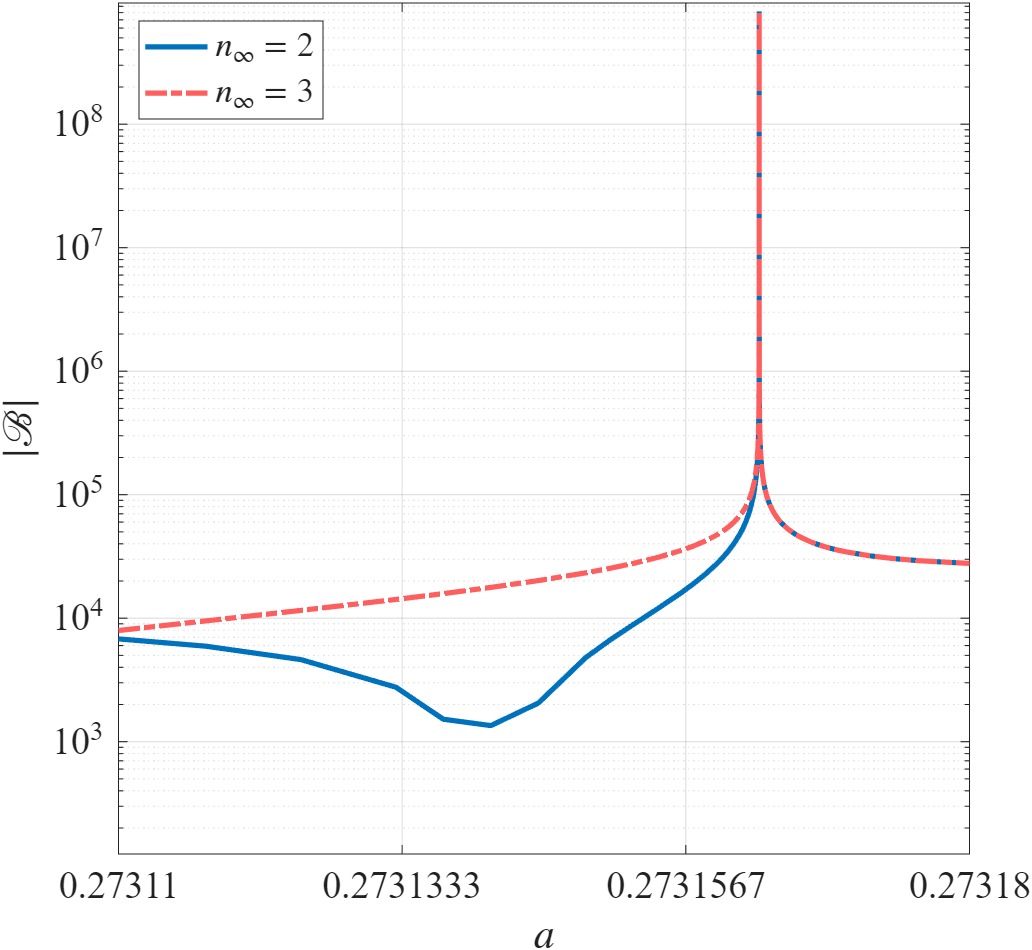}}\,\,
	\subfloat[Global evolution of the excitation factor\label{subfig:TTM_ExcFac_L3_global}]{\includegraphics[width=2.35in]{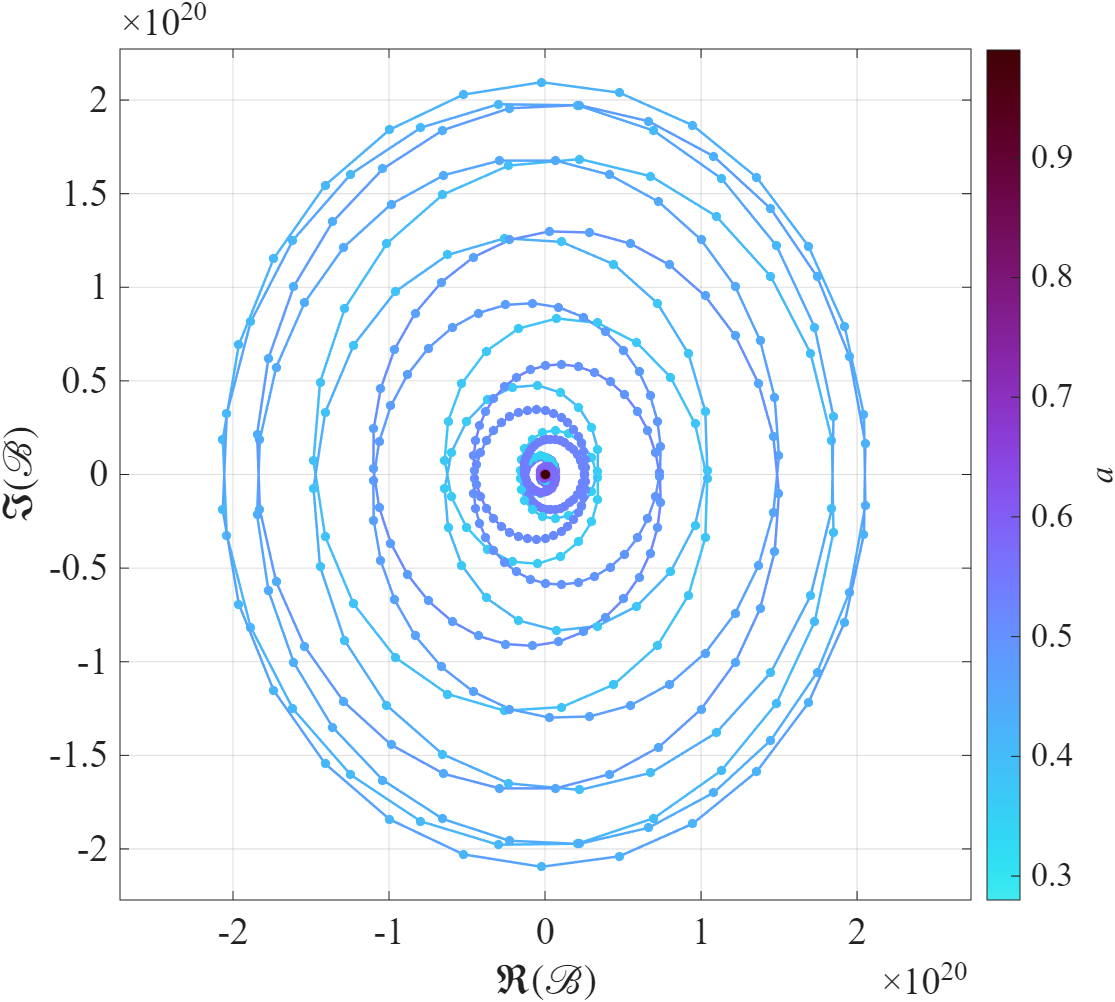}}\,\,
	\subfloat[Detailed view near the exceptional point\label{subfig:TTM_ExcFac_L3_detail}]{\includegraphics[width=2.35in]{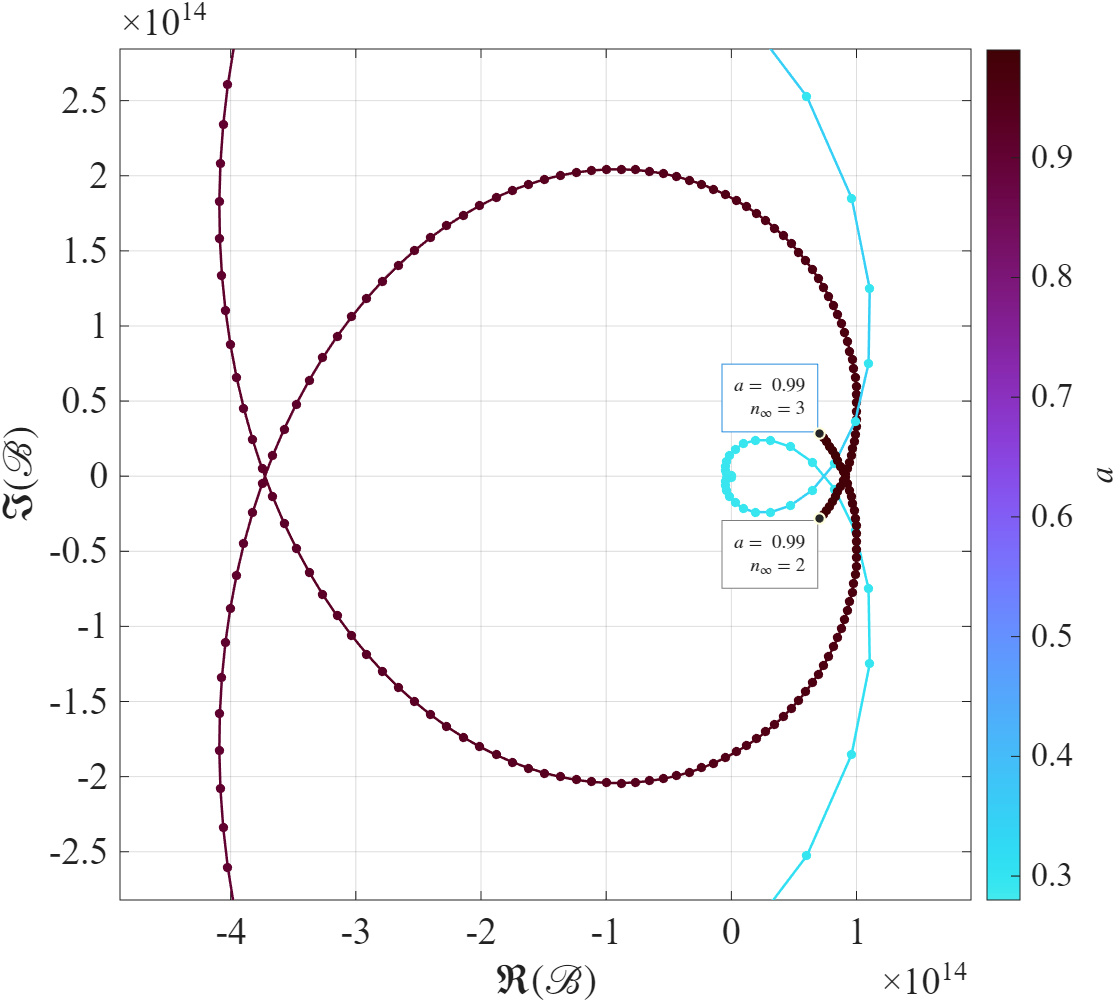}}
	\caption{TTM excitation factors for the $n_{\infty}=2$ and $n_{\infty}=3$ branches with $m=0$ and $\ell=3$ over $a\in[0.266,0.99]$.}
	\label{fig:TTM_M0_ExcFac_L3}
\end{figure*}

\subsection{Complete Kerr TTM}

Starting from the TTM roots identified at the extreme-Kerr limit, each branch is continued toward smaller values of $a$, thereby tracing the four $n_\infty$ families across the Kerr parameter range.

\subsubsection{$m=0$ TTM spectrum and exceptional points}
Figure~\ref{fig:TTM_M0} shows the evolution of the $m=0$ TTM spectrum for the four $n_\infty$ families. The $n_\infty=1$ and $n_\infty=4$ trajectories are symmetric with respect to the imaginary axis. For the $n_\infty=2$ and $n_\infty=3$ families, the portions of the trajectories with $\Re(\omega)\neq0$ are likewise symmetric with respect to the imaginary axis. As $a$ decreases, the two trajectories approach the imaginary axis from opposite sides. They first reach $\Re(\omega)=0$ at the same frequency, defining a unique coincidence point on the imaginary axis. We denote the corresponding spin and frequency by $a_{\rm EP}$ and $\omega_{\rm EP}$, respectively. Such a frequency degeneracy suggests the presence of an exceptional point in the non-Hermitian TTM spectrum~\cite{Cavalcante:2024swt,Cavalcante:2025abr,Cavalcante:2026vgr,PanossoMacedo:2025xnf,Rossi:2026als}. Different values of $\ell$ lead to coincidence points at different locations on the imaginary axis, which move progressively farther from the origin as $\ell$ increases. For $a<a_{\rm EP}$, both continuations remain on the imaginary axis. The $n_\infty=2$ family subsequently evolves toward complex infinity as $a\to0$, whereas the $n_\infty=3$ family ultimately approaches the finite Schwarzschild frequency~\eqref{eq:ASmode}.

The local structure near the coincidence point is shown more clearly in \cref{fig:TTM_M0_L2_4_8} for $\ell=2,4,8$. In each case, the $n_\infty=2$ and $n_\infty=3$ branches meet at the same frequency on the imaginary axis rather than remaining separated, distinguishing this behavior from an avoided crossing.
At $a=a_{\rm EP}$, the two branches cease to be locally distinguishable, while their global family assignments are retained through continuation from regions where the branches are well separated.
After passing through the exceptional point, the $n_\infty=2$ branch remains on the imaginary axis and moves progressively farther from the origin as $a$ decreases, eventually evolving toward complex infinity in the limit $a\to0$.
The subsequent evolution of the $n_\infty=3$ family is illustrated in \cref{fig:TTM_M0_4_8}.
When $|\omega_{\rm EP}|>|\omega_{\rm AS}|$,
the branch proceeds directly toward the finite Schwarzschild frequency~\eqref{eq:ASmode}. Otherwise, it initially moves away from this frequency, reaches a turning point, and then reverses direction before approaching it.

To independently verify the exceptional-point nature of the mode coalescence, we evaluate the radial scattering amplitudes developed in Appendix~\ref{sec:TTM-Scatter-Amp}.
Taking the spin weight $s=+2$ (${\rm TTM}_{\rm R}$) representation as a concrete realization, the spectral function $\mathcal{F}_{\rm R}(\omega)$ coincides with the reflection amplitude \eqref{eq:TTMR-spectral-ratio}.
Because the coalescence occurs on the imaginary axis where the radial scattering amplitudes encounter a branch cut, $\mathcal{F}_{\rm R}(\omega)$ is evaluated via the two lateral limits
\begin{equation}
\mathcal{F}_{\rm R}^{\pm}(\omega_{\rm EP})
=
\lim_{\epsilon\to0^{+}}
\mathcal{F}_{\rm R}(\omega_{\rm EP}\pm\epsilon).
\end{equation}
In numerical implementations, we take $\epsilon = 10^{-10}$, and the frequency derivatives are evaluated using an eighth-order finite difference without sampling the branch cut directly.
Numerically, both lateral limits and their corresponding frequency derivatives approach identical values within the truncation error.
As summarized in Table~\ref{tab:TTMR_Zero} for $\ell \in \{2, 3, 4\}$, the modulus of the reflection amplitude $|\mathcal{F}_{\rm R}^{\pm}|$ and that of its first frequency derivative $|\partial_\omega \mathcal{F}_{\rm R}^\pm|$ are numerically suppressed to $\lesssim 10^{-14}$ at the coincidence point $(a_{\rm EP},\omega_{\rm EP})$, whereas the second derivative $|\partial^2_\omega \mathcal{F}_{\rm R}^\pm|$ remains finite and nonzero.
Hence, the scattering condition satisfies
\begin{equation}\label{eq:EP_scattering_derivatives}
|\mathcal{F}_{\rm R}^{\pm}(\omega_{\rm EP})| \simeq 0,
\quad
\left|\frac{{\rm d}\mathcal{F}_{\rm R}^{\pm}}{{\rm d}\omega}\right| \simeq 0,
\quad
\left|\frac{{\rm d}^{2}\mathcal{F}_{\rm R}^{\pm}}{{\rm d}\omega^{2}}\right| \neq 0.
\end{equation}
This demonstrates that the ${\rm TTM}_{\rm R}$ scattering condition develops a genuine second-order zero, providing direct scattering-matrix evidence for the existence of an exceptional point, in exact analogy with the coalescing secular roots encountered in QNM spectra~\cite{Cavalcante:2024swt,Cavalcante:2025abr,Cavalcante:2026vgr,PanossoMacedo:2025xnf,Rossi:2026als}.
\begin{table}[htbp]
  \centering
\caption{Numerical verification of the second-order zero of $F_R^\pm$ via the moduli of the function and its frequency derivatives at the exceptional point $(a_{\rm EP},\omega_{\rm EP})$ for $m=0$ and $\ell\in\{2,3,4\}$. The numerical tolerance used in the computation is listed in the last column.}
    \begin{tabular}{ccccc}
    \toprule
    $\ell$  & $|\mathcal{F}^{\pm}_{\rm R}|$    & $|\partial_\omega \mathcal{F}_{\rm R}^\pm|$   & $|\partial^2_\omega \mathcal{F}_{\rm R}^\pm|$  & Numerical tolerance \\
    \midrule
    2   & $2.88 \times 10^{-16}$ & $3.04 \times 10^{-15}$ & $3.21 \times 10^{-2}$ & $10^{-200}$ \\
    3   & $7.92 \times 10^{-16}$ & $9.91 \times 10^{-15}$ & $7.94 \times 10^{-5}$ & $10^{-350}$ \\
    4   & $1.13 \times 10^{-18}$ & $3.89 \times 10^{-17}$ & $1.68 \times 10^{-6}$ & $10^{-450}$ \\
    \bottomrule
    \end{tabular}%
  \label{tab:TTMR_Zero}%
\end{table}

Expanding $\mathcal{F}_{\rm R}(\omega,a)$ locally on a fixed lateral sheet around the exceptional point yields
\begin{equation}
\mathcal{F}_{\rm R}(\omega,a)
\simeq
\frac{1}{2}\partial_\omega^2\mathcal{F}_{\rm R}
(\omega-\omega_{\rm EP})^2
+\partial_a\mathcal{F}_{\rm R}(a-a_{\rm EP})
+\cdots .
\end{equation}
Imposing the TTM condition $\mathcal{F}_{\rm R}=0$ immediately leads to the characteristic square-root branch-point behavior:
\begin{equation}\label{eq:EP_squareroot}
\omega_{n_\infty=2,3}(a)-\omega_{\rm EP}
\propto
\pm\sqrt{a-a_{\rm EP}}.
\end{equation}
Equation~\eqref{eq:EP_squareroot} accounts for the continuous coalescence of the $n_\infty=2$ and $n_\infty=3$ mode frequencies at $a_{\rm EP}$, where the local distinction between the two branches is lost.
Together with the second-order zero of the scattering spectral function, this square-root behavior identifies the merger as a genuine exceptional-point coalescence rather than an overtone-multiplet splitting~\cite{Cook:2022kbb}.

\begin{figure*}[htbp]
	\centering
	\subfloat[$n_{\infty}=1$\label{subfig:TTM_M1_N1}]{\includegraphics[width=3.4in]{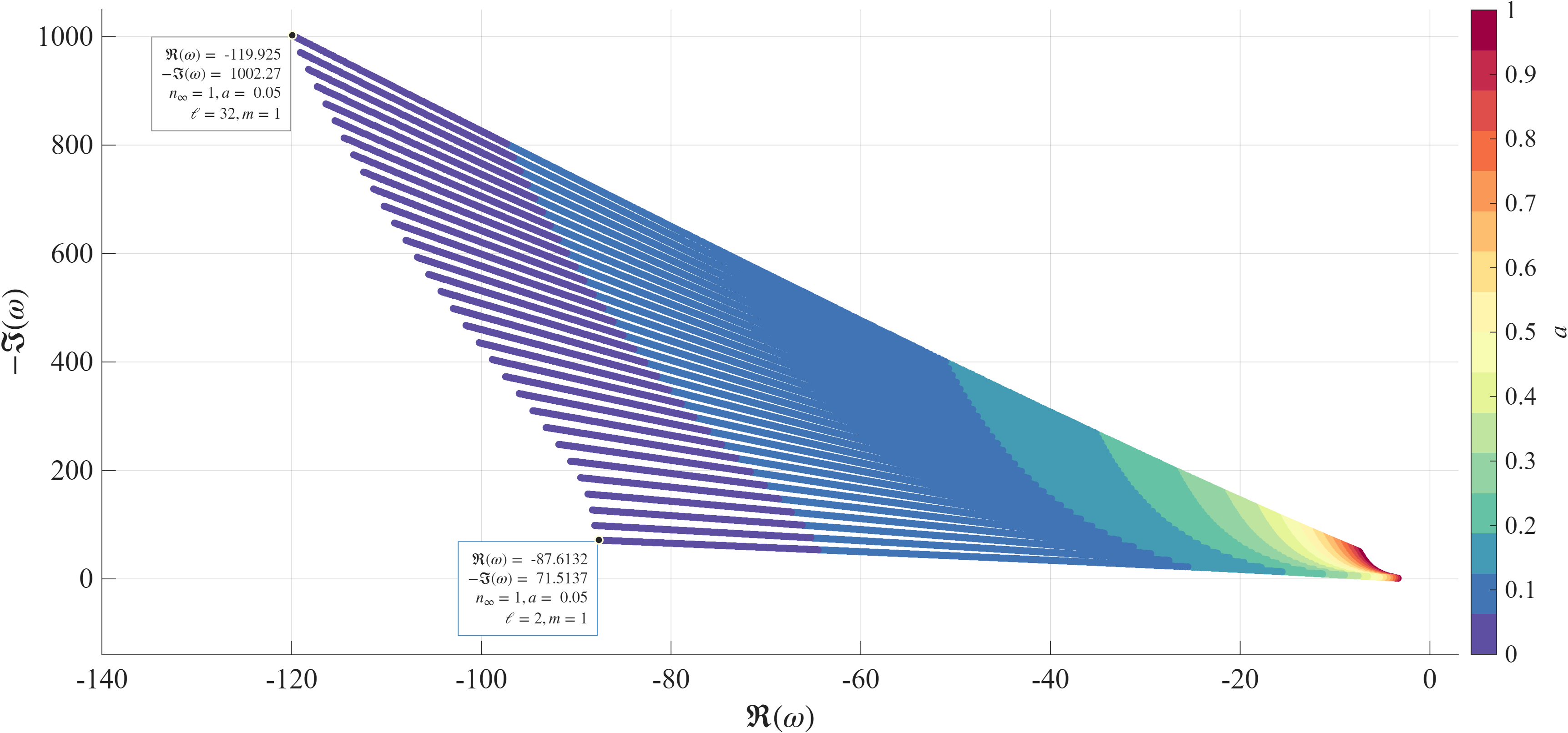}}\,\,
	\subfloat[Logarithmic scale for  $n_{\infty}=1$\label{subfig:TTM_M1_LogN1}]{\includegraphics[width=3.4in]{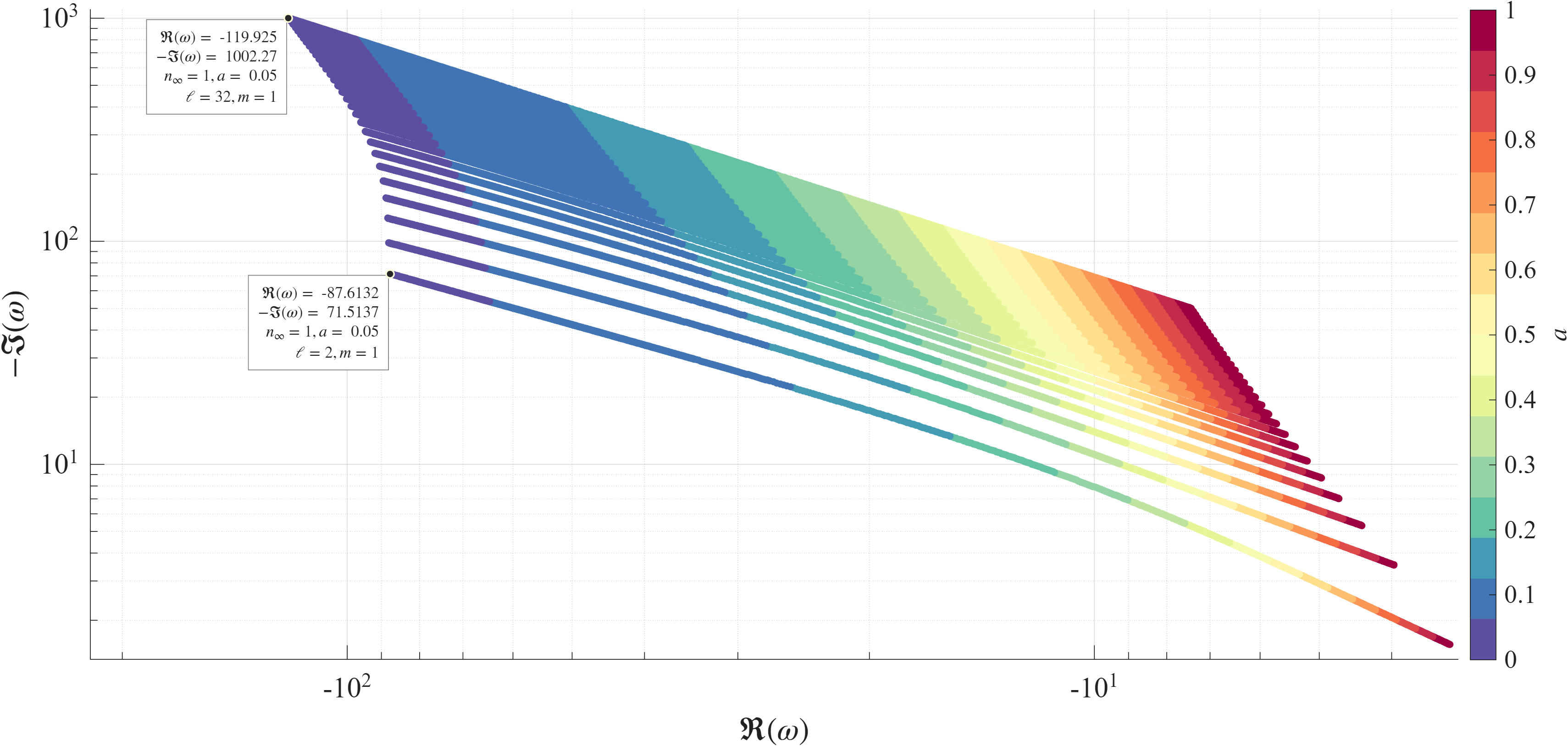}}\,\,
	\subfloat[$n_{\infty}=2$\label{subfig:TTM_M1_N2}]{\includegraphics[width=3.4in]{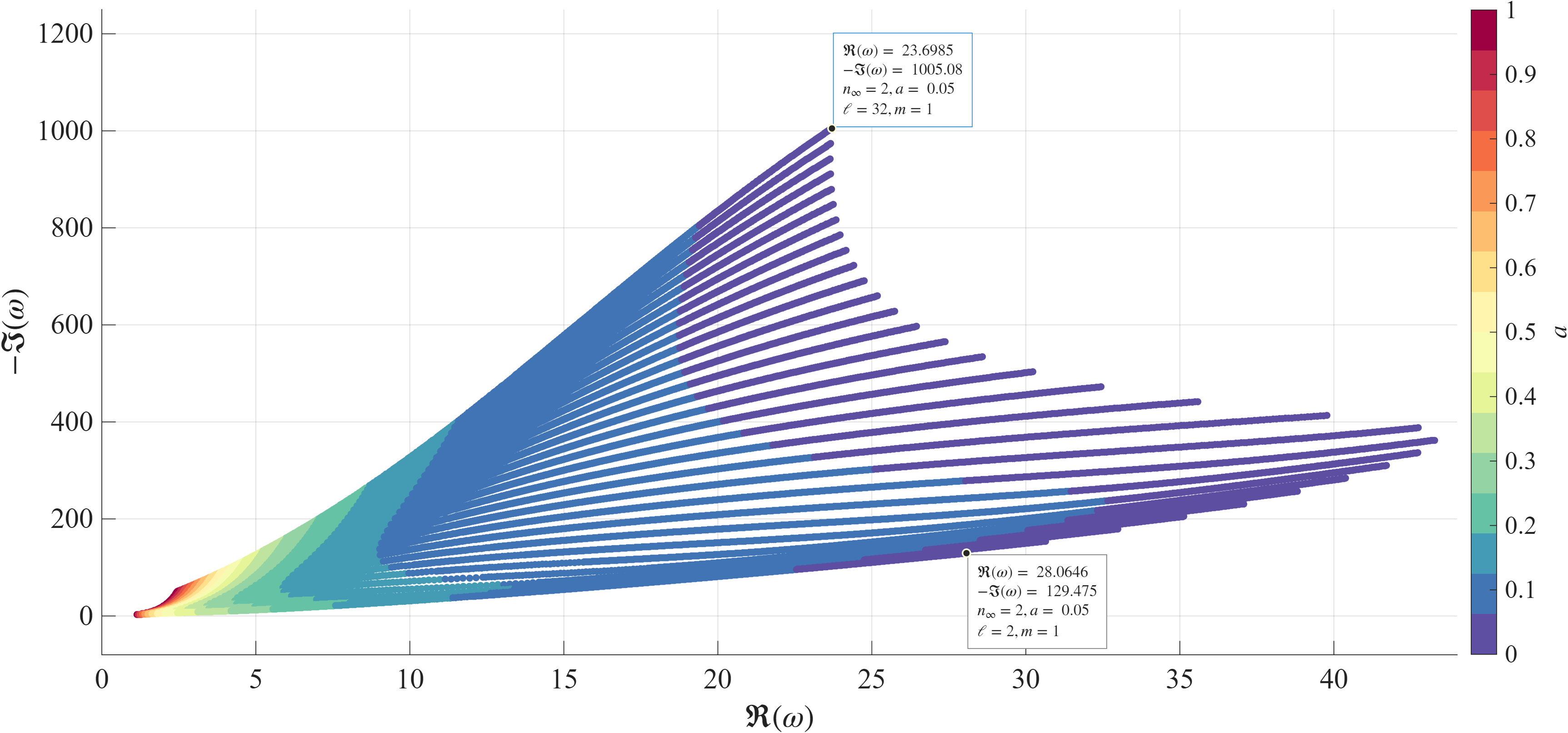}}\,\,
	\subfloat[Logarithmic scale for  $n_{\infty}=2$\label{subfig:TTM_M1_LogN2}]{\includegraphics[width=3.4in]{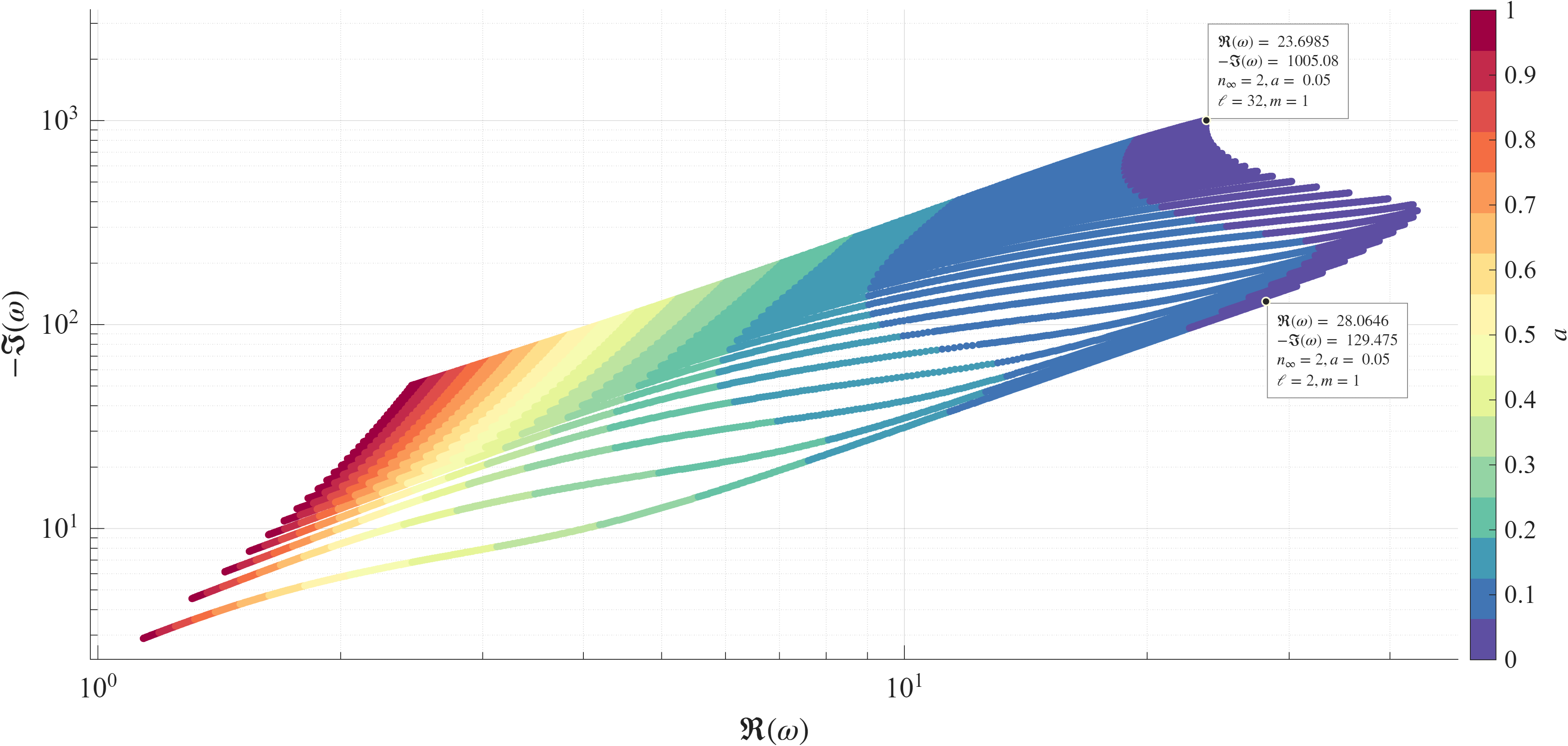}}\,\,
	\subfloat[$n_{\infty}=3$\label{subfig:TTM_M1_N3}]{\includegraphics[width=3.4in]{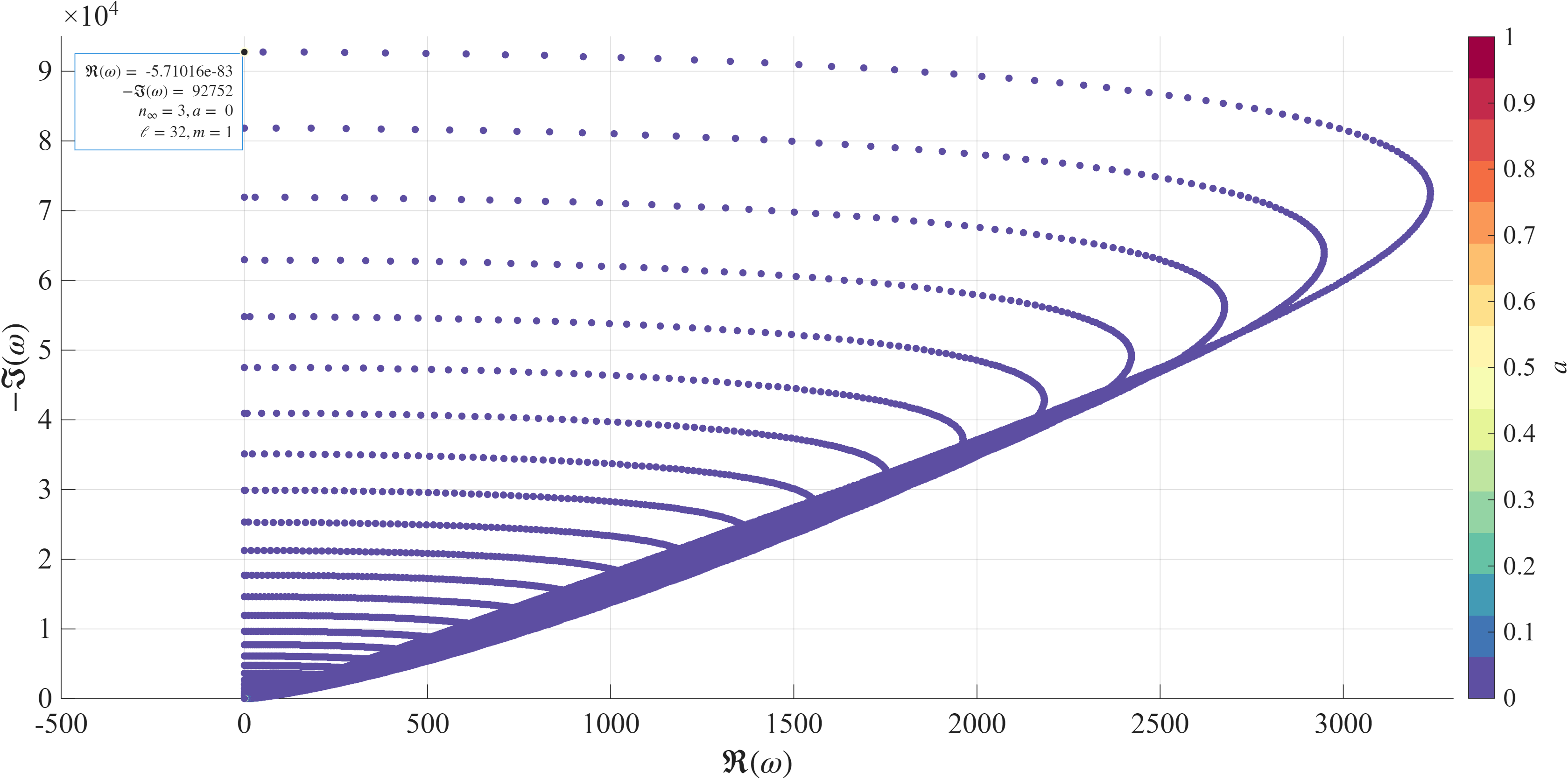}}\,\,
	\subfloat[Logarithmic scale for  $n_{\infty}=3$\label{subfig:TTM_M1_LogN3}]{\includegraphics[width=3.4in]{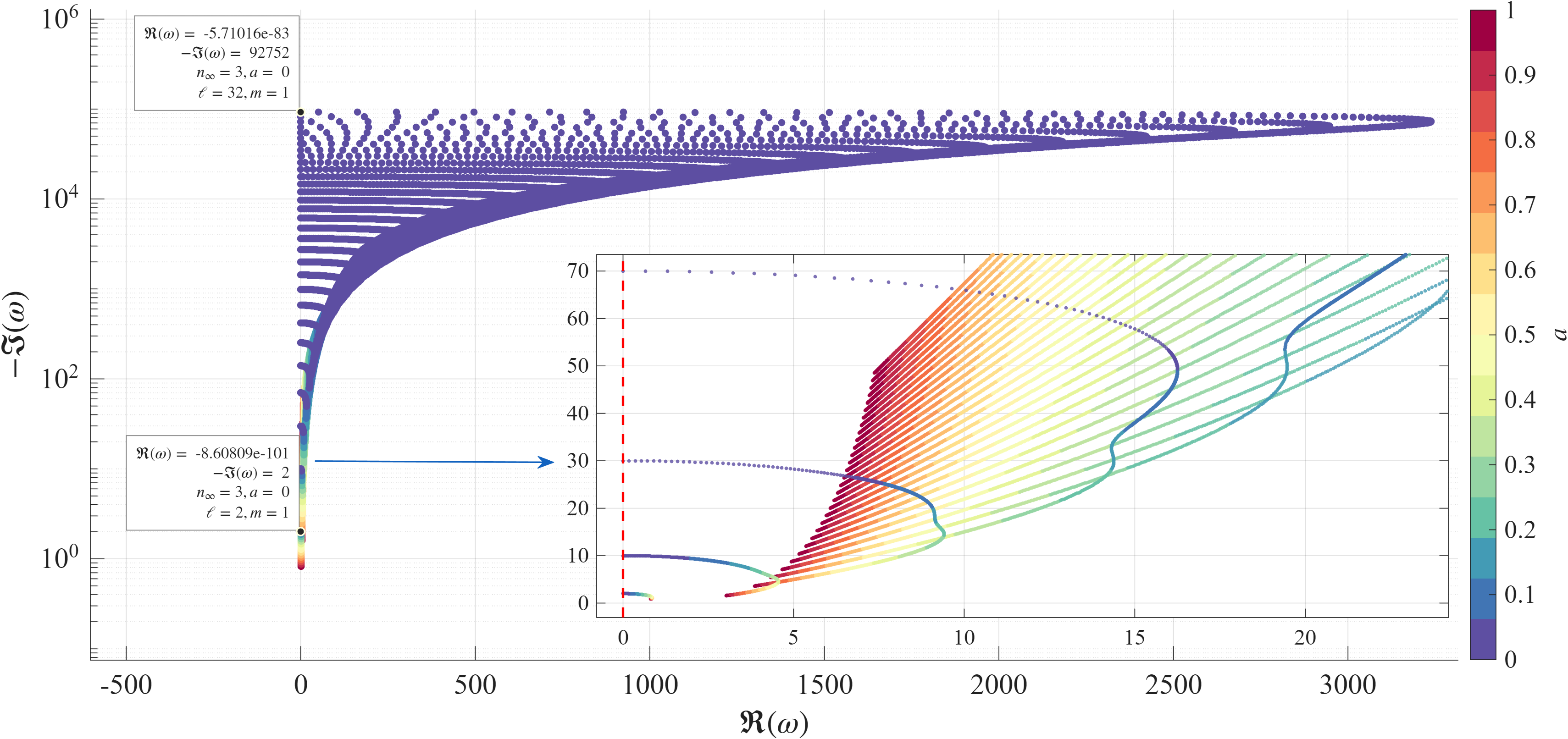}}\,\,
	\subfloat[$n_{\infty}=4$\label{subfig:TTM_M1_N4}]{\includegraphics[width=3.4in]{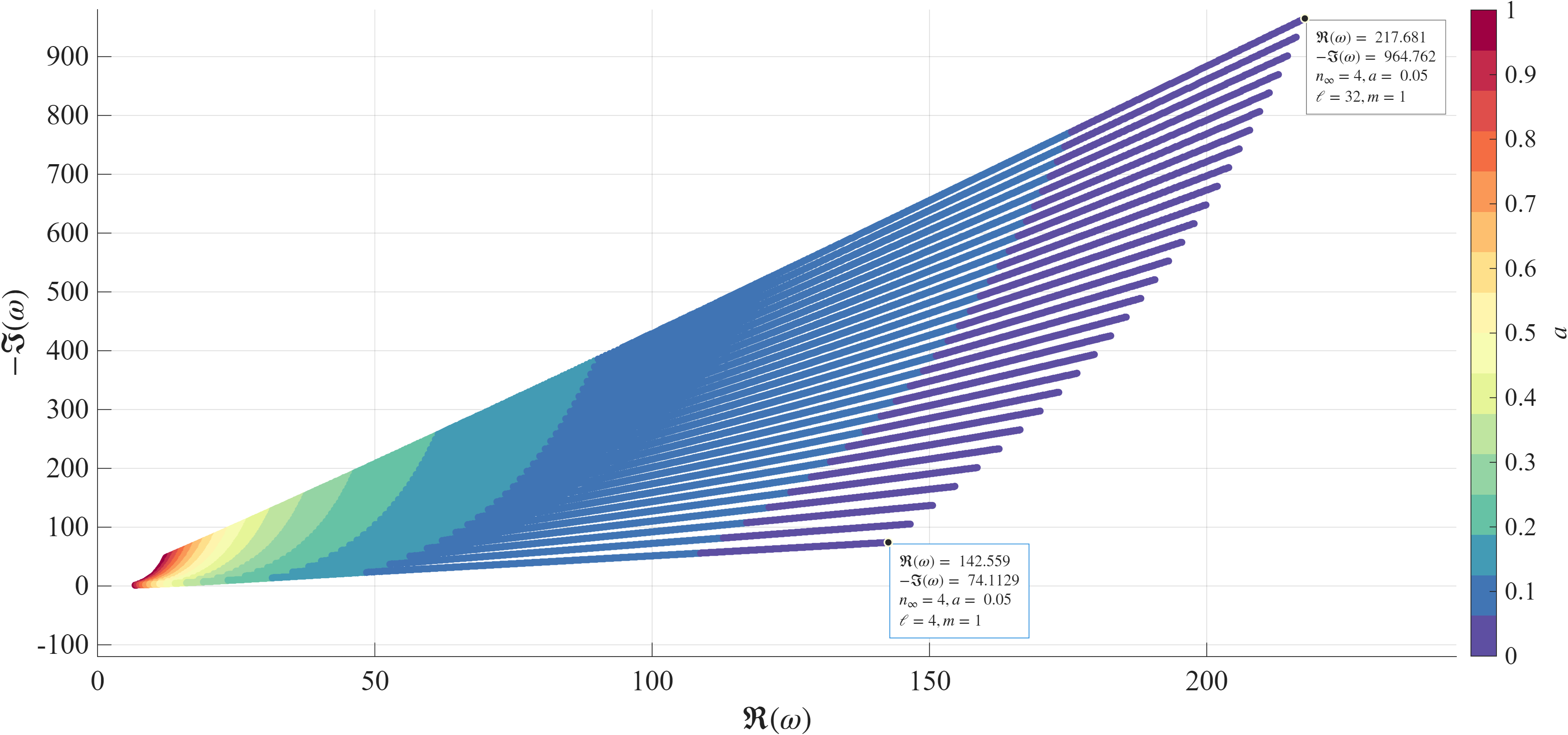}}\,\,
	\subfloat[Logarithmic scale for  $n_{\infty}=4$\label{subfig:TTM_M1_LogN4}]{\includegraphics[width=3.4in]{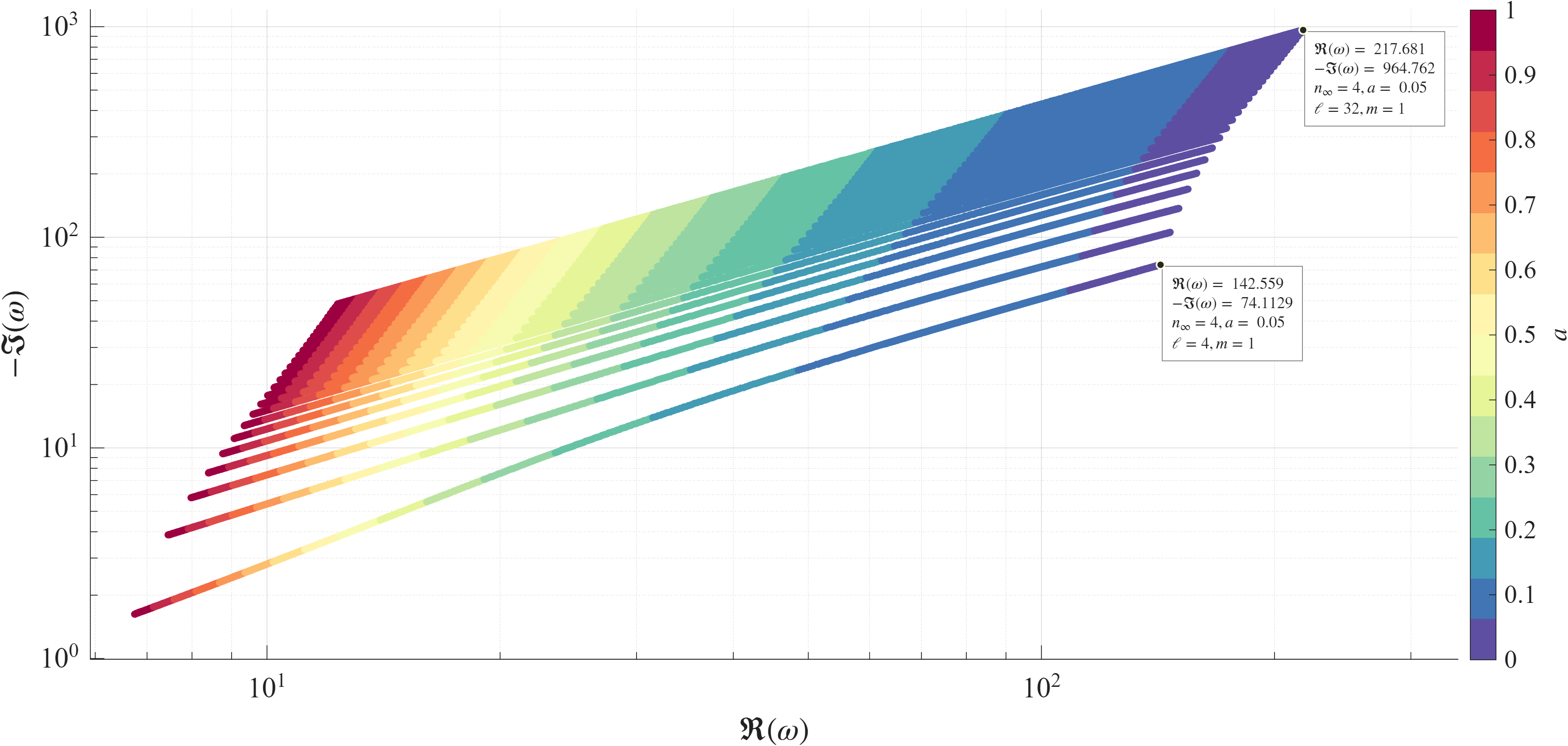}}
	\caption{Kerr TTM frequency trajectories for $m=1$ and $2\leq \ell \leq 32$. Panels on the left present the trajectories on a linear scale, while panels on the right display the corresponding trajectories on a logarithmic scale. For the $n_{\infty}=3$ family, the trajectories are shown over the full range $a\in[0,1]$, since its TTM frequencies remain finite as $a\to0$. For the other families, the small-$a$ limit is singular: their frequencies diverge as $a\to0$, so the trajectories are shown only for $a\in[0.05,1]$.}
	\label{fig:TTM_M1}
\end{figure*}

\begin{figure*}[htbp]
	\centering
	\subfloat[$n_{\infty}=1$\label{subfig:TTM_M5_N1}]{\includegraphics[width=3.4in]{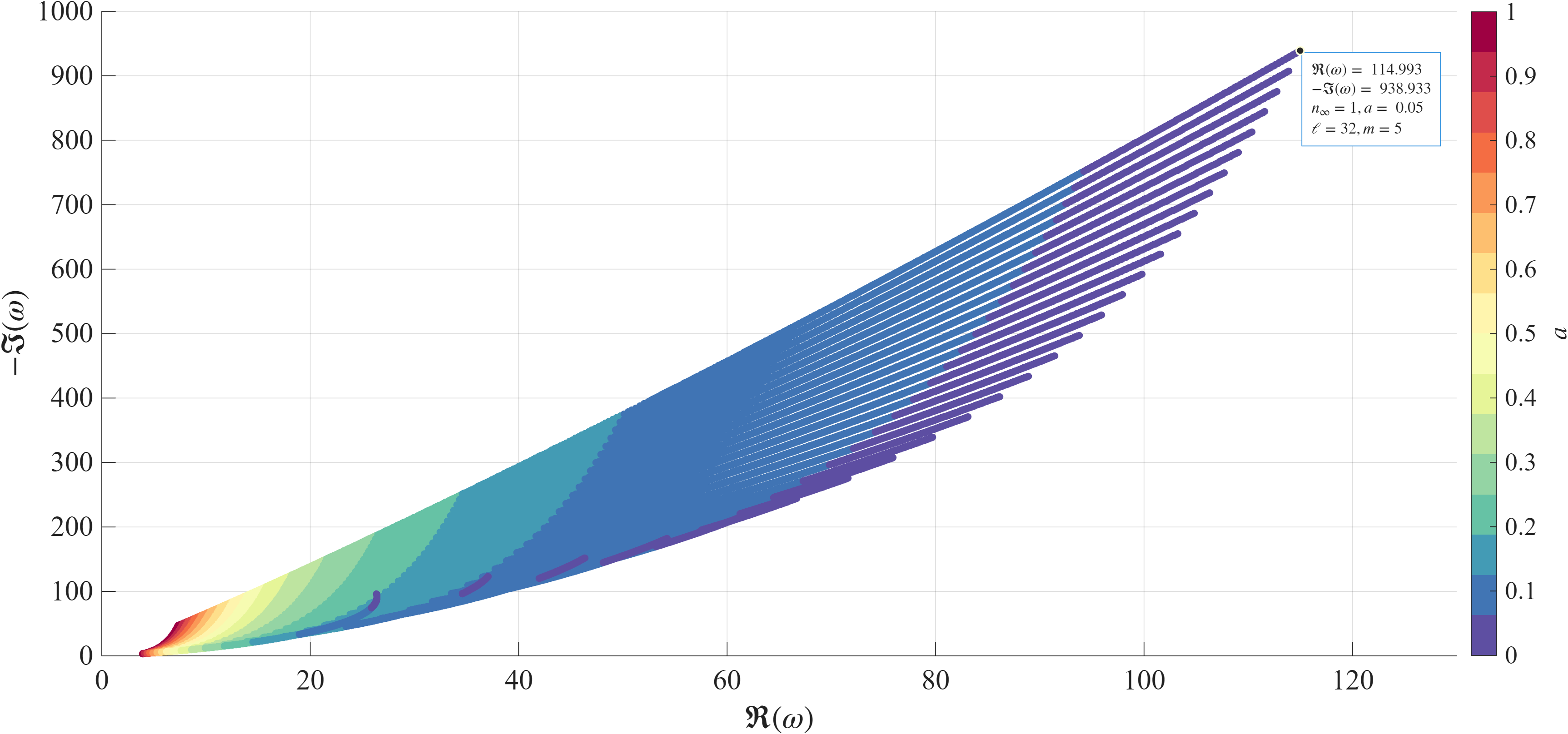}}\,\,
	\subfloat[Logarithmic scale for  $n_{\infty}=1$\label{subfig:TTM_M5_LogN1}]{\includegraphics[width=3.4in]{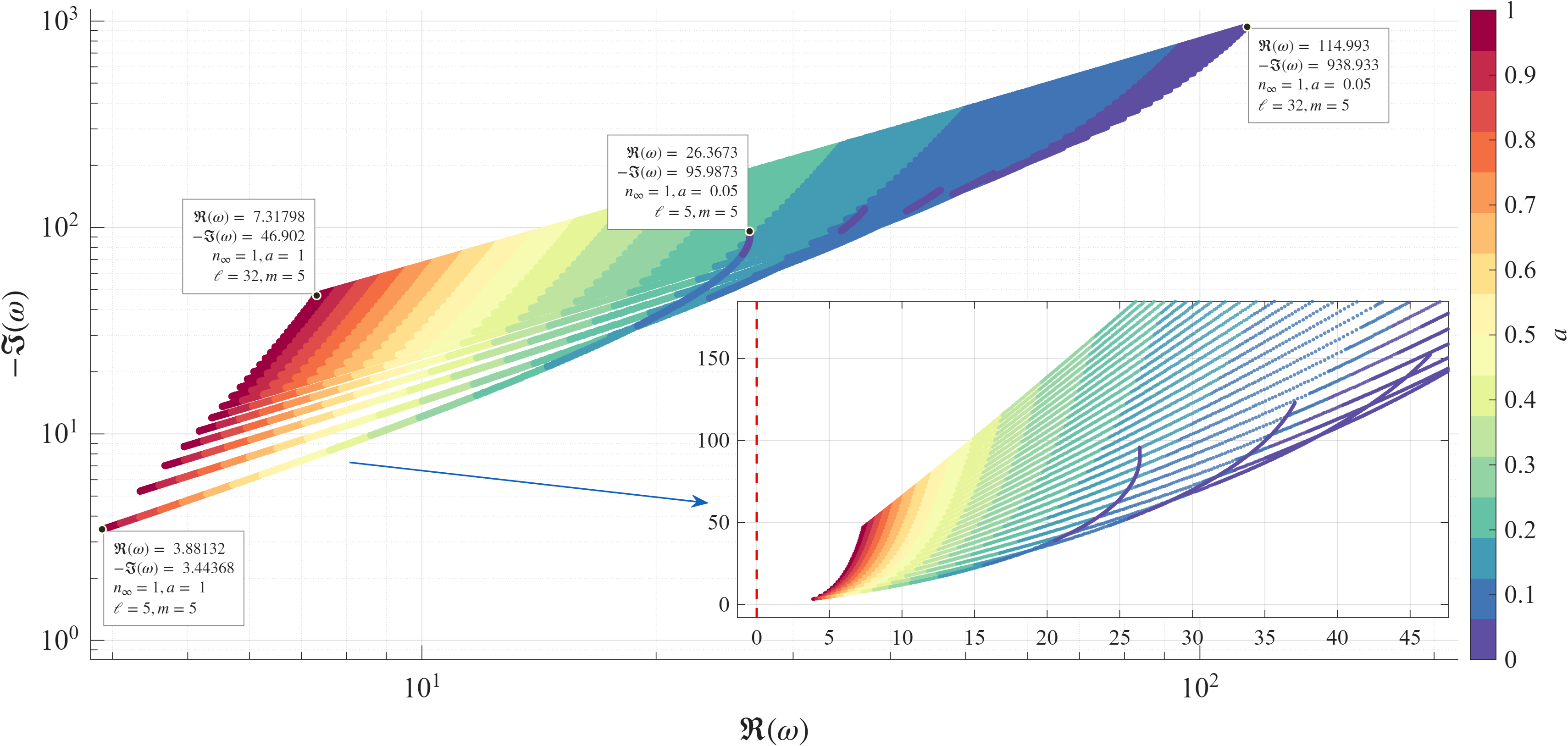}}\,\,
	\subfloat[$n_{\infty}=2$\label{subfig:TTM_M5_N2}]{\includegraphics[width=3.4in]{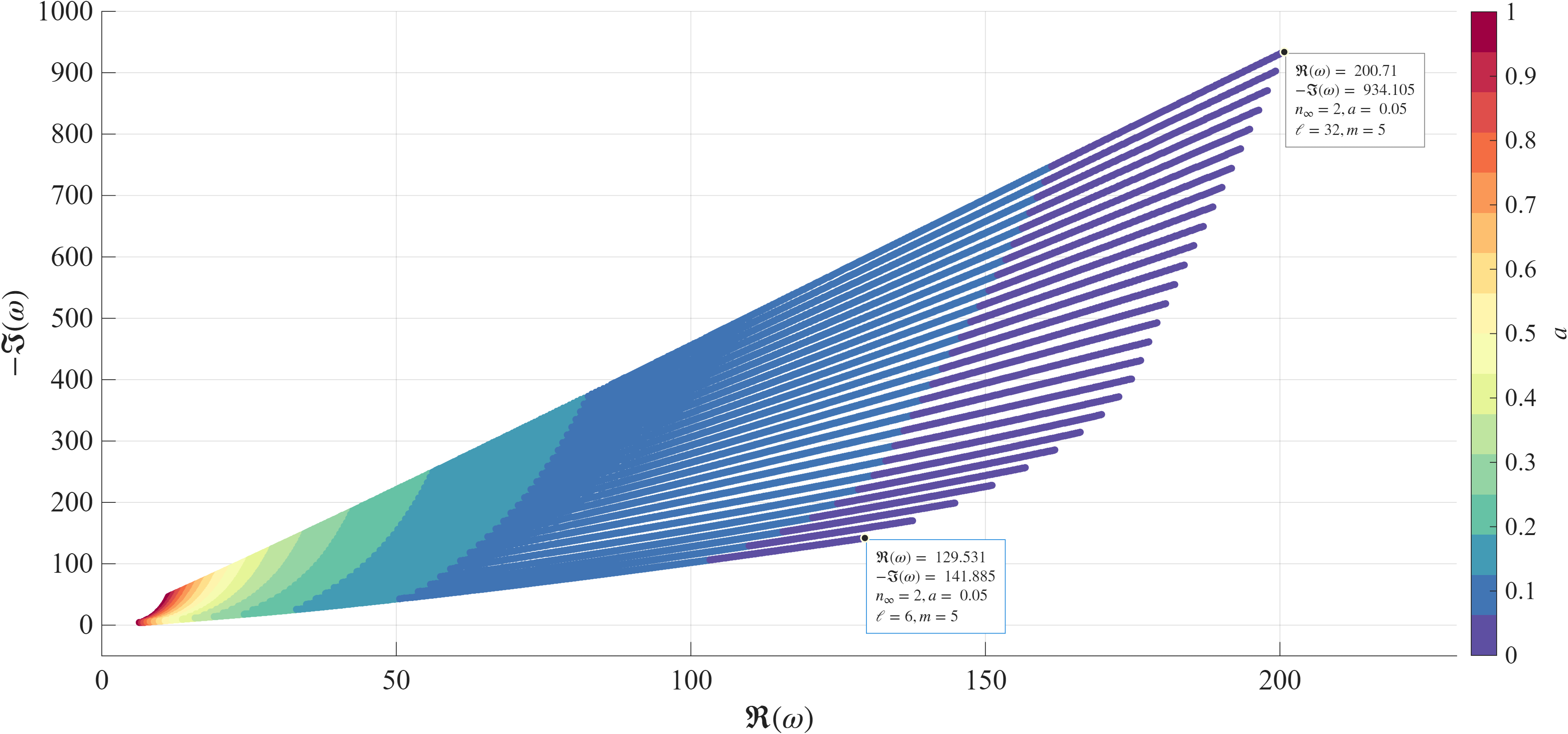}}\,\,
	\subfloat[Logarithmic scale for  $n_{\infty}=2$\label{subfig:TTM_M5_LogN2}]{\includegraphics[width=3.4in]{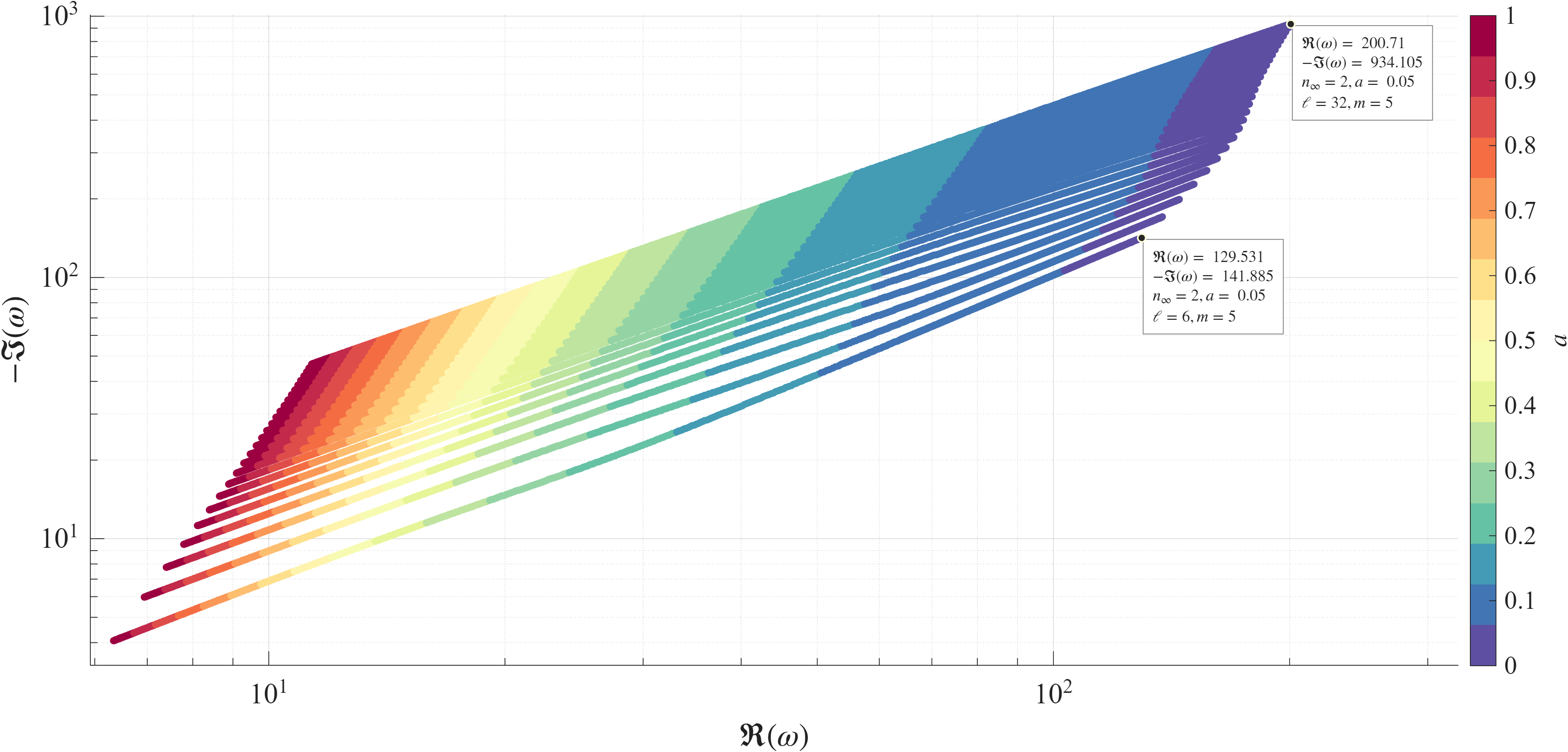}}\,\,
	\subfloat[$n_{\infty}=3$\label{subfig:TTM_M5_N3}]{\includegraphics[width=3.4in]{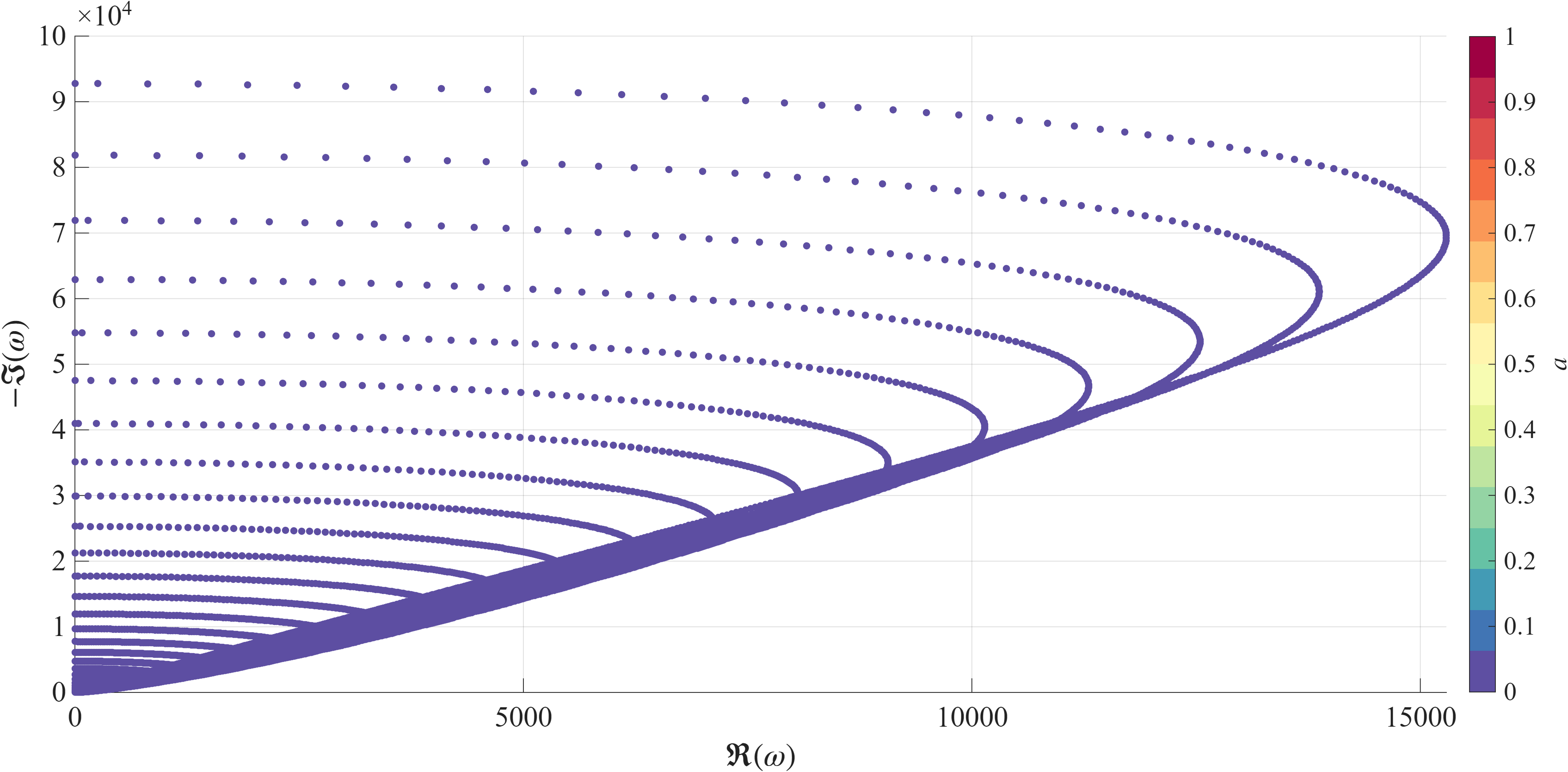}}\,\,
	\subfloat[Logarithmic scale for  $n_{\infty}=3$\label{subfig:TTM_M5_LogN3}]{\includegraphics[width=3.4in]{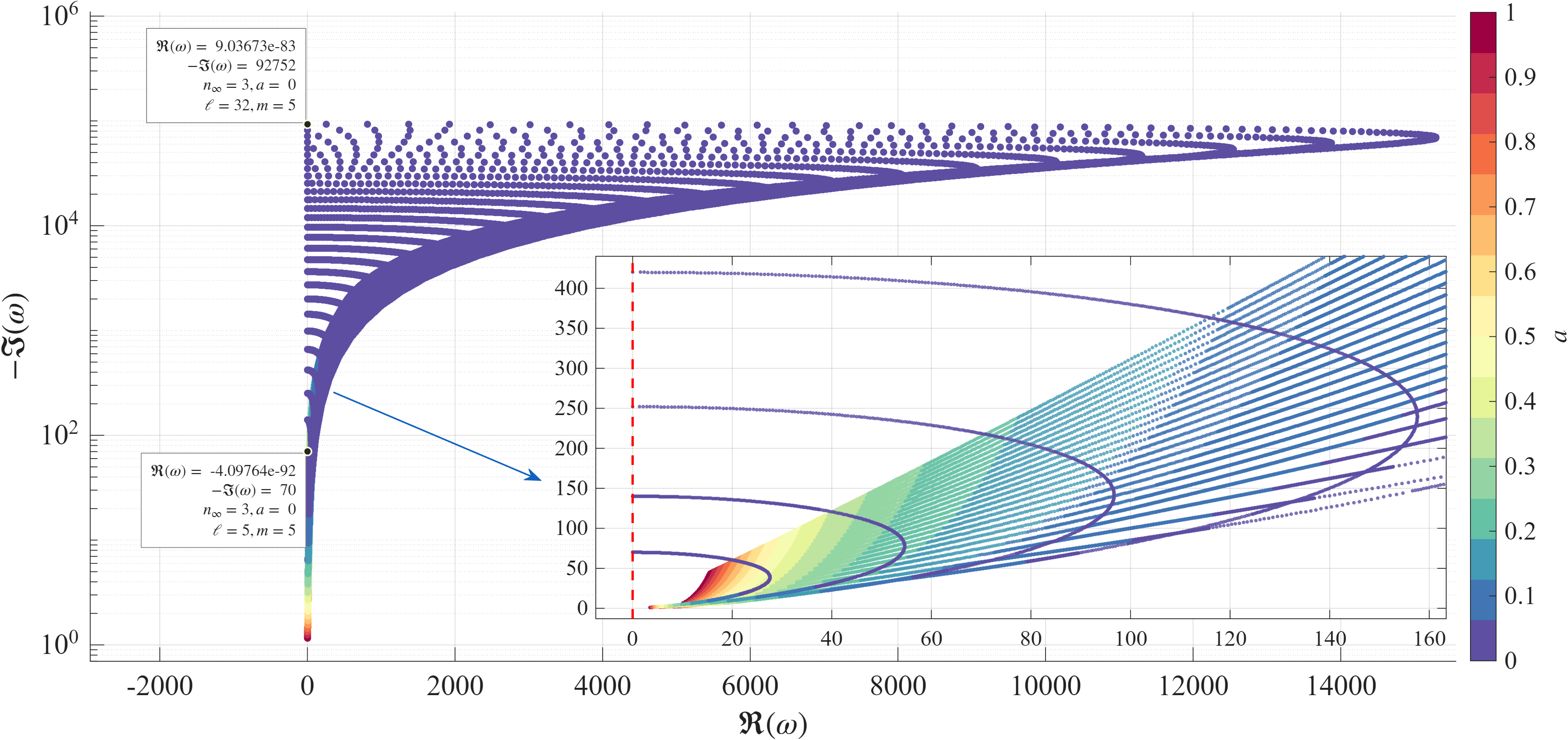}}\,\,
	\subfloat[$n_{\infty}=4$\label{subfig:TTM_M5_N4}]{\includegraphics[width=3.4in]{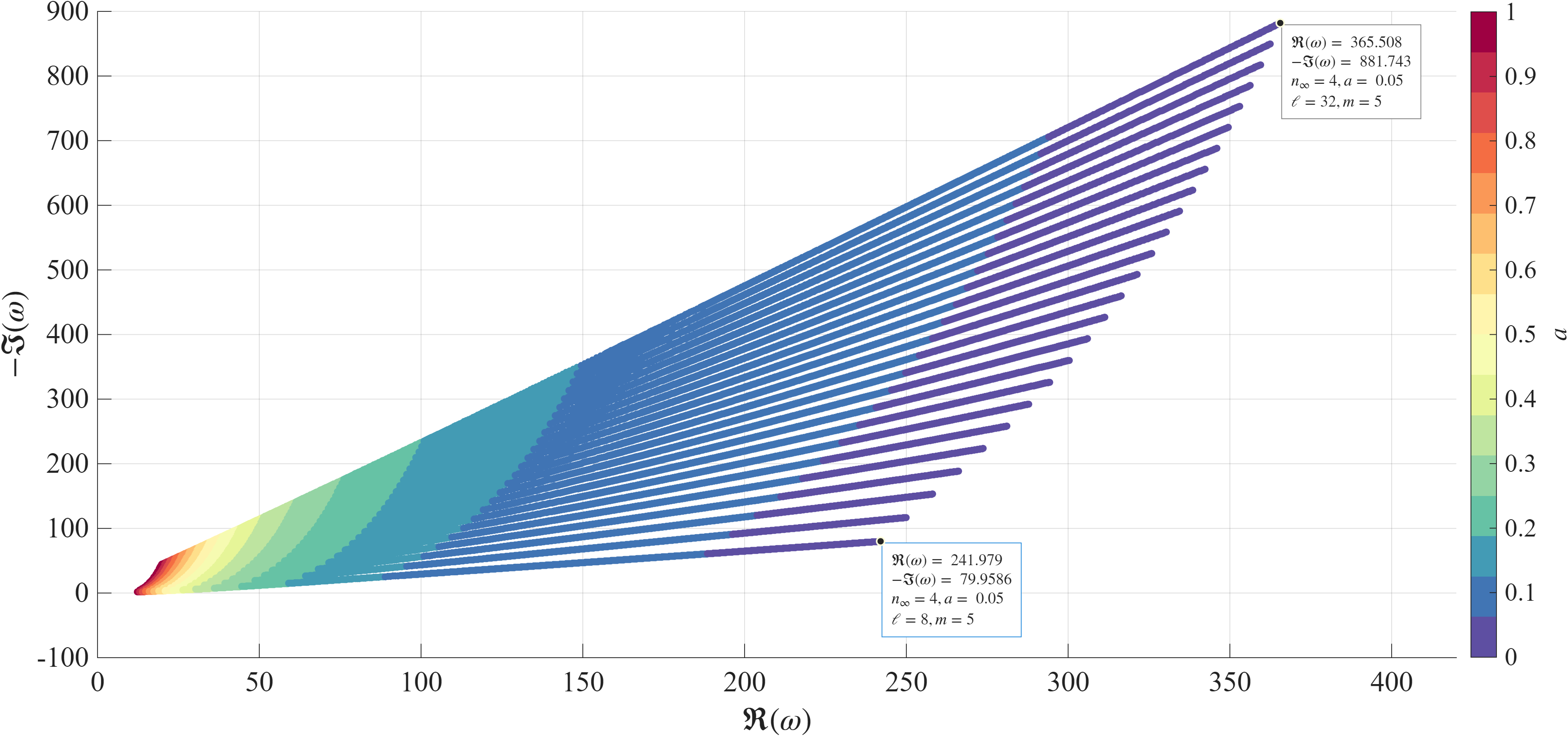}}\,\,
	\subfloat[Logarithmic scale for  $n_{\infty}=4$\label{subfig:TTM_M5_LogN4}]{\includegraphics[width=3.4in]{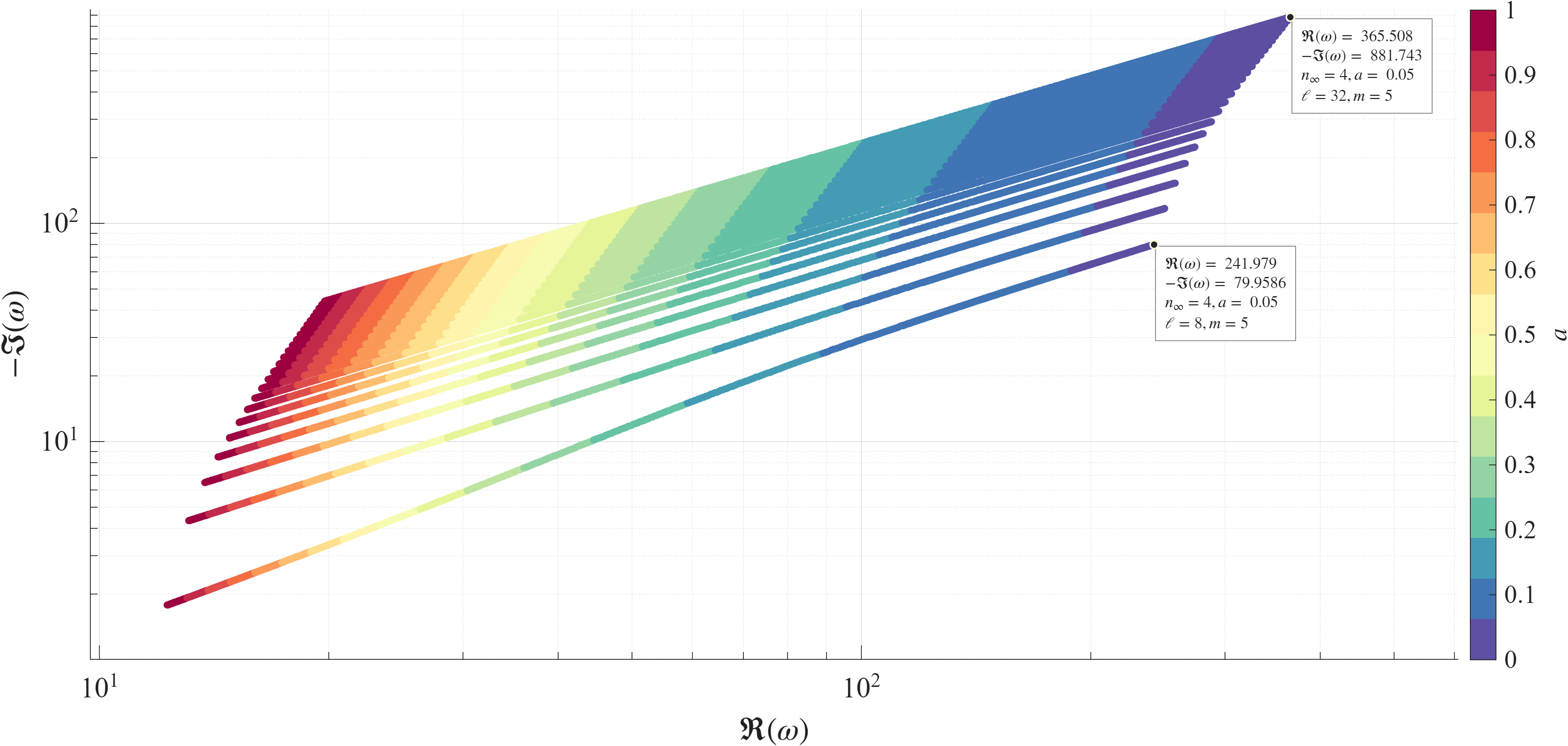}}
	\caption{Kerr TTM frequency trajectories for $m=5$ and $5\leq \ell \leq 32$. Panels on the left present the trajectories on a linear scale, while panels on the right display the corresponding trajectories on a logarithmic scale. For the $n_{\infty}=3$ family, the trajectories are shown over the full range $a\in[0,1]$, since its TTM frequencies remain finite as $a\to0$. For the other families, the small-$a$ limit is singular: their frequencies diverge as $a\to0$, so the trajectories are shown only for $a\in[0.05,1]$.}
	\label{fig:TTM_M5}
\end{figure*}

\begin{figure*}[htbp]
	\centering
	\subfloat[$n_{\infty}=1$\label{subfig:TTM_M10_N1}]{\includegraphics[width=3.4in]{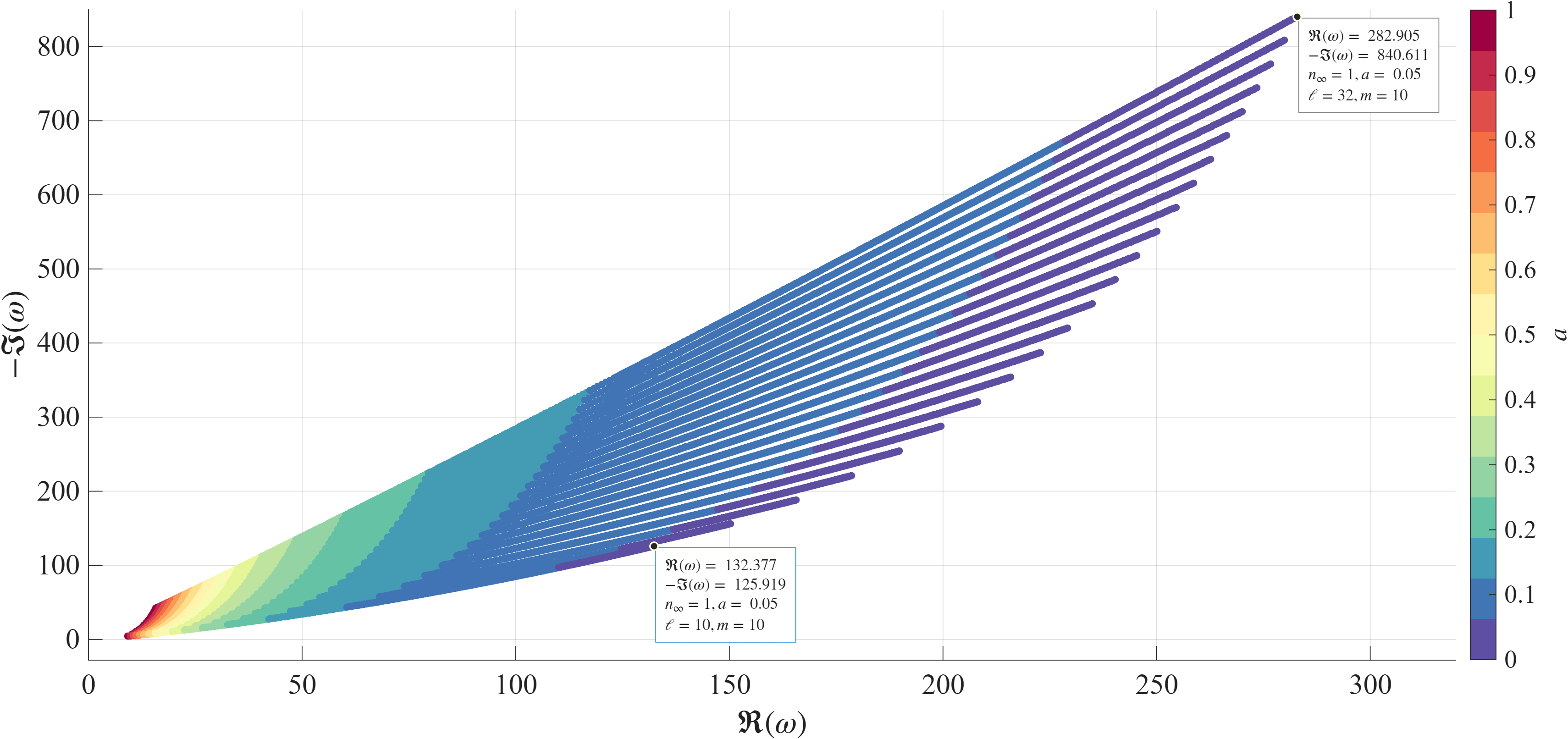}}\,\,
	\subfloat[Logarithmic scale for  $n_{\infty}=1$\label{subfig:TTM_M10_LogN1}]{\includegraphics[width=3.4in]{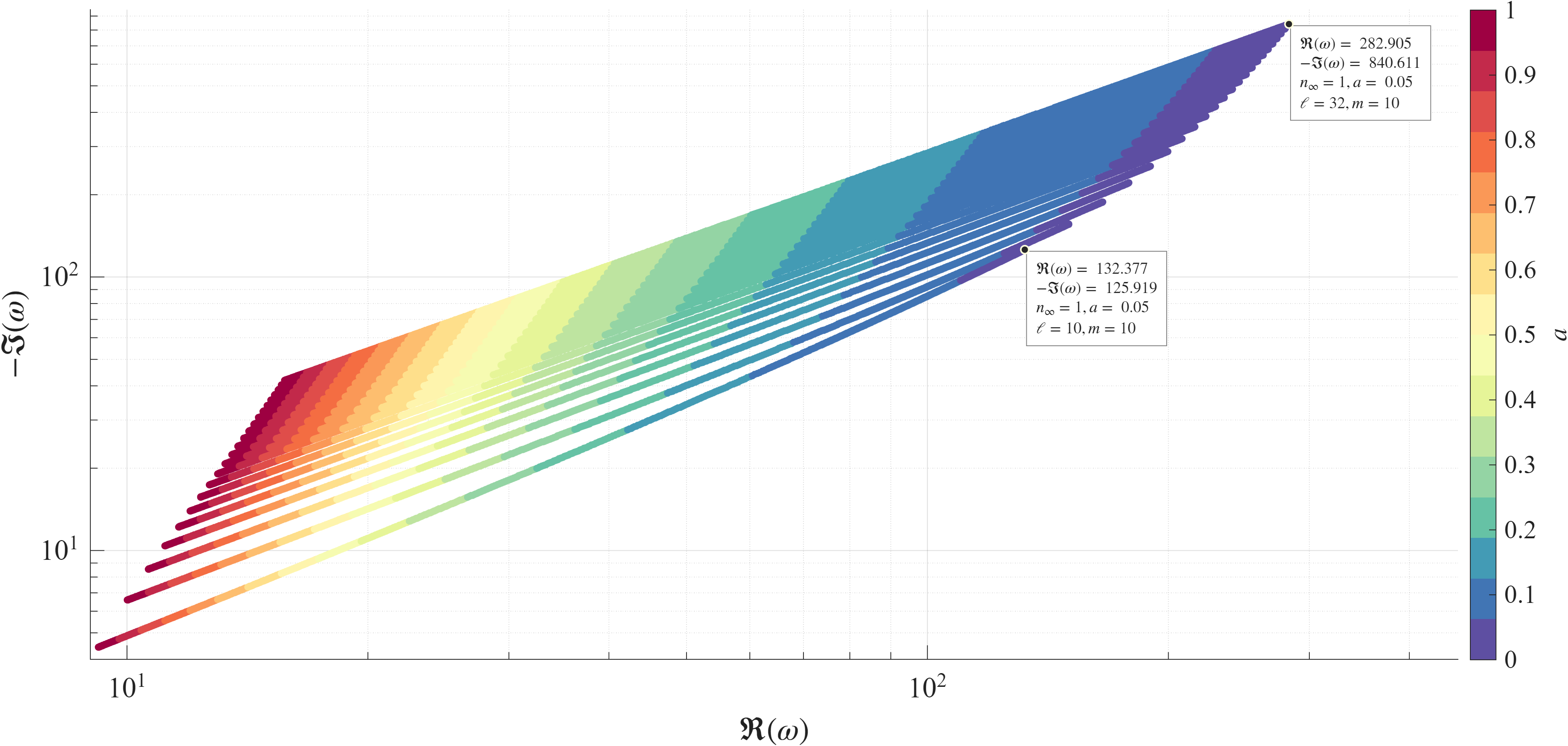}}\,\,
	\subfloat[$n_{\infty}=2$\label{subfig:TTM_M10_N2}]{\includegraphics[width=3.4in]{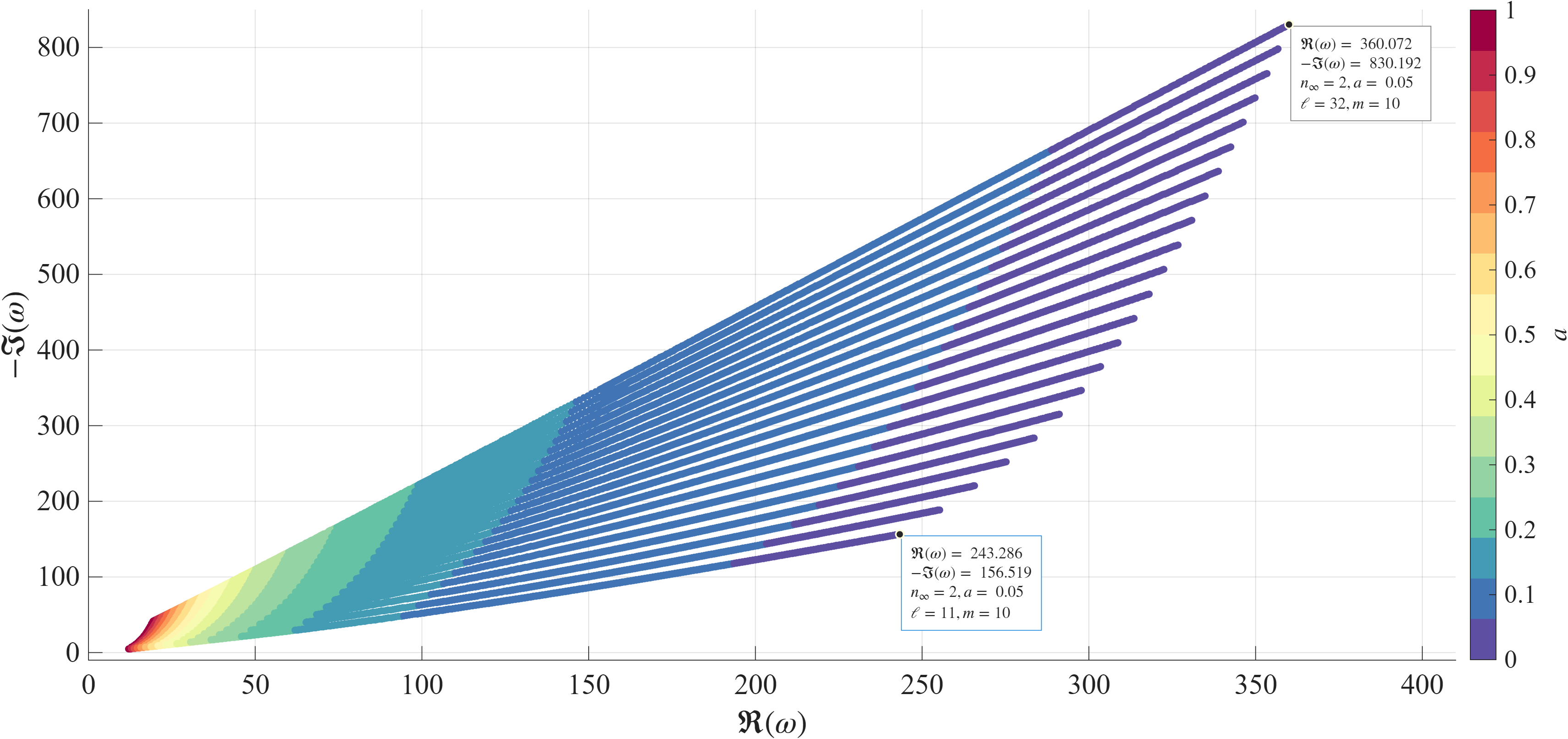}}\,\,
	\subfloat[Logarithmic scale for  $n_{\infty}=2$\label{subfig:TTM_M10_LogN2}]{\includegraphics[width=3.4in]{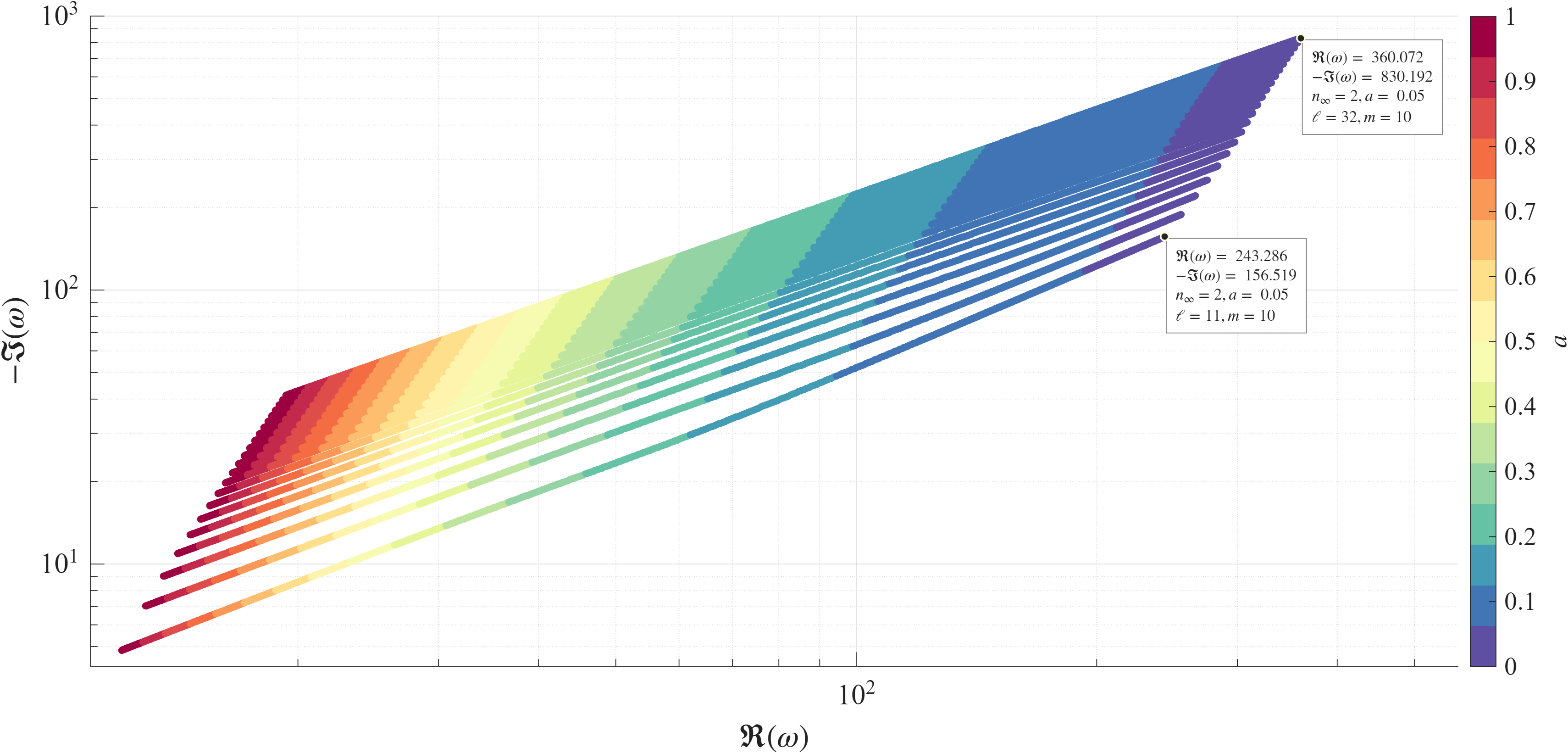}}\,\,
	\subfloat[$n_{\infty}=3$\label{subfig:TTM_M10_N3}]{\includegraphics[width=3.4in]{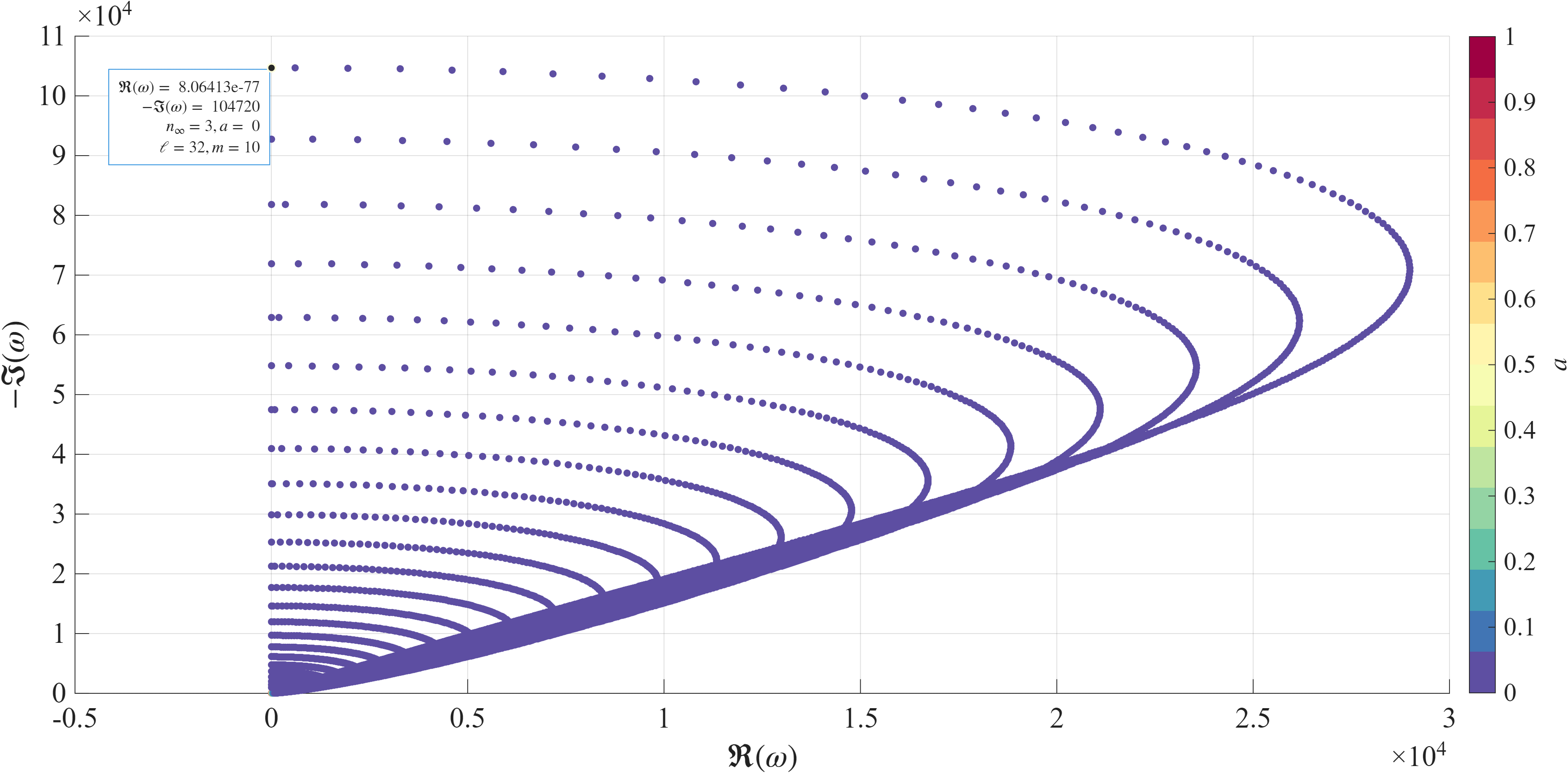}}\,\,
	\subfloat[Logarithmic scale for  $n_{\infty}=3$\label{subfig:TTM_M10_LogN3}]{\includegraphics[width=3.4in]{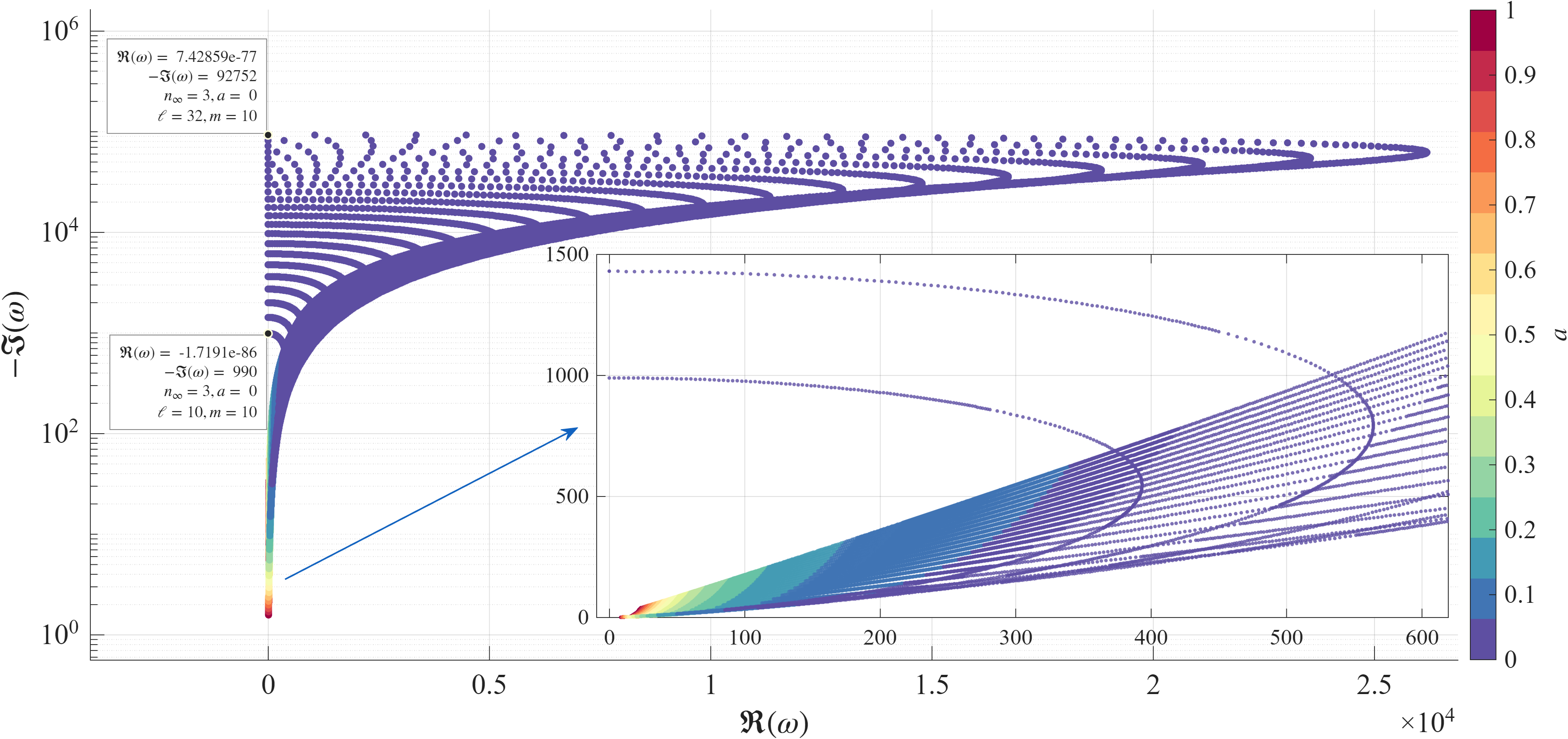}}\,\,
	\subfloat[$n_{\infty}=4$\label{subfig:TTM_M10_N4}]{\includegraphics[width=3.4in]{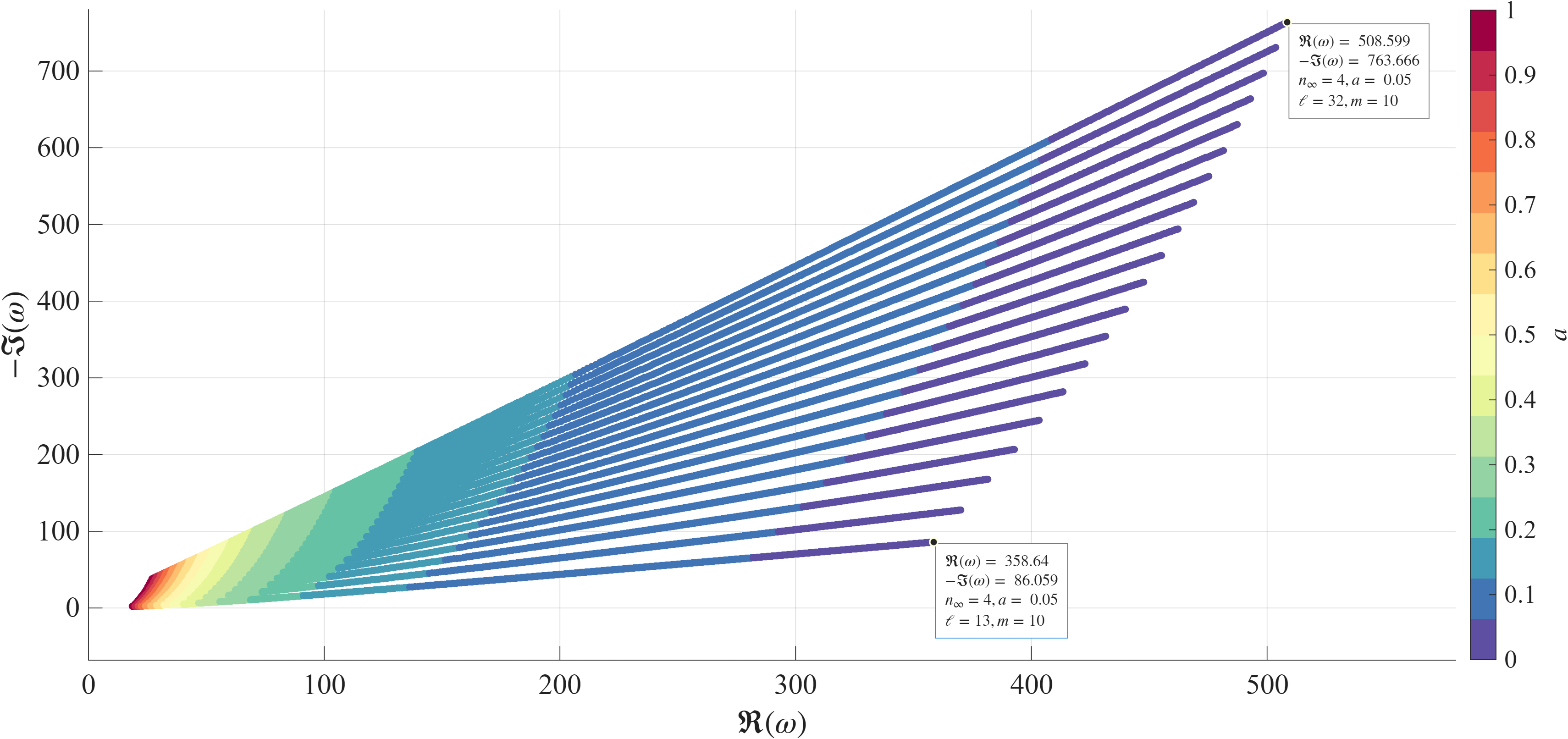}}\,\,
	\subfloat[Logarithmic scale for  $n_{\infty}=4$\label{subfig:TTM_M10_LogN4}]{\includegraphics[width=3.4in]{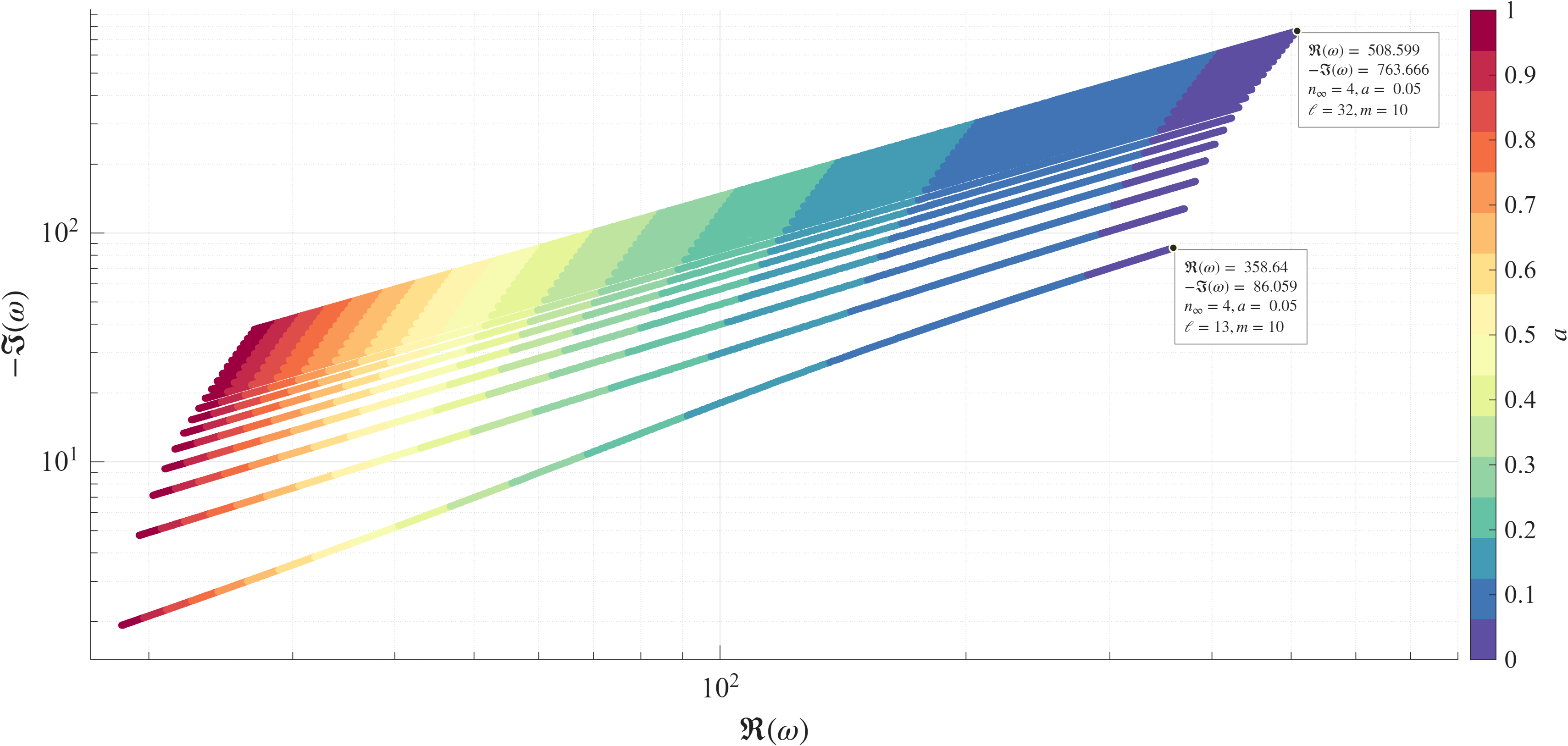}}
	\caption{Kerr TTM frequency trajectories for $m=10$ and $10\leq \ell \leq 32$. Panels on the left present the trajectories on a linear scale, while panels on the right display the corresponding trajectories on a logarithmic scale. For the $n_{\infty}=3$ family, the trajectories are shown over the full range $a\in[0,1]$, since its TTM frequencies remain finite as $a\to0$. For the other families, the small-$a$ limit is singular: their frequencies diverge as $a\to0$, so the trajectories are shown only for $a\in[0.05,1]$.}
	\label{fig:TTM_M10}
\end{figure*}

\begin{figure*}[htbp]
	\centering
	\subfloat[$n_{\infty}=1, m=1$\label{subfig:TTM_sa_N1_M1}]{\includegraphics[width=3.4in]{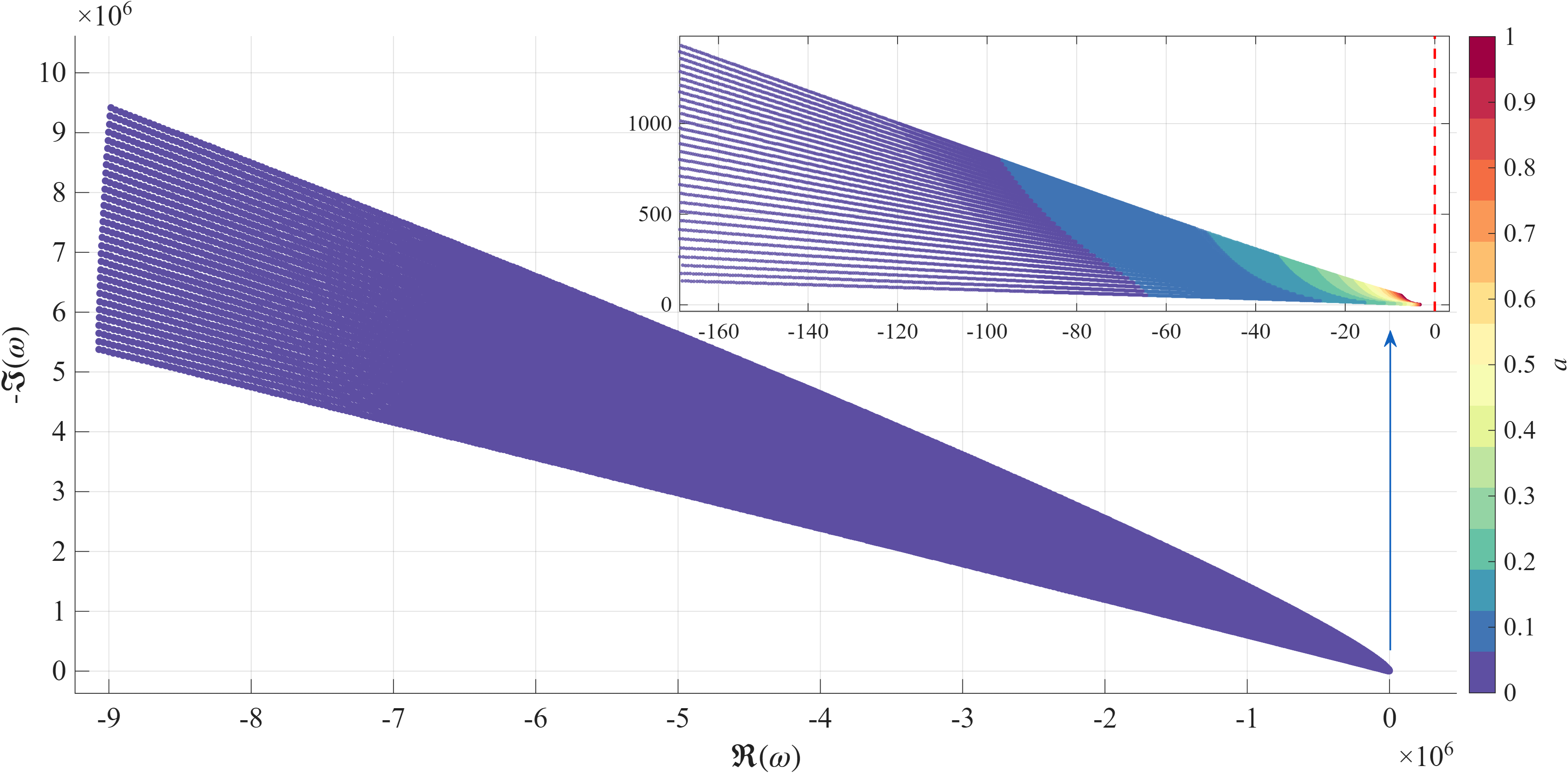}}\,\,
	\subfloat[$n_{\infty}=1, m=10$\label{subfig:TTM_sa_N1_M10}]{\includegraphics[width=3.4in]{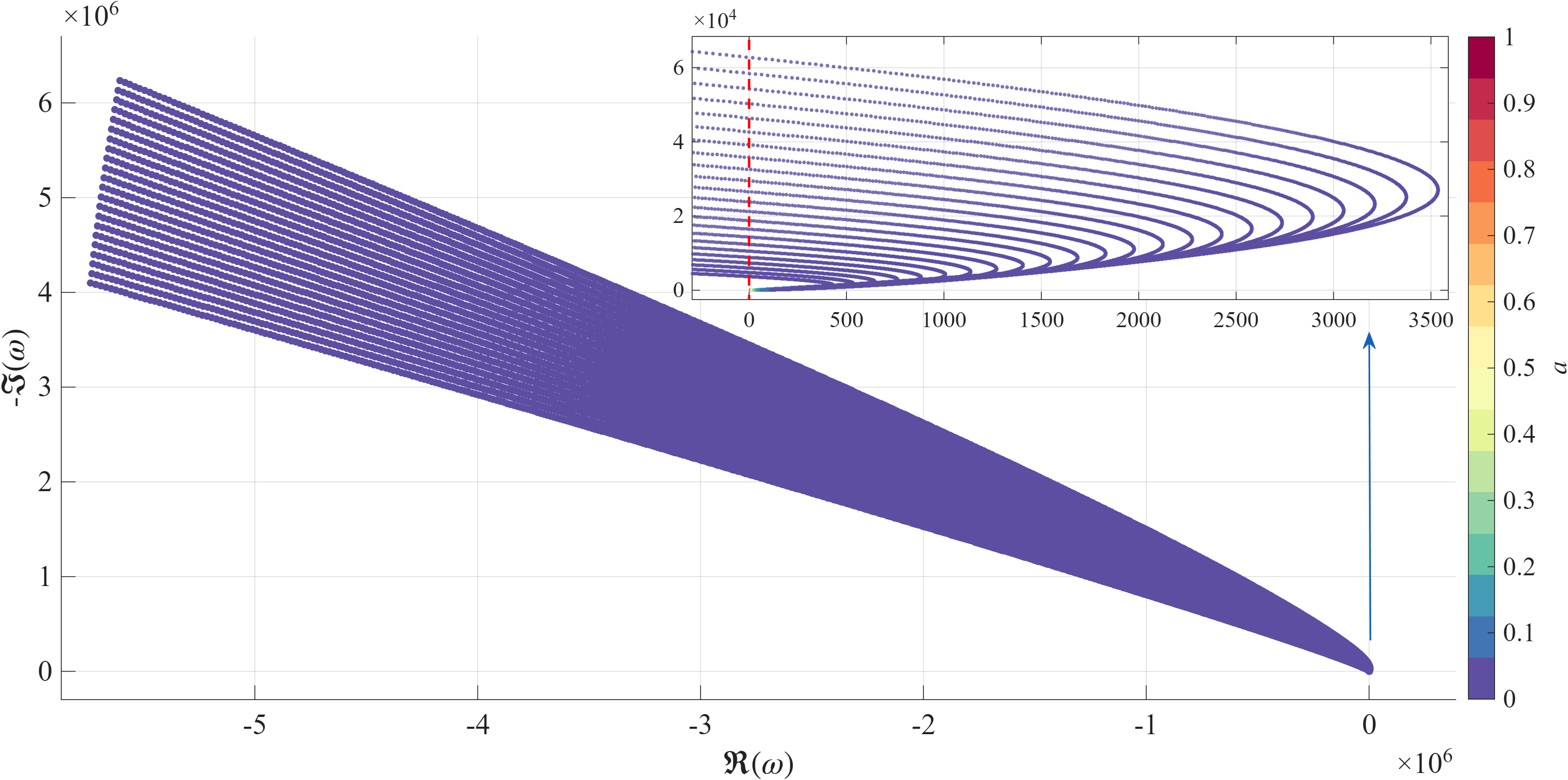}}\,\,
	\subfloat[$n_{\infty}=2, m=1$\label{subfig:TTM_sa_N2_M1}]{\includegraphics[width=3.4in]{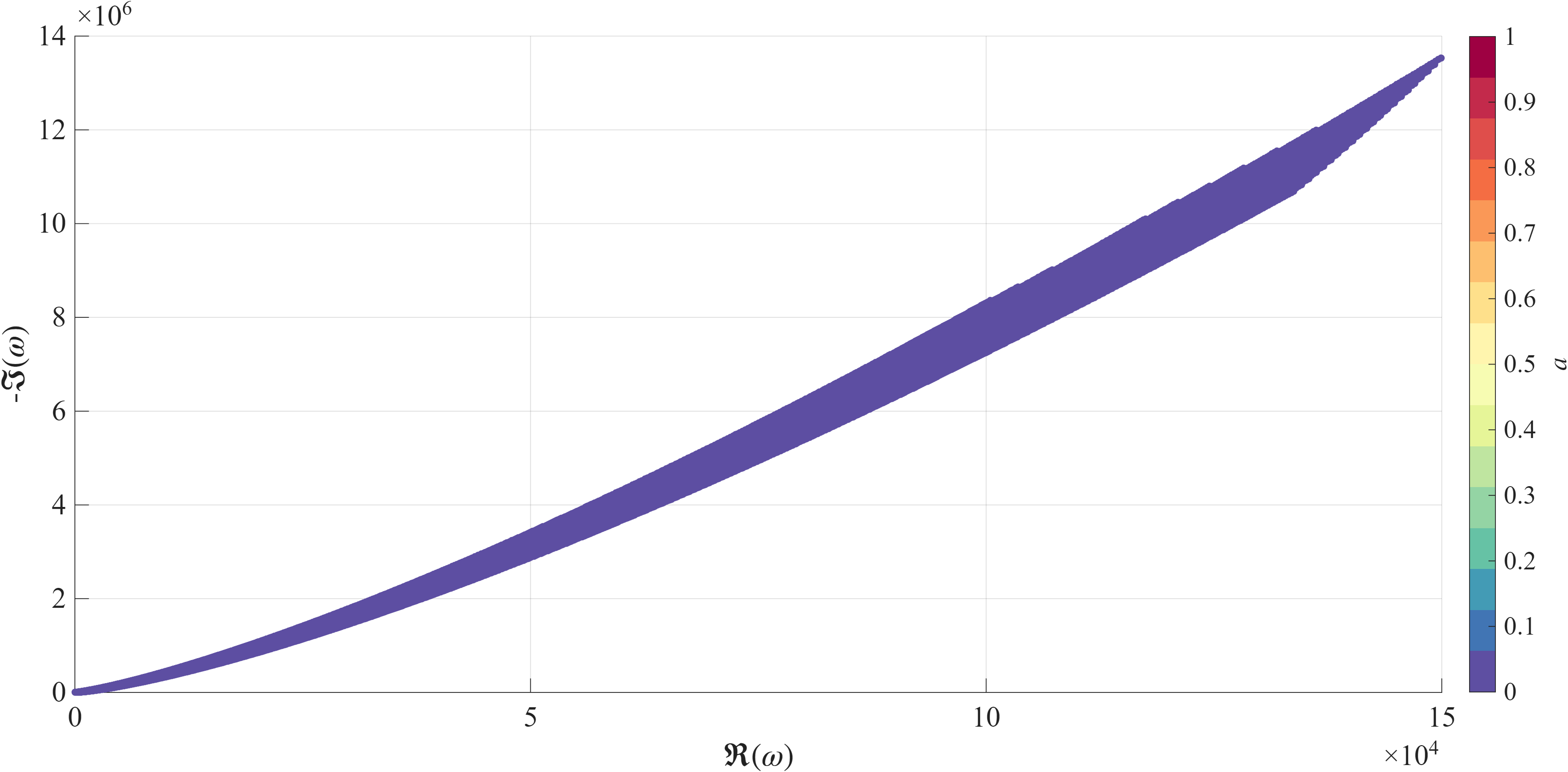}}\,\,
	\subfloat[$n_{\infty}=2, m=10$\label{subfig:TTM_sa_N2_M10}]{\includegraphics[width=3.4in]{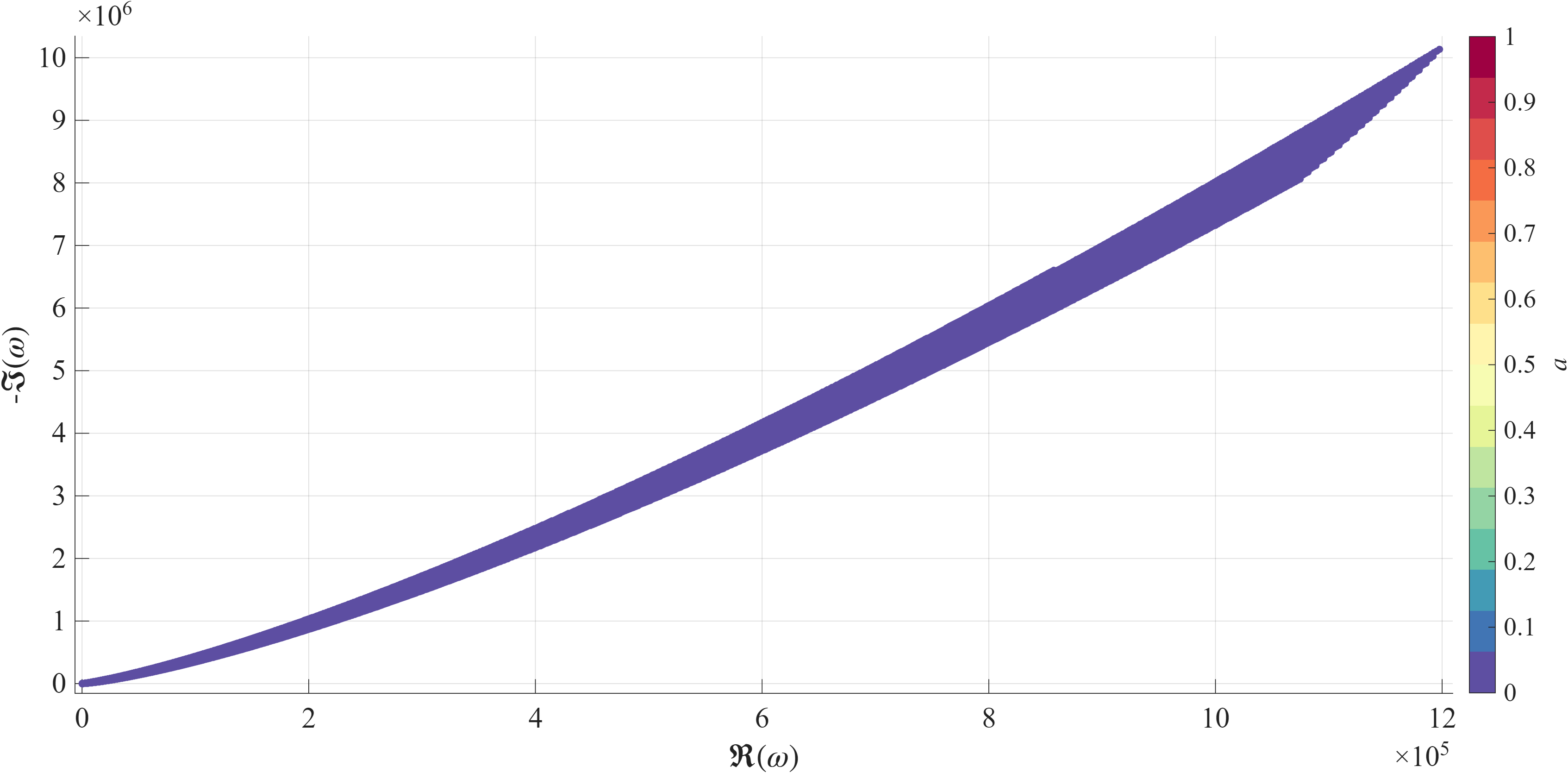}}\,\,
	\subfloat[$n_{\infty}=4, m=1$\label{subfig:TTM_sa_N4_M1}]{\includegraphics[width=3.4in]{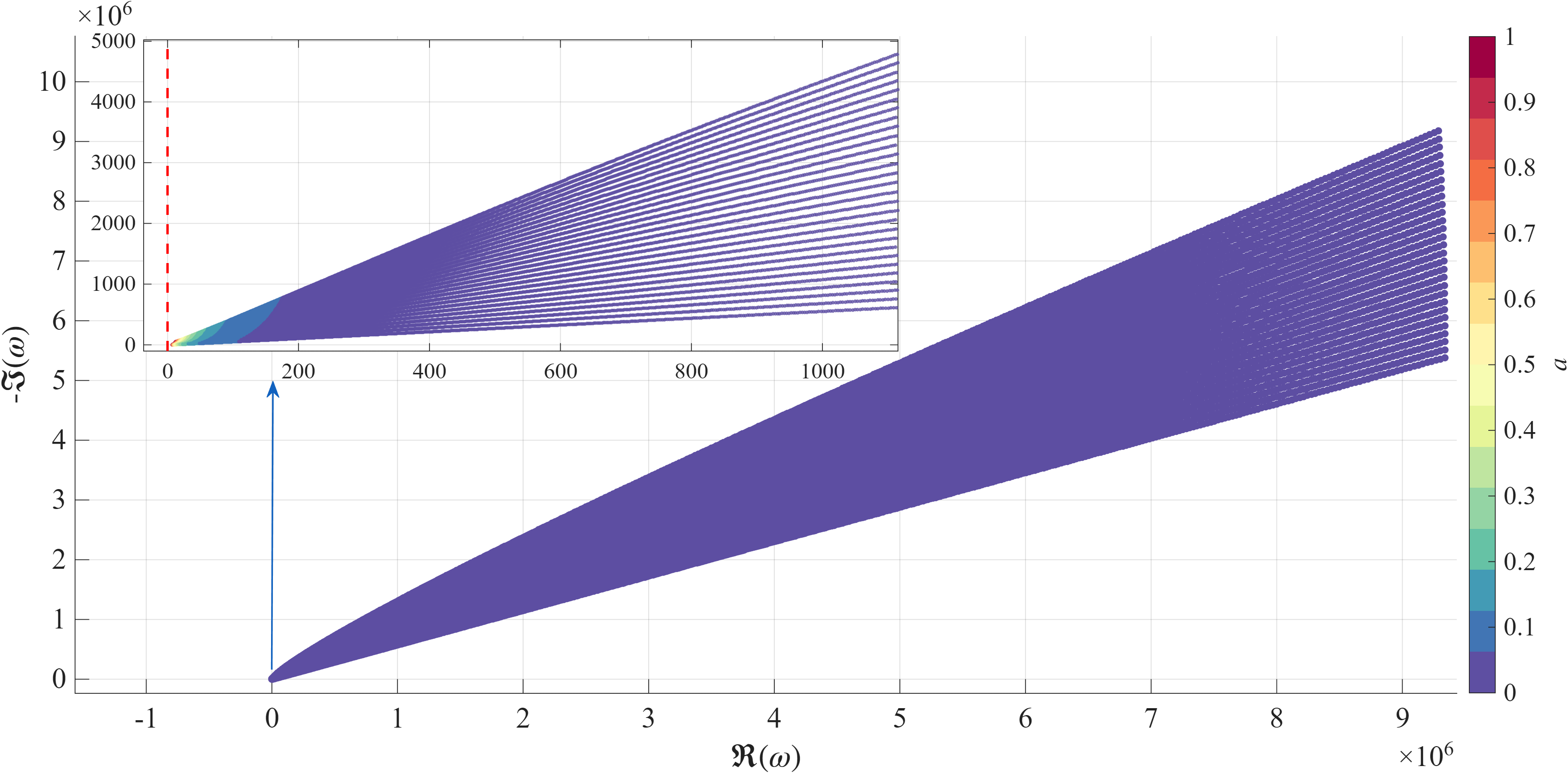}}\,\,
	\subfloat[$n_{\infty}=4, m=10$\label{subfig:TTM_sa_N4_M10}]{\includegraphics[width=3.4in]{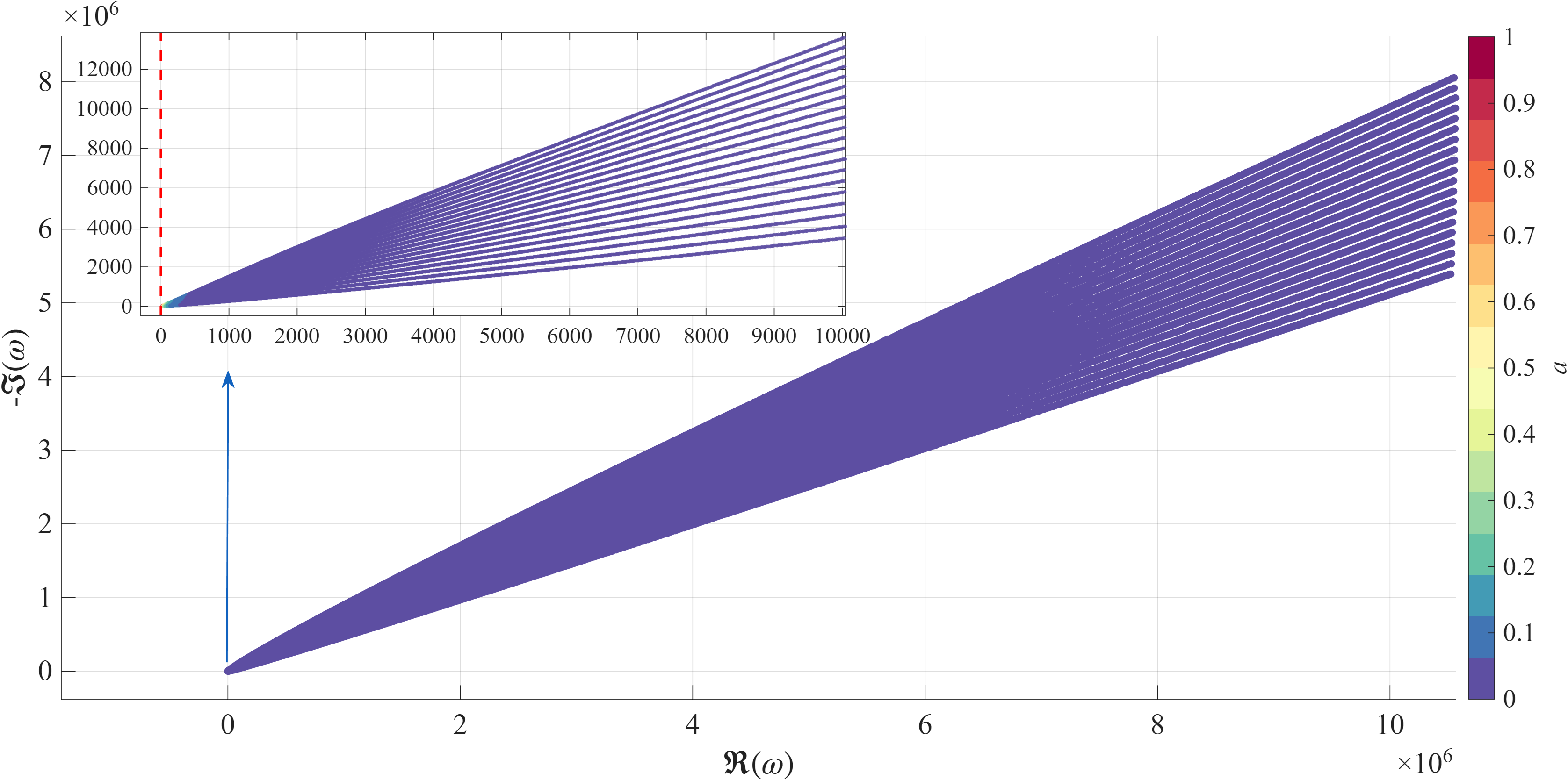}}
	\caption{Kerr TTM frequency trajectories for the three divergent families $n_\infty=1,2,$ and $4$ for $m=1$ and $m=10$, including all available angular branches up to $\ell=32$. The trajectories are shown over $a\in[10^{-5},1]$.}\label{fig:TTM_M1M10_small_spin}
\end{figure*}
The second-order zero condition~\eqref{eq:EP_scattering_derivatives} has direct and dramatic consequences for the unified excitation factor $\mathscr{B}_{\rm R}$ defined in \cref{eq:unified-excitation-factor}.
Because $\alpha_{\rm R} = i\,d\mathcal{F}_{\rm R}/d\omega \to 0$ as $a \to a_{\rm EP}$, the simple-pole residue formula exhibits a sharp resonant enhancement.
Figures~\ref{fig:TTM_M0_ExcFac_L2} and \ref{fig:TTM_M0_ExcFac_L3} illustrate the evolution of the excitation factors for the coalescing $n_\infty=2$ and $n_\infty=3$ branches with $\ell=2$ and $\ell=3$, respectively.
As shown in \Cref{subfig:TTM_ExcFac_L2_mag,subfig:TTM_ExcFac_L3_mag}, the magnitudes $|\mathscr{B}|$ of both branches undergo a pronounced divergence as the black hole spin approaches the critical value $a_{\rm EP}$ ($a_{\rm EP} \approx 0.4944459549144958$ for $\ell=2$ and $a_{\rm EP} \approx 0.2731627005644386$ for $\ell=3$).
The peak magnitude exceeds $10^7$ for $\ell=2$ and surpasses $10^{14}$ for $\ell=3$, clearly reflecting the collapse of the simple-zero assumption.

Figures~\ref{fig:TTM_M0_ExcFac_L2} and \ref{fig:TTM_M0_ExcFac_L3} depict the trajectories of $\mathscr{B}_{\rm R}$ in the complex plane as $a$ is varied.
From a global perspective (\Cref{subfig:TTM_ExcFac_L2_global,subfig:TTM_ExcFac_L3_global}), the excitation factors form nested spiral orbits.
In the regime far from the exceptional point, the trajectories wind tightly inward toward the origin into compact loops.
As $a$ decreases and approaches the exceptional point $a_{\rm EP}$, the excitation factors rapidly unwind and shoot outward, expanding into enormous spiral orbits across multiple orders of magnitude.
In the detailed views near the exceptional point (\Cref{subfig:TTM_ExcFac_L2_detail,subfig:TTM_ExcFac_L3_detail}), the $n_\infty=2$ and $n_\infty=3$ branches exhibit clear geometric mirror symmetry across the coordinate axes, directly manifesting the square-root branch-point topology and the characteristic phase rotation associated with the exceptional point.

\subsubsection{$m\geq1$ TTM spectrum}
\label{sec:m_geq_1_TTM}

For nonaxisymmetric perturbations with $m\geq1$, we compute the Kerr TTM spectra across $m\leq\ell\leq32$ for the representative cases $m=1,5,10$.
The frequency trajectories of the four families, $n_\infty \in \{1, 2, 3, 4\}$, are shown in \Cref{fig:TTM_M1,fig:TTM_M5,fig:TTM_M10}.
In each figure, the left column displays the trajectories on a linear scale, while the right column shows the corresponding curves on a logarithmic scale.
These figures illustrate the distinct distribution and small-$a$ behavior of the four families.
The $n_\infty = 3$ family remains regular over the full spin range $a \in [0, 1]$.
As $a\to0$, its real part approaches zero, while the frequency approaches the finite Schwarzschild frequency~\eqref{eq:ASmode}.
The insets in the logarithmic panels of \Cref{fig:TTM_M1,fig:TTM_M5,fig:TTM_M10} highlight this smooth convergence onto the negative imaginary axis for each $\ell$.

At finite spin, the three divergent families follow visibly different paths in the complex-frequency plane. In particular, the $n_\infty=1$ and $n_\infty=4$ trajectories approach an approximate reflection symmetry about the imaginary axis as $a$ decreases. The $n_\infty=2$ trajectories instead cross the imaginary axis at finite spin and then continue toward complex infinity along a distinct path.
The small-$a$ asymptotic behavior of all three divergent families is analyzed in \cref{subsec:asymptotic}.

\begin{figure*}[htbp]
  \centering
  \subfloat[Spectral proximity between the $n_\infty=3$ TTM family and the $n=8$ QNM sequence for $\ell=2$ and $m=\pm1,\pm2$\label{subfig:proximity_traj}]{\includegraphics[width=3.4in]{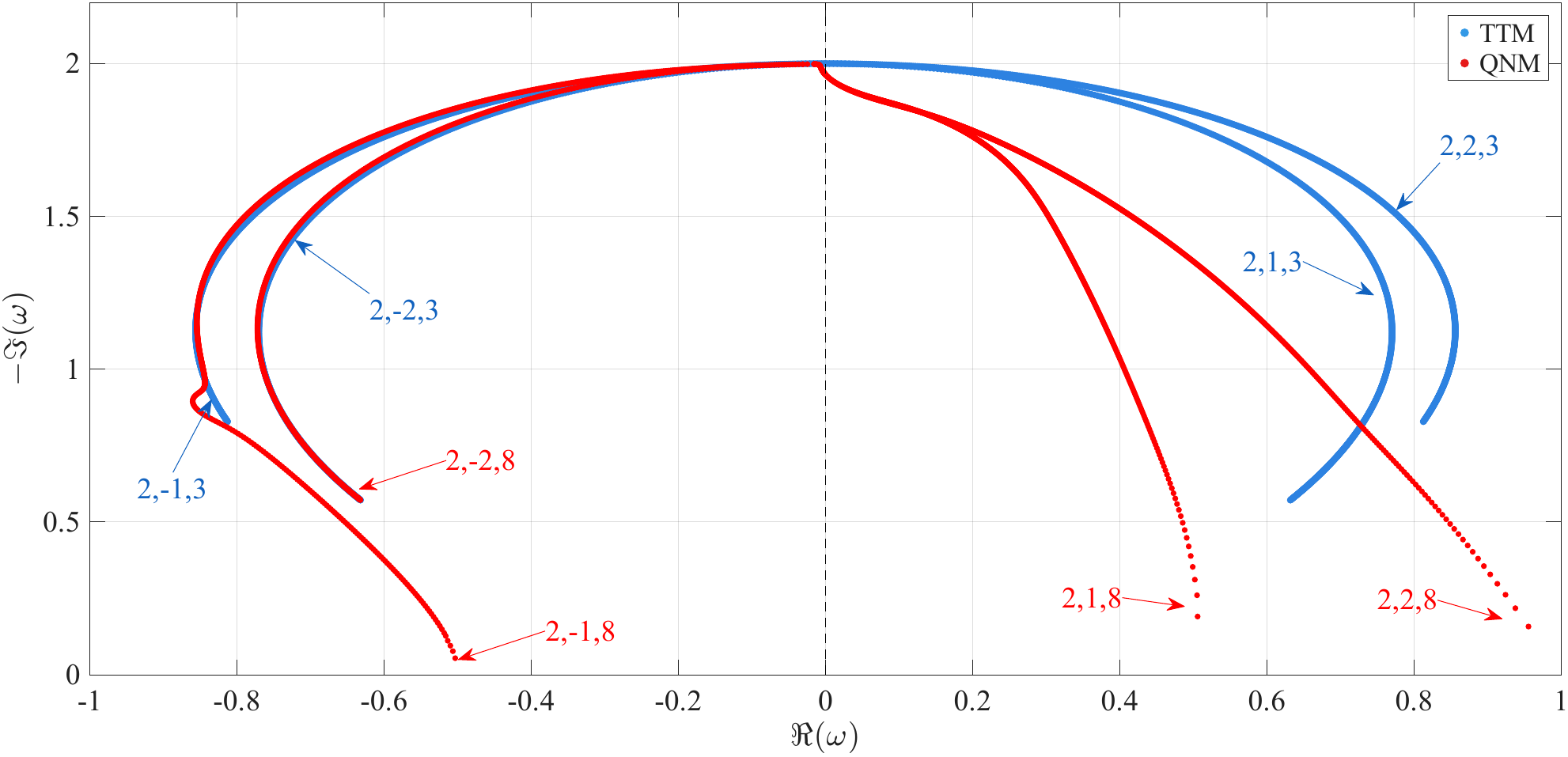}}\hfill
  \subfloat[Absolute frequency deviation between the $\{\ell,m,n_\infty\}=\{2,-2,3\}$ TTM and the $\{\ell,m,n\}=\{2,-2,8\}$ QNM\label{subfig:proximity_err}]{\includegraphics[width=3.4in]{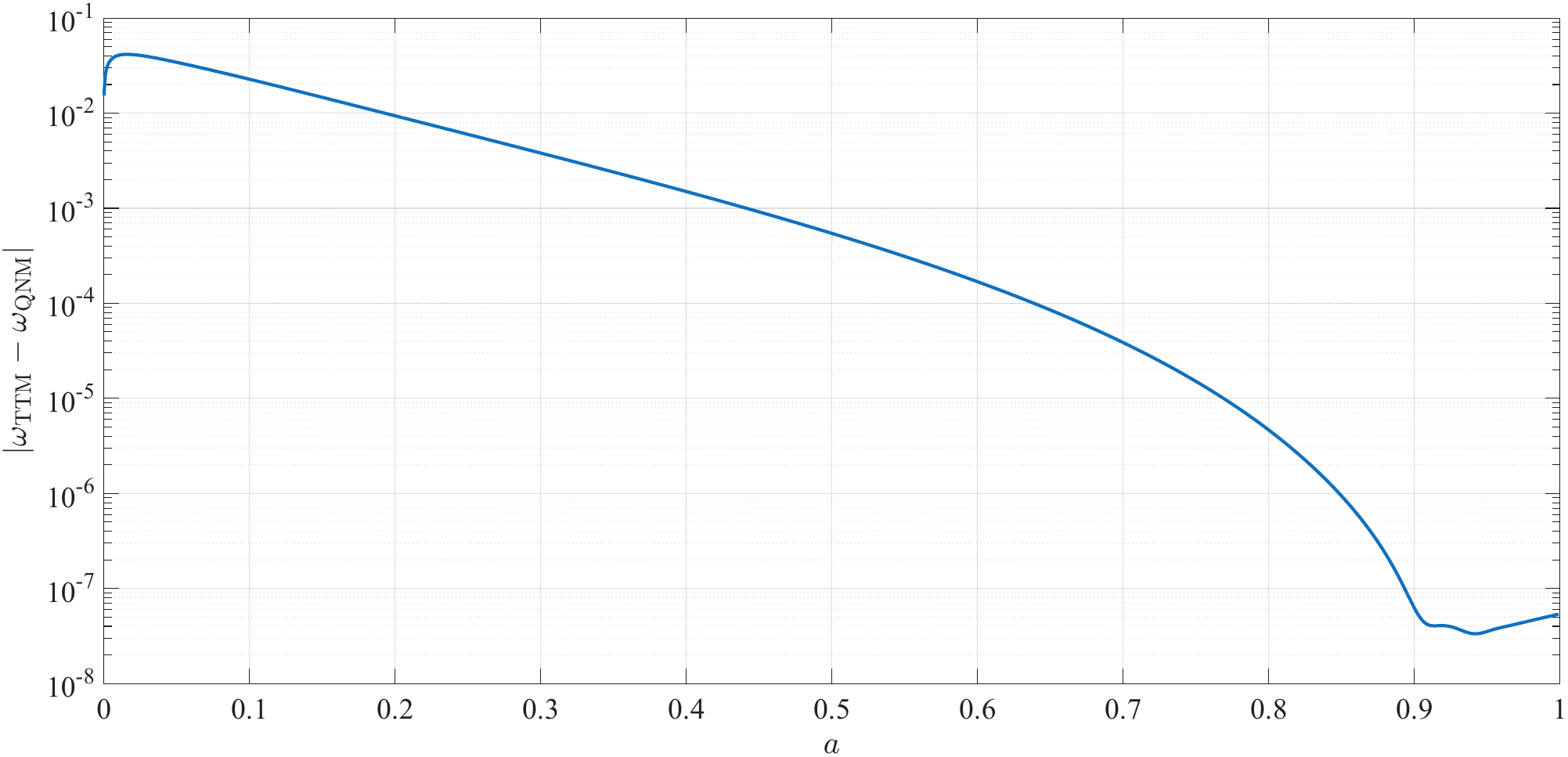}}\\[1ex]
  \subfloat[Residual of the $\{\ell,m,n\}=\{2,-2,8\}$ QNM\label{subfig:proximity_qnm_res}]{\includegraphics[width=3.4in]{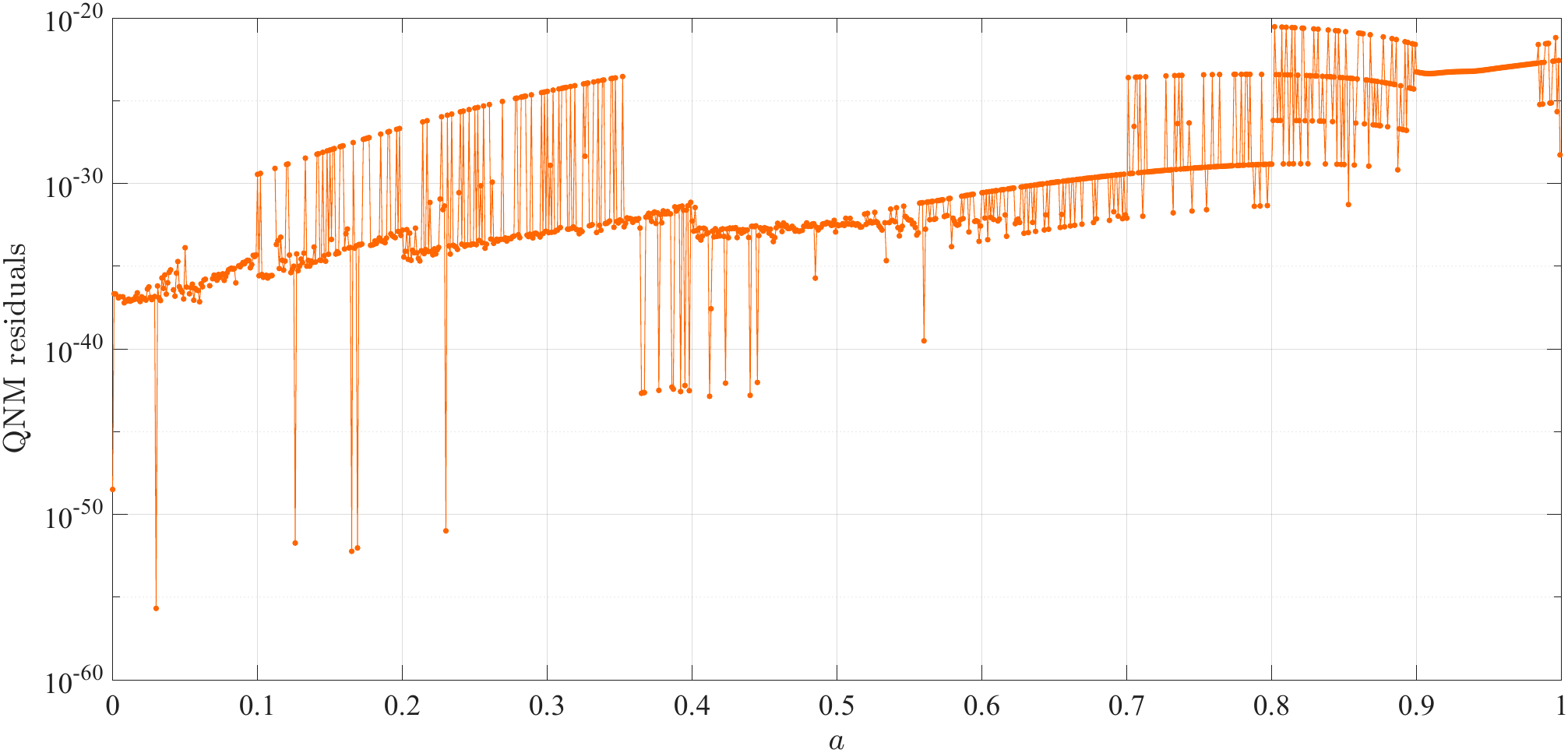}}\hfill
  \subfloat[Residual of the $\{\ell,m,n_\infty\}=\{2,-2,3\}$ TTM\label{subfig:proximity_ttm_res}]{\includegraphics[width=3.4in]{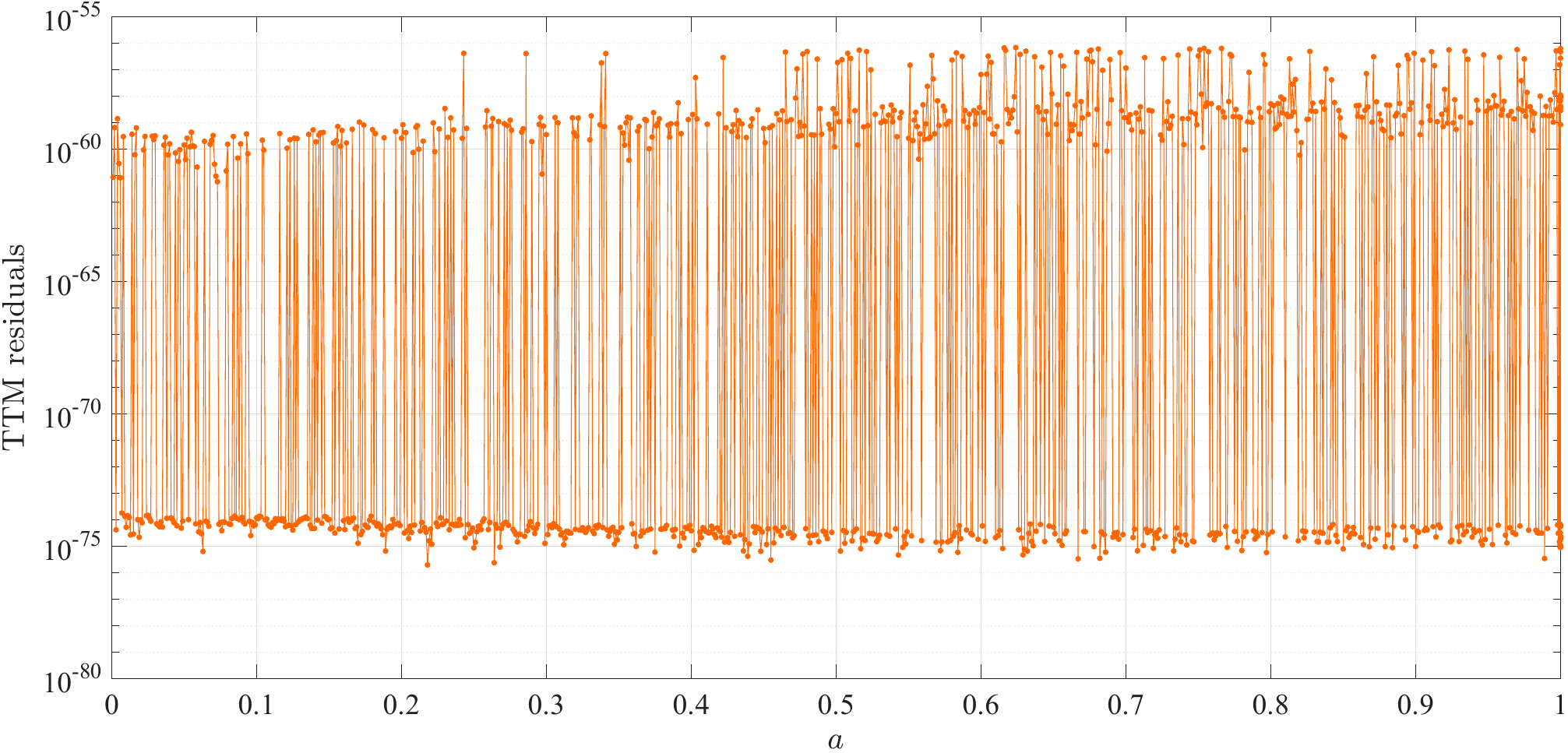}}
  \caption{Spectral proximity between the $n_\infty=3$ TTM family and the unconventional $n=8$ QNM sequence for $\ell=2$.}
  \label{fig:Proximity_TTM_QNM}
\end{figure*}

\subsubsection{Anomalous proximity between QNM and TTM}

In the complex-frequency plane, the $n_\infty=3$ TTM family and the unconventional $n=8$ QNM sequence exhibit an anomalously close spectral approach.
As shown in \cref{subfig:proximity_traj}, this behavior is visible for the $\ell=2$ trajectories with $m=\pm1,\pm2$.
To examine the proximity quantitatively, we focus on the representative pair consisting of the $\{\ell,m,n_\infty\}=\{2,-2,3\}$ TTM and the $\{\ell,m,n\}=\{2,-2,8\}$ QNM over the full spin range $a\in[0,1]$.

At $a=0$, the $\{\ell,m,n_\infty\}=\{2,-2,3\}$ TTM originates from the Schwarzschild algebraically special frequency $\omega_{\rm AS}=-2i$, whereas the $\{\ell,m,n\}=\{2,-2,8\}$ QNM originates from the Schwarzschild unconventional mode
\begin{equation}
\omega\approx -0.015324504760650-1.998411841858943i.
\end{equation}
As the rotation is turned on, the two trajectories approach each other remarkably closely.
The absolute frequency deviation,
\begin{equation}
\left|\omega_{\rm TTM}-\omega_{\rm QNM}\right|,
\end{equation}
is shown in \cref{subfig:proximity_err}.
The deviation decreases as the spin increases and reaches a minimum of order $10^{-8}$ around $a\sim0.94$.

To determine whether this minimum separation is numerically resolved rather than a numerical artifact, we compare it with the residuals of both spectra.
As shown in \Cref{subfig:proximity_qnm_res,subfig:proximity_ttm_res}, the numerical residual of the QNM sequence remains below $10^{-20}$, while the STC residual of the TTM sequence remains below $10^{-55}$ across the entire spin range.
Furthermore, independent evaluations using the radial scattering amplitudes (Appendix~\ref{sec:TTM-Scatter-Amp}) confirm that the boundary-condition reflection residuals for the TTM branch are of the same order of magnitude as the STC residuals.
These residuals are many orders of magnitude smaller than the minimum frequency separation.
The observed separation is therefore well resolved numerically, and the two branches do not coincide.
The QNM and TTM sequences remain distinct for all $a>0$.

We emphasize that this anomalous proximity is not restricted to the particular $\{\ell,m\}=\{2,-2\}$ example analyzed quantitatively here.
More generally, such close spectral pairing occurs between the $n_\infty=3$ TTM family and the QNM sequence originating from the unconventional Schwarzschild mode for $m=-\ell$.

\subsection{Small-spin asymptotics of the TTMs}
\label{subsec:asymptotic}

The trajectories displayed in \Cref{fig:TTM_M1,fig:TTM_M5,fig:TTM_M10} separate the four $n_\infty$ families into two qualitatively different classes as $a\rightarrow0$. The $n_\infty=3$ family remains finite and approaches the algebraically special frequency~\eqref{eq:ASmode}, whereas the $n_\infty=1,2,$ and $4$ families diverge toward complex infinity. The latter behavior is shown explicitly for $m=1$ and $m=10$ in \cref{fig:TTM_M1M10_small_spin}.

The fractional-power scaling of the three divergent families can be obtained directly from the STC~\eqref{eq:Starobinsky_const}. For the $s=-2$ representation considered here, $\lambdabar=\lambda$ according to \cref{eq:lambdabar_neg2}. Introducing $c=a\omega$, the numerical separation constants satisfy
\begin{equation}
\lambda\sim c^2
\end{equation}
along all three divergent TTM families. Suppose $\omega\sim a^{-p}$. Then
\begin{equation}
c\sim a^{1-p},\quad \lambda\sim c^2\sim a^{2-2p}.
\end{equation}
The leading contributions to the STC~\eqref{eq:Starobinsky_const} are $\lambda^4$ and $144\omega^2$, which scale as $a^{8-8p}$ and $a^{-2p}$, respectively. Balancing these two terms gives
\begin{equation}
8-8p=-2p,
\end{equation}
and hence $p=\frac{4}{3}$. Therefore,
\begin{equation}
\omega\sim a^{-\frac{4}{3}},\quad c\sim a^{-\frac{1}{3}},\quad \lambda\sim a^{-\frac{2}{3}}.
\label{eq:leadingScalings}
\end{equation}
The corresponding Puiseux expansions may therefore be written directly in powers of $a^{\frac{1}{3}}$ as
\begin{align}
\omega &= a^{-\frac{4}{3}}(W_0+W_1a^{\frac{1}{3}}+W_2a^{\frac{2}{3}}+\mathcal{O}(a)),
\label{eq:omegaPuiseux}\\
\lambda &= a^{-\frac{2}{3}}(\mathcal{L}_0+\mathcal{L}_1a^{\frac{1}{3}}+\mathcal{L}_2a^{\frac{2}{3}}+\mathcal{O}(a)).
\label{eq:lambdaPuiseux}
\end{align}
The coefficients $W_k$ and $\mathcal{L}_k$ depend on the TTM family $n_\infty$; to avoid clutter, the family label is suppressed in the general derivation and restored as a left subscript only when individual families need to be distinguished. For large $|c|$, the separation constant admits the expansion
\begin{equation}\label{eq:lambdaComplexInfinity}
\lambda(c)=c^2+\mu c+\nu+\mathcal{O}(c^{-1}),
\end{equation}
where $\mu$ and $\nu$ denote the coefficients of the linear and constant terms, respectively, in the large-$|c|$ expansion. Since $c=a\omega$, \cref{eq:omegaPuiseux} gives
\begin{equation}\label{eq:cPuiseux}
c=W_0a^{-\frac{1}{3}}+W_1+W_2a^{\frac{1}{3}}+\mathcal{O}(a^{\frac{2}{3}}).
\end{equation}
Substituting \cref{eq:cPuiseux} into \cref{eq:lambdaComplexInfinity} and comparing equal powers of $a^{\frac{1}{3}}$ with \cref{eq:lambdaPuiseux} gives
\begin{subequations}\label{eq:LfromW}
  \begin{align}
\mathcal{L}_0 &= W_0^2,\quad \mathcal{L}_1=W_0(2W_1+\mu),\\
\mathcal{L}_2 &= W_1^2+2W_0W_2+\mu W_1+\nu.
\end{align}
\end{subequations}
Substituting these expansions into the STC~\eqref{eq:Starobinsky_const} determines the frequency coefficients order by order. At leading order, $\mathcal{O}(a^{-\frac{8}{3}})$, one obtains $W_0^8+144W_0^2=0.$ Discarding the trivial solution $W_0=0$, the divergent branches therefore satisfy
\begin{equation}\label{eq:W0six}
W_0^6=-144.
\end{equation}
For the lower-half-plane branch $\omega_{\rm I}$, the three divergent families select three different lower-half-plane roots of \cref{eq:W0six}. Restoring the family label explicitly,
\begin{align}\label{eq:threeW0}
{}_{1}W_0 &= 144^{\frac{1}{6}}{\rm e}^{-\frac{5i}{6}\pi}\nonumber\\
&\approx-1.98270322825-1.14471424255\,i,\nonumber\\
{}_{2}W_0 &= 144^{\frac{1}{6}}{\rm e}^{-\frac{i}{2}\pi}\nonumber\\
&\approx-2.28942848511\,i,\nonumber\\
{}_{4}W_0 &= 144^{\frac{1}{6}}{\rm e}^{-\frac{i}{6}\pi}\nonumber\\
&\approx1.98270322825-1.14471424255\,i.
\end{align}
The corresponding asymptotic directions are $-150^\circ$, $-90^\circ$, and $-30^\circ$ for $n_\infty=1,2,$ and $4$, respectively. The corresponding roots for the upper-half-plane branch $\omega_{\rm II}$ follow from exact complex conjugation.

At the next order, $\mathcal{O}(a^{-\frac{7}{3}})$, the STC~\eqref{eq:Starobinsky_const} gives
\begin{equation}
8W_0^7W_1+4W_0^7\mu+288W_0W_1=0.
\end{equation}
Using \cref{eq:W0six}, this reduces to
\begin{equation}\label{eq:W1mu}
W_1=-\frac{2}{3}\mu.
\end{equation}
At order $\mathcal{O}(a^{-2})$, substitution of \cref{eq:W1mu} gives
\begin{equation}
-864W_0W_2+96\mu^2-576\nu+5184=0,
\end{equation}
and hence
\begin{equation}\label{eq:W2munu}
W_2=\frac{\mu^2-6\nu+54}{9W_0}.
\end{equation}
Substituting \Cref{eq:W1mu,eq:W2munu} into \cref{eq:LfromW} gives the corresponding coefficients of the separation constant,
\begin{equation}\label{eq:L12munu}
\mathcal{L}_1=\frac{1}{2}W_0W_1=-\frac{1}{3}\mu W_0,\quad
\mathcal{L}_2=12-\frac{\nu}{3}.
\end{equation}

The coefficients $\mu$ and $\nu$ can be expressed in terms of a nonnegative integer $j$ associated with the large-$|c|$ angular asymptotics. Throughout this work, however, the TTM sequences remain labeled by the polar index $\ell$ determined from the spherical-limit continuation described above; the index $j$ is introduced only as an auxiliary asymptotic label. For a given $n_\infty$ family and azimuthal number $m$, the two indices are related by
\begin{equation}\label{eq:defj}
\ell=j+\ell_0(n_\infty,m).
\end{equation}
Here $j=0,1,2,\ldots$, and $\ell_0(n_\infty,m)$ is the first polar index of that TTM family, corresponding to $j=0$. For the divergent families $n_\infty=(1,2,4)$, the numerical continuations give $\ell_0=(2,2,4)$ for $m=1$ and $\ell_0=(10,11,13)$ for $m=10$, with the entries ordered by $n_\infty=(1,2,4)$.
The numerical data are consistent with the empirical relation
\begin{equation}
\ell_0(n_\infty,m)=\max\left[2,|m|+n_\infty-1\right],
\label{eq:ell0_relation}
\end{equation}
for $n_\infty\in\{1,2,4\}$. Thus, for example, the $m=10$ families $n_\infty=1$, $2$, and $4$ begin at the polar indices $\ell=10$, $11$, and $13$, respectively. In terms of the auxiliary index $j$, the large-$|c|$ angular coefficients are
\begin{equation}\label{eq:muAsym}
\mu=-2m+i(2j+1),
\end{equation}
and
\begin{equation}\label{eq:nuAsym}
\nu=-\frac{1}{4}[2j(j+1)-4m^2-21].
\end{equation}
Together with \Cref{eq:W1mu,eq:W2munu,eq:L12munu}, these relations determine the Puiseux coefficients through the first two subleading orders.

The numerical fits confirm this expansion for all three divergent families.
The fitted $W_0$ coefficients agree with the predicted roots of $W_0^6=-144$ with relative errors below $5\times10^{-5}$, while the $W_1$ and $W_2$ coefficients agree with the analytical values at the $10^{-5}$--$10^{-4}$ and $10^{-3}$--$10^{-2}$ levels, respectively.
Independent fits of $\lambda$ reproduce the corresponding $\mathcal{L}_1$ and $\mathcal{L}_2$ coefficients.
These results provide an independent numerical verification of the small-$a$ asymptotics for the $m=1$ and $m=10$ data sets over angular branches extending to $\ell=32$.

The fractional-power expansion itself is consistent with that obtained by Cook and Lu~\cite{Cook:2022kbb},
\begin{equation}\label{eq:CookPuiseux}
M\omega=\sum_{p=0}^{\infty}B_p\, a^{-\frac{4-p}{3}}.
\end{equation}
In the present notation, $B_0=W_0$, $B_1=W_1$, and $B_2=W_2$. The new information provided here lies in how these local small-$a$ branches are embedded in the global TTM structure. The Cook--Lu finite $(n=0)$ and divergent $(n=1)$ families correspond to the present $n_\infty=3$ and $n_\infty=2$ continuations, respectively, while their $n=2$ family separates into the two distinct mirror-related continuations $n_\infty=1$ and $n_\infty=4$ under global tracking.

A further distinction concerns the angular labeling. Throughout the present work, the TTM sequences are labeled by the polar index $\ell$ obtained from spherical-limit continuation. The auxiliary index $j$ enters only in the large-$|c|$ angular asymptotics and is related to $\ell$ through the family-dependent onset $\ell_0(n_\infty,m)$. For $m=1$ and $m=10$, the numerical results give $\ell_0=(2,2,4)$ and $\ell_0=(10,11,13)$, respectively, with the entries ordered by $n_\infty=(1,2,4)$, consistent with the empirical relation~\eqref{eq:ell0_relation}. In particular, the three divergent families at $m=10$ begin at the polar indices $\ell=10,11,$ and $13$, respectively.
Thus, the Cook--Lu fractional-power structure is recovered within a finer four-family global classification, while the large-$|c|$ asymptotic index is related explicitly to the polar labeling used throughout the present work. Table~\ref{tab:asymptoticComparison} summarizes these correspondences.

\begin{table*}[t]
\caption{Small-$a$ behavior of the four $n_\infty$ families, their first polar indices, and their relation to the Cook--Lu classification. The listed asymptotic frequencies refer to the lower-half-plane branch $\omega_{\rm I}$; the conjugate branch is obtained by complex conjugation.}
\label{tab:asymptoticComparison}
\begin{ruledtabular}
\begin{tabular}{cccc}
Present family & Leading small-$a$ behavior & First polar index $\ell_0$ & Cook--Lu classification \\
$n_\infty=1$ & $\omega\sim144^{\frac{1}{6}}{\rm e}^{-\frac{5\pi i}{6}}a^{-\frac{4}{3}}$ & $\ell_0=\max(2,|m|)$ & One mirror-related sector of $(n=2)$ \\
$n_\infty=2$ & $\omega\sim144^{\frac{1}{6}}{\rm e}^{-\frac{\pi i}{2}}a^{-\frac{4}{3}}$ & $\ell_0=\max(2,|m|+1)$ & $(n=1)$ \\
$n_\infty=3$ & $\omega\rightarrow -\frac{(\ell-1)\ell(\ell+1)(\ell+2)}{12}i$ & $\ell_0=\max(2,|m|)$ & $(n=0)$ \\
$n_\infty=4$ & $\omega\sim144^{\frac{1}{6}}{\rm e}^{-\frac{\pi i}{6}}a^{-\frac{4}{3}}$ & $\ell_0=\max(2,|m|+3)$ & The other mirror-related sector of $(n=2)$
\end{tabular}
\end{ruledtabular}
\end{table*}

\section{CONCLUSION}
\label{sec:Conclusion}

In this work, we have constructed the complete gravitational total-transmission-mode spectrum of Kerr black holes using the HeunC formulation of the Teukolsky equations together with the Starobinsky--Teukolsky condition. At the extreme-Kerr limit, a dense root search reveals eight root sequences forming four complex-conjugate pairs, and an enlargement of the search region produces no additional roots. Continuously tracing these roots toward smaller spin organizes the spectrum into four globally continuous families, $n_\infty=1,2,3,4$. This four-family classification differs from the Cook--Lu classification~\cite{Cook:2022kbb} in the global assignment of roots among the families, rather than through the appearance of additional symmetry-unrelated TTM solutions.

A central result is the uniform symmetry structure of the complete TTM spectrum. For each family, complex conjugation relates the two frequency branches, $\omega_{\rm II}=\omega_{\rm I}^*$, which accounts for the eight-sequence structure in the full complex-frequency plane. The independent mirror transformation $m\rightarrow-m$ and $\omega\rightarrow-\omega^*$ connects spectra with opposite azimuthal numbers. Separating these two symmetries resolves the mirror-related sectors of the Cook--Lu $n=2$ family into independent continuations, while for $m=0$ the two operations act within the same spectrum to generate the expected fourfold root symmetry.

The four families also display sharply different evolutions toward the Schwarzschild limit. The $n_\infty=3$ family remains finite and approaches the algebraically special Schwarzschild frequency, whereas the $n_\infty=1,2,$ and $4$ families evolve toward complex infinity. The three divergent families share the fractional-power scaling $\omega\propto a^{-4/3}$, with the lower-half-plane branches approaching asymptotic directions $-150^\circ$, $-90^\circ$, and $-30^\circ$ for $n_\infty=1,2,$ and $4$, respectively, and with a common leading magnitude $144^{1/6}$.
Fits to the high-precision data for $m=1$ and $m=10$, extending to $\ell=32$, verify the leading and subleading coefficients in the corresponding expansions of both the frequency and the separation constant.
The resulting Puiseux expansion reproduces the local small-$a$ structure obtained by Cook and Lu, while the present global continuation reveals how those asymptotic branches are embedded in the four-family spectrum.

The angular labeling provides an additional distinction. The polar index $\ell$ used here is fixed by continuously tracing each angular eigensolution to the spherical limit, while the large-$|c|$ angular ordering is retained separately through the asymptotic index $j$. Their relation is family dependent, with $\ell=j+\ell_0(n_\infty,m)$. In particular, for $m=10$ the three divergent families begin at $\ell=10$, $11$, and $13$, respectively. This distinction separates the spherical-limit polar labeling from the asymptotic angular ordering and reveals the family-dependent structure of the singular TTM branches.

For axisymmetric perturbations, the $n_\infty=2$ and $n_\infty=3$ trajectories coalesce on the imaginary axis at exceptional points.
The scattering spectral function develops a second-order zero there: both the function and its first frequency derivative vanish, while the second derivative remains finite and nonzero.
Consequently, the two frequencies exhibit the characteristic square-root splitting about the exceptional point, and their excitation factors undergo a strong resonant enhancement. This provides an independent scattering-matrix characterization of the mode coalescence beyond the frequency trajectories alone.

Finally, we have examined the unusually close spectral approach between the $n_\infty=3$ TTM family and the unconventional Kerr QNM sequence for $m=-\ell$. Although the two trajectories can approach each other to separations of order $10^{-8}$, this separation remains many orders of magnitude larger than the numerical residuals.
The two spectra therefore remain distinct for nonzero spin.

Taken together, these results establish the global four-family continuation, exact frequency symmetries, spherical-limit polar labeling, exceptional-point structure, and singular small-spin asymptotics of the Kerr TTM spectrum.
The Cook--Lu local asymptotic structure is recovered within this broader classification, while the separation between the polar index $\ell$ and the large-$|c|$ angular index $j$ exposes additional family-dependent structure that is not apparent from the local small-$a$ asymptotics alone.
\section*{DATA AVAILABILITY}

The data and numerical programs supporting the findings of this work are openly available in Ref.~\cite{ChenTTM}.

\begin{acknowledgments}
This work was supported by the National Key Research and Development Program of China (No.~2021YFC2203001) and the National Natural Science Foundation of China (No.~12475049).
\end{acknowledgments}

%
%
%
%%%%%%%%%%%%%%%%%%%%%%%%%%%%%%%%%%%%%%%%%%%%%%%%%%%%%%%%%%%%%%%%%
%%%%%%%%%%%%%%%%%%%%%%%%%%%%%%%%%%%%%%%%%%%%%%%%%%%%%%%%%%%%%%%%%
%
\appendix

\section{Scattering Amplitudes of TTMs}
\label{sec:TTM-Scatter-Amp}

The calculation of scattering amplitudes in the Teukolsky formalism can be formulated as a two-point connection problem for the confluent Heun equation. A local solution satisfying a prescribed wave condition at the event horizon must be analytically connected to the asymptotic basis at spatial infinity, where the coefficients of the ingoing and outgoing waves determine the corresponding scattering amplitudes.

In our previous work, we developed two complementary methods for computing the relevant connection coefficients. The first method~\cite{Chen:2023ese,Chen:2023lsa} is based on an MST-type series expansion in special-function bases and allows the connection coefficients to be evaluated with arbitrary numerical precision. The second method~\cite{Chen:2026bym} employs a hybrid analytic-continuation algorithm. At machine precision, the latter is typically more efficient and can provide highly accurate connection coefficients at lower computational cost. Its efficiency, however, decreases when substantially higher numerical precision is required. Since the TTM calculations considered in this work require precision far beyond standard machine precision, we adopt the first method. The essential ingredients needed for the present scattering calculation are summarized below.

\subsection{${\rm TTM}_{\rm L}$ Frequencies}

For the time dependence $\psi\sim e^{-i\omega t}$, the solution that is outgoing at the event horizon is written as
\begin{equation}
{\cal R}_{\ell m}^{{\rm up}}
=
S_r(\alpha_r^+,\beta_r^-,\gamma_r^+)
{\rm HeunC}
(\alpha_r^+,\beta_r^-,\gamma_r^+,\delta_r,\eta_r;x_r).
\end{equation}
The corresponding HeunC parameters for Kerr scattering are
\begin{subequations}\label{eq:RTE_HCparameters_Kerr}
\begin{align}
&{\alpha^\pm _r} = \pm 2i\omega ({r_ + } - {r_ - }),\\
&{\beta^\pm _r} = \mp\left(s + \frac{{2i\omega (r_ + ^2 + {a^2}) - 2iam}}{{{r_ + } - {r_ - }}}\right),\\
&{\gamma^\pm _r} = \pm\left(s - \frac{{2i\omega (r_ - ^2 + {a^2}) - 2iam}}{{{r_ + } - {r_ - }}}\right),\\
&{\delta _r} = 2is\omega ({r_ - } - {r_ + }) + 2{\omega ^2}(r_{-}^2-r_+^2),\\
&\eta_r  = 2is\omega {r_ + } - \tfrac{1}{2}{s^2} - s - \lambda \\
&\quad
-\frac{{2\left[ {\omega (r_ + ^2 + {a^2}) - am} \right]
\left[ { ({a^2} + 2{r_ - }{r_ + } - r_ + ^2)\omega - am} \right]}}
{{{{({r_ + } - {r_ - })}^2}}},\nonumber\\
&{r_\pm}=M\pm\sqrt{M^2-a^2}.
\end{align}
\end{subequations}

Near the event horizon, ${\cal R}_{\ell m}^{\rm up}$ is normalized to the outgoing wave solution. After analytic continuation to spatial infinity, it becomes a linear combination of the outgoing and ingoing asymptotic solutions,
\begin{equation}
\label{eq:BoundaryCondition-TTML}
{\cal R}_{\ell m}^{{\rm up}}
\to
\left\{
\begin{array}{ll}
{\cal B}_{\ell m}^{{\rm trans}}R_{{\rm up}}^{r_+},
& r\to r_+,\\[1mm]
{\cal B}_{\ell m}^{{\rm ref}}R_{{\rm up}}^\infty
+
{\cal B}_{\ell m}^{{\rm inc}}R_{{\rm in}}^\infty,
& r\to+\infty.
\end{array}
\right.
\end{equation}
Thus, ${\cal B}_{\ell m}^{\rm inc}$ and ${\cal B}_{\ell m}^{\rm ref}$ are the two connection amplitudes at infinity, while ${\cal B}_{\ell m}^{\rm trans}$ fixes the horizon normalization. A ${\rm TTM}_{\rm L}$ contains no ingoing wave at infinity and therefore satisfies
\begin{equation}
{\cal B}_{\ell m}^{\rm inc}=0.
\end{equation}

\subsection{${\rm TTM}_{\rm R}$ Frequencies}
\label{TTM_R_Eq}

For the same time dependence $\psi\sim e^{-i\omega t}$, the solution that is ingoing at the event horizon is
\begin{equation}
R_{\ell m}^{{\rm in}}
=
S_r(\alpha_r^+,\beta_r^+,\gamma_r^+)
{\rm HeunC}
(\alpha_r^+,\beta_r^+,\gamma_r^+,\delta_r,\eta_r;x_r).
\end{equation}
Its asymptotic form is
\begin{equation}
R_{\ell m}^{{\rm in}}
\to
\left\{
\begin{array}{ll}
B_{\ell m}^{{\rm trans}}R_{{\rm in}}^{r_+},
& r\to r_+,\\[1mm]
B_{\ell m}^{{\rm ref}}R_{{\rm up}}^\infty
+
B_{\ell m}^{{\rm inc}}R_{{\rm in}}^\infty,
& r\to+\infty.
\end{array}
\right.
\end{equation}
The ${\rm TTM}_{\rm R}$ condition requires the outgoing component at infinity to vanish,
\begin{equation}
B_{\ell m}^{\rm ref}=0.
\end{equation}
The TTM conditions are therefore expressed directly in terms of the vanishing of one of the two connection amplitudes generated by continuation from the horizon to infinity.
\begin{widetext}
\subsection{Connection Coefficients and Scattering Amplitudes}
To evaluate these amplitudes, consider the confluent Heun function
\begin{equation}
\mathbf{H}(x_r)
=
{\rm HeunC}
(\alpha_r,\beta_r,\gamma_r,\delta_r,\eta_r;x_r).
\end{equation}
At large $|x_r|$, its asymptotic form consists of two independent contributions,
\begin{equation}
\lim_{|x_r|\to\infty}\mathbf{H}(x_r)
=
D_\odot^{\beta_r}
x_r^{-\frac{\beta_r+\gamma_r+2}{2}-\frac{\delta_r}{\alpha_r}}
+
D_\otimes^{\beta_r}
e^{-\alpha_r x_r}
x_r^{-\frac{\beta_r+\gamma_r+2}{2}+\frac{\delta_r}{\alpha_r}}.
\end{equation}
The coefficients $D_\odot^{\beta_r}$ and $D_\otimes^{\beta_r}$ therefore encode the connection between the local HeunC solution at the horizon and the two independent asymptotic wave solutions at infinity. After restoring the $S$-homotopic prefactor $S_r$, these two terms can be identified with the ingoing and outgoing Teukolsky-wave amplitudes.
For the MST-type representation used here, the connection coefficients are
\begin{align}
D_\odot ^{\beta_r}
&=
\Xi _{{\rm n},\nu }^{\beta_r}D_{\odot,{\rm n},\nu}^{\beta_r}
+
e^{-i\pi(\nu+\frac12)}
\frac{\sin\pi\left(\nu+\frac{\delta_r}{\alpha_r}\right)}
{\sin\pi\left(\nu-\frac{\delta_r}{\alpha_r}\right)}
\Xi _{-{\rm n},-\nu-1}^{\beta_r}
D_{\odot,-{\rm n},-\nu-1}^{\beta_r},
\\
D_\otimes ^{\beta_r}
&=
\Xi _{{\rm n},\nu }^{\beta_r}D_{\otimes,{\rm n},\nu}^{\beta_r}
+
e^{i\pi(\nu+\frac12)}
\Xi _{-{\rm n},-\nu-1}^{\beta_r}
D_{\otimes,-{\rm n},-\nu-1}^{\beta_r},
\end{align}
with
\begin{align}
D_{ \odot ,{\rm n},\nu }^{\beta_r}  &= {\left( { - 1} \right)^{\frac{{\gamma_r  + {\beta_r}  + 2}}{2} + \frac{\delta_r }{\alpha_r }}}{\big( {\frac{\alpha_r }{2}} \big)^\tau } {\big( -{\frac{{i\alpha_r }}{2}} \big)^{ - \frac{{\gamma_r  + {\beta_r}  + 2}}{2} - \frac{\delta_r }{\alpha_r }}}{{\rm{e}}^{ - \frac{{i\pi \tau +\alpha_r}}{2}}}\times
{2^{ - 1 - \frac{\delta_r }{\alpha_r }}}{{\rm{e}}^{\frac{{i\pi }}{2}\big( {\nu  + 1 + \frac{\delta_r }{\alpha_r }} \big)}} \Xi _{{\rm n},\nu }^{\beta_r}\frac{{\Gamma \big( {\nu  + 1 + \frac{\delta_r }{\alpha_r }} \big)}}{{\Gamma \big( {\nu  + 1 - \frac{\delta_r }{\alpha_r }} \big)}},
\\
 D_{  \otimes,{\rm n},\nu }^{\beta_r}  &= {\big( { - 1} \big)^{\frac{{\gamma_r  + {\beta_r}  + 2}}{2} - \frac{\delta_r }{\alpha_r }}}{\left( {\frac{\alpha_r }{2}} \right)^\tau } {\left(- {\frac{{i\alpha_r }}{2}} \right)^{ - \frac{{\gamma_r  + {\beta_r}  + 2}}{2} + \frac{\delta_r }{\alpha_r }}} \nonumber \\
&\times {{\rm{e}}^{ - \frac{{i\pi \tau - \alpha_r}}{2}}}\Xi _{{\rm n},\nu }^{\beta_r}  \frac{ {2^{ - 1 + \frac{\delta_r }{\alpha_r }}}{{\rm{e}}^{ - \frac{{i\pi }}{2}\big( {\nu  + 1 - \frac{\delta_r }{\alpha_r }} \big)}}} {\sum\limits_{{\rm n} =  - \infty }^{ + \infty } {f_{\rm n}^\nu } }   \times\Big( \sum\limits_{{\rm n} =  - \infty }^{ + \infty } {{{( - 1)}^{\rm n}}} \frac{{{\big( {\nu  + 1 - \frac{\delta_r }{\alpha_r }} \big)}_{\rm n}}}{{{{\big( {\nu  + 1 + \frac{\delta_r }{\alpha_r }} \big)}_{\rm n}}}}f_{\rm n}^\nu  \Big),
\end{align}
The normalization factor $\Xi_{{\rm n},\nu}^{\beta_r}$ is
\begin{align}\label{eq:Xinv}
&{\Xi _{{\rm{n}},\nu }^{{\beta _r}} = \frac{{{2^{ - \nu }}{{\left( {\frac{{{\alpha _r}}}{2}} \right)}^{ - \hat \tau }}{{\rm{e}}^{\frac{{i\pi \hat \tau  + {\alpha _r}}}{2}}}\Gamma \left( {{\beta _r} + 1} \right)\Gamma \left( {2\nu  + 2} \right)}}{{\Gamma \left( {\nu  + 1 + \frac{{{\delta _r}}}{{{\alpha _r}}}} \right)\Gamma \left( {\nu  + 1 - \frac{{{\beta _r} + {\gamma _r}}}{2}} \right)\Gamma \left( {\nu  + 1 + \frac{{{\gamma _r} - {\beta _r}}}{2}} \right)}} \times {{(\sum\limits_{{\rm{n}} =  - \infty }^0 {\frac{{{{( - 1)}^{\rm{n}}}{{\left( {\nu  + 1 - \frac{{{\delta _r}}}{{{\alpha _r}}}} \right)}_{\rm{n}}}}}{{( - {\rm{n}})!{{\left( {2\nu  + 2} \right)}_{\rm{n}}}{{\left( {\nu  + 1 + \frac{{{\delta _r}}}{{{\alpha _r}}}} \right)}_{\rm{n}}}}}} f_{\rm{n}}^\nu )}^{ - 1}}}
\nonumber\\
& \times \Big( {\sum\limits_{{\rm n} = 0}^\infty  {{{\left( { - 1} \right)}^{\rm n}\frac{{\Gamma \left( {{\rm n} + 2\nu  + 1} \right)\Gamma \left( {{\rm n} + \nu  + 1 + \frac{{\gamma_r  - {\beta_r} }}{2}} \right)\Gamma \left( {{\rm n} + \nu  + 1 - \frac{{{\beta_r}  + \gamma_r }}{2}} \right)}}{{( {\rm n}!)\Gamma \left( {{\rm n} + \nu  + 1 + \frac{{{\beta_r}  - \gamma_r }}{2}} \right)\Gamma \left( {{\rm n} + \nu  + 1 + \frac{{{\beta_r}  + \gamma_r }}{2}} \right)}}f_{\rm n}^\nu } } \Big)},
\end{align}
\end{widetext}
where $\tau  = \frac{1}{4}\left( {3{\beta_r}  + \gamma_r  + \frac{{2\delta_r }}{\alpha_r }} \right)$ and $\hat \tau  = \frac{{\beta_r  - \gamma_r }}{4} + \nu  + \frac{\delta_r }{{2\alpha_r }}$.
Here $\nu$ is the renormalized angular momentum entering the MST-type expansion, and $f_{\rm n}^{\nu}$ are the corresponding series coefficients. Their determination follows the recurrence construction described in Refs.~\cite{Chen:2023ese,Chen:2023lsa}. Once $\nu$ and $f_{\rm n}^{\nu}$ are known, the connection coefficients $D_\odot^{\beta_r}$ and $D_\otimes^{\beta_r}$ can be evaluated to arbitrary precision.

The scattering amplitudes follow by matching the two large-$r$ HeunC asymptotic contributions to the Teukolsky wave basis. Because the identification of the two contributions depends on the choice of $\beta_r$ and on the time dependence, four cases are relevant.

For $\psi\sim e^{-i\omega t}$ and ${\rm TTM}_{\rm L}$,
\begin{subequations}\label{eq:TTM_L_Amplitudes_1}
\begin{align}
{ {\cal B}_{\ell m}^{{\rm{inc}}}}&{ = {{( - 1)}^{ - \frac{{{\beta _r} + {\gamma _r} + 2}}{2} - \frac{{{\delta _r}}}{{{\alpha _r}}}}}{{\left( {{r_ + } - {r_ - }} \right)}^{s + 1 + \frac{{{\delta _r}}}{{{\alpha _r}}}}}} \nonumber\\
{}&{ \times {{\left( {{r_ + } + {r_ - }} \right)}^{ - 2iM\omega }}{{\rm{e}}^{i\omega {r_ + }}}D_ \odot ^{{\beta _r}},}\\
{ {\cal B}_{\ell m}^{{\rm{ref}}}}&{ = {{( - 1)}^{ - \frac{{{\beta _r} + {\gamma _r} + 2}}{2} + \frac{{{\delta _r}}}{{{\alpha _r}}}}}{{\left( {{r_ + } - {r_ - }} \right)}^{s + 1 - \frac{{{\delta _r}}}{{{\alpha _r}}}}}}\nonumber\\
{}& \times {\left( {{r_ + } + {r_ - }} \right)^{2iM\omega }}{{\rm{e}}^{ - i\omega {r_ + }}}D_ \otimes ^{{\beta _r}},\\
&\quad\quad\quad\quad\text{with}\quad\quad {\alpha _r} = \alpha _r^+ ,\quad {\beta _r} = \beta _r^- .
\end{align}
\end{subequations}

For $\psi\sim e^{i\omega^*t}$ and ${\rm TTM}_{\rm L}$,
\begin{subequations}\label{eq:TTM_L_Amplitudes_2}
\begin{align}
{ {\cal B}_{\ell m}^{{\rm{inc}}}}&{ = {{( - 1)}^{ - \frac{{{\beta _r} + {\gamma _r} + 2}}{2} + \frac{{{\delta _r}}}{{{\alpha _r}}}}}{{\left( {{r_ + } - {r_ - }} \right)}^{s + 1 - \frac{{{\delta _r}}}{{{\alpha _r}}}}}}\nonumber\\
{}& \times {\left( {{r_ + } + {r_ - }} \right)^{2iM\omega }}{{\rm{e}}^{ - i\omega {r_ + }}}D_ \otimes ^{{\beta _r}},\\
{ {\cal B}_{\ell m}^{{\rm{ref}}}}&{ = {{( - 1)}^{ - \frac{{{\beta _r} + {\gamma _r} + 2}}{2} - \frac{{{\delta _r}}}{{{\alpha _r}}}}}{{\left( {{r_ + } - {r_ - }} \right)}^{s + 1 + \frac{{{\delta _r}}}{{{\alpha _r}}}}}} \nonumber\\
{}&{ \times {{\left( {{r_ + } + {r_ - }} \right)}^{ - 2iM\omega }}{{\rm{e}}^{i\omega {r_ + }}}D_ \odot ^{{\beta _r}},}\\
&\quad\quad\quad\quad\text{with}\quad\quad {\alpha _r} = \alpha _r^+ ,\quad {\beta _r} = \beta _r^+ .
\end{align}
\end{subequations}

For $\psi\sim e^{-i\omega t}$ and ${\rm TTM}_{\rm R}$,
\begin{subequations}\label{eq:TTM_R_Amplitudes_1}
\begin{align}
{ B_{\ell m}^{{\rm{inc}}}}&{ = {{( - 1)}^{ - \frac{{{\beta _r} + {\gamma _r} + 2}}{2} - \frac{{{\delta _r}}}{{{\alpha _r}}}}}{{\left( {{r_ + } - {r_ - }} \right)}^{s + 1 + \frac{{{\delta _r}}}{{{\alpha _r}}}}}} \nonumber\\
{}&{ \times {{\left( {{r_ + } + {r_ - }} \right)}^{ - 2iM\omega }}{{\rm{e}}^{i\omega {r_ + }}}D_ \odot ^{{\beta _r}},}\\
{ B_{\ell m}^{{\rm{ref}}}}&{ = {{( - 1)}^{ - \frac{{{\beta _r} + {\gamma _r} + 2}}{2} + \frac{{{\delta _r}}}{{{\alpha _r}}}}}{{\left( {{r_ + } - {r_ - }} \right)}^{s + 1 - \frac{{{\delta _r}}}{{{\alpha _r}}}}}}\nonumber\\
{}& \times {\left( {{r_ + } + {r_ - }} \right)^{2iM\omega }}{{\rm{e}}^{ - i\omega {r_ + }}}D_ \otimes ^{{\beta _r}},\\
&\quad\quad\quad\quad\text{with}\quad\quad {\alpha _r} = \alpha _r^+ ,\quad {\beta _r} = \beta _r^+ .
\end{align}
\end{subequations}

Finally, for $\psi\sim e^{i\omega^*t}$ and ${\rm TTM}_{\rm R}$,
\begin{subequations}\label{eq:TTM_R_Amplitudes_2}
\begin{align}
{ B_{\ell m}^{{\rm{inc}}}}&{ = {{( - 1)}^{ - \frac{{{\beta _r} + {\gamma _r} + 2}}{2} + \frac{{{\delta _r}}}{{{\alpha _r}}}}}{{\left( {{r_ + } - {r_ - }} \right)}^{s + 1 - \frac{{{\delta _r}}}{{{\alpha _r}}}}}} \nonumber\\
{}& \times {\left( {{r_ + } + {r_ - }} \right)^{2iM\omega }}{{\rm{e}}^{ - i\omega {r_ + }}}D_ \otimes ^{{\beta _r}},\\
{ B_{\ell m}^{{\rm{ref}}}}&{ = {{( - 1)}^{ - \frac{{{\beta _r} + {\gamma _r} + 2}}{2} - \frac{{{\delta _r}}}{{{\alpha _r}}}}}{{\left( {{r_ + } - {r_ - }} \right)}^{s + 1 + \frac{{{\delta _r}}}{{{\alpha _r}}}}}}\nonumber\\
{}&{ \times {{\left( {{r_ + } + {r_ - }} \right)}^{ - 2iM\omega }}{{\rm{e}}^{i\omega {r_ + }}}D_ \odot ^{{\beta _r}},}\\
&\quad\quad\quad\quad\text{with}\quad\quad {\alpha _r} = \alpha _r^+ ,\quad {\beta _r} = \beta _r^- .
\end{align}
\end{subequations}
Either $\gamma_r^+$ or $\gamma_r^-$ may be used for $\gamma_r$.

These expressions provide the scattering amplitudes directly from the HeunC connection coefficients. In practice, for a given $(a,\omega,\ell,m,s)$, the angular problem is first solved to determine the separation constant $\lambda$. The corresponding radial HeunC parameters are then constructed from \cref{eq:RTE_HCparameters_Kerr}, after which $\nu$ and the coefficients $f_{\rm n}^{\nu}$ are obtained from the MST-type recurrence relations. Evaluation of $D_\odot^{\beta_r}$ and $D_\otimes^{\beta_r}$ then determines the asymptotic amplitudes through Eqs.~\eqref{eq:TTM_L_Amplitudes_1}--\eqref{eq:TTM_R_Amplitudes_2}.

The TTM condition can therefore be checked independently of the STC calculation through the corresponding scattering amplitudes. For ${\rm TTM}_{\rm L}$, one verifies ${\cal B}_{\ell m}^{\rm inc}=0$, whereas for ${\rm TTM}_{\rm R}$ one verifies $B_{\ell m}^{\rm ref}=0$. For comparison, the QNM condition is $B_{\ell m}^{\rm inc}=0$. Thus, the QNM and ${\rm TTM}_{\rm R}$ conditions differ by which asymptotic amplitude vanishes at infinity. These amplitudes are also used to construct the reflection and transmission amplitudes entering the exceptional-point analysis and the TTM excitation factors in the main text.

\bibliography{mybibfile}

%apsrev4-2.bst 2019-01-14 (MD) hand-edited version of apsrev4-1.bst
%Control: key (0)
%Control: author (8) initials jnrlst
%Control: editor formatted (1) identically to author
%Control: production of article title (0) allowed
%Control: page (0) single
%Control: year (1) truncated
%Control: production of eprint (0) enabled
\begin{thebibliography}{38}%
\makeatletter
\providecommand \@ifxundefined [1]{%
 \@ifx{#1\undefined}
}%
\providecommand \@ifnum [1]{%
 \ifnum #1\expandafter \@firstoftwo
 \else \expandafter \@secondoftwo
 \fi
}%
\providecommand \@ifx [1]{%
 \ifx #1\expandafter \@firstoftwo
 \else \expandafter \@secondoftwo
 \fi
}%
\providecommand \natexlab [1]{#1}%
\providecommand \enquote  [1]{``#1''}%
\providecommand \bibnamefont  [1]{#1}%
\providecommand \bibfnamefont [1]{#1}%
\providecommand \citenamefont [1]{#1}%
\providecommand \href@noop [0]{\@secondoftwo}%
\providecommand \href [0]{\begingroup \@sanitize@url \@href}%
\providecommand \@href[1]{\@@startlink{#1}\@@href}%
\providecommand \@@href[1]{\endgroup#1\@@endlink}%
\providecommand \@sanitize@url [0]{\catcode `\\12\catcode `\$12\catcode
  `\&12\catcode `\#12\catcode `\^12\catcode `\_12\catcode `\%12\relax}%
\providecommand \@@startlink[1]{}%
\providecommand \@@endlink[0]{}%
\providecommand \url  [0]{\begingroup\@sanitize@url \@url }%
\providecommand \@url [1]{\endgroup\@href {#1}{\urlprefix }}%
\providecommand \urlprefix  [0]{URL }%
\providecommand \Eprint [0]{\href }%
\providecommand \doibase [0]{https://doi.org/}%
\providecommand \selectlanguage [0]{\@gobble}%
\providecommand \bibinfo  [0]{\@secondoftwo}%
\providecommand \bibfield  [0]{\@secondoftwo}%
\providecommand \translation [1]{[#1]}%
\providecommand \BibitemOpen [0]{}%
\providecommand \bibitemStop [0]{}%
\providecommand \bibitemNoStop [0]{.\EOS\space}%
\providecommand \EOS [0]{\spacefactor3000\relax}%
\providecommand \BibitemShut  [1]{\csname bibitem#1\endcsname}%
\let\auto@bib@innerbib\@empty
%</preamble>
\bibitem [{\citenamefont {Sasaki}\ and\ \citenamefont
  {Tagoshi}(2003)}]{Sasaki:2003xr}%
  \BibitemOpen
  \bibfield  {author} {\bibinfo {author} {\bibfnamefont {M.}~\bibnamefont
  {Sasaki}}\ and\ \bibinfo {author} {\bibfnamefont {H.}~\bibnamefont
  {Tagoshi}},\ }\bibfield  {title} {\bibinfo {title} {{Analytic black hole
  perturbation approach to gravitational radiation}},\ }\href
  {https://doi.org/10.12942/lrr-2003-6} {\bibfield  {journal} {\bibinfo
  {journal} {Living Rev. Rel.}\ }\textbf {\bibinfo {volume} {6}},\ \bibinfo
  {pages} {6} (\bibinfo {year} {2003})},\ \Eprint
  {https://arxiv.org/abs/gr-qc/0306120} {arXiv:gr-qc/0306120} \BibitemShut
  {NoStop}%
\bibitem [{\citenamefont {Berti}\ \emph {et~al.}(2009)\citenamefont {Berti},
  \citenamefont {Cardoso},\ and\ \citenamefont {Starinets}}]{Berti:2009kk}%
  \BibitemOpen
  \bibfield  {author} {\bibinfo {author} {\bibfnamefont {E.}~\bibnamefont
  {Berti}}, \bibinfo {author} {\bibfnamefont {V.}~\bibnamefont {Cardoso}},\
  and\ \bibinfo {author} {\bibfnamefont {A.~O.}\ \bibnamefont {Starinets}},\
  }\bibfield  {title} {\bibinfo {title} {{Quasinormal modes of black holes and
  black branes}},\ }\href {https://doi.org/10.1088/0264-9381/26/16/163001}
  {\bibfield  {journal} {\bibinfo  {journal} {Class. Quant. Grav.}\ }\textbf
  {\bibinfo {volume} {26}},\ \bibinfo {pages} {163001} (\bibinfo {year}
  {2009})},\ \Eprint {https://arxiv.org/abs/0905.2975} {arXiv:0905.2975
  [gr-qc]} \BibitemShut {NoStop}%
\bibitem [{\citenamefont {Teukolsky}(1973)}]{Teukolsky:1973ha}%
  \BibitemOpen
  \bibfield  {author} {\bibinfo {author} {\bibfnamefont {S.~A.}\ \bibnamefont
  {Teukolsky}},\ }\bibfield  {title} {\bibinfo {title} {{Perturbations of a
  rotating black hole. 1. Fundamental equations for gravitational
  electromagnetic and neutrino field perturbations}},\ }\href
  {https://doi.org/10.1086/152444} {\bibfield  {journal} {\bibinfo  {journal}
  {Astrophys. J.}\ }\textbf {\bibinfo {volume} {185}},\ \bibinfo {pages} {635}
  (\bibinfo {year} {1973})}\BibitemShut {NoStop}%
\bibitem [{\citenamefont {Teukolsky}\ and\ \citenamefont
  {Press}(1974)}]{Teukolsky:1974yv}%
  \BibitemOpen
  \bibfield  {author} {\bibinfo {author} {\bibfnamefont {S.~A.}\ \bibnamefont
  {Teukolsky}}\ and\ \bibinfo {author} {\bibfnamefont {W.~H.}\ \bibnamefont
  {Press}},\ }\bibfield  {title} {\bibinfo {title} {{Perturbations of a
  rotating black hole. III - Interaction of the hole with gravitational and
  electromagnetic radiation}},\ }\href {https://doi.org/10.1086/153180}
  {\bibfield  {journal} {\bibinfo  {journal} {Astrophys. J.}\ }\textbf
  {\bibinfo {volume} {193}},\ \bibinfo {pages} {443} (\bibinfo {year}
  {1974})}\BibitemShut {NoStop}%
\bibitem [{\citenamefont {Berti}\ \emph {et~al.}(2026)\citenamefont {Berti}
  \emph {et~al.}}]{Berti:2025hly}%
  \BibitemOpen
  \bibfield  {author} {\bibinfo {author} {\bibfnamefont {E.}~\bibnamefont
  {Berti}} \emph {et~al.},\ }\bibfield  {title} {\bibinfo {title} {{Black hole
  spectroscopy: from theory to experiment}},\ }\href
  {https://doi.org/10.1088/1361-6382/ae59e2} {\bibfield  {journal} {\bibinfo
  {journal} {Class. Quant. Grav.}\ }\textbf {\bibinfo {volume} {43}},\ \bibinfo
  {pages} {123001} (\bibinfo {year} {2026})},\ \Eprint
  {https://arxiv.org/abs/2505.23895} {arXiv:2505.23895 [gr-qc]} \BibitemShut
  {NoStop}%
\bibitem [{Ber()}]{BertiWeb}%
  \BibitemOpen
  \href@noop {} {\bibinfo {title} {{Webpages with Mathematica notebooks and
  numerical quasinormal mode tables}}},\ \bibinfo {howpublished}
  {\href{https://pages.jh.edu/eberti2/ringdown/}{https://pages.jh.edu/eberti2/ringdown/}}\BibitemShut
  {NoStop}%
\bibitem [{\citenamefont {Cook}\ and\ \citenamefont
  {Zalutskiy}(2014)}]{Cook:2014cta}%
  \BibitemOpen
  \bibfield  {author} {\bibinfo {author} {\bibfnamefont {G.~B.}\ \bibnamefont
  {Cook}}\ and\ \bibinfo {author} {\bibfnamefont {M.}~\bibnamefont
  {Zalutskiy}},\ }\bibfield  {title} {\bibinfo {title} {{Gravitational
  perturbations of the Kerr geometry: High-accuracy study}},\ }\href
  {https://doi.org/10.1103/PhysRevD.90.124021} {\bibfield  {journal} {\bibinfo
  {journal} {Phys. Rev. D}\ }\textbf {\bibinfo {volume} {90}},\ \bibinfo
  {pages} {124021} (\bibinfo {year} {2014})},\ \Eprint
  {https://arxiv.org/abs/1410.7698} {arXiv:1410.7698 [gr-qc]} \BibitemShut
  {NoStop}%
\bibitem [{\citenamefont {Chen}\ \emph {et~al.}(2025)\citenamefont {Chen},
  \citenamefont {Jing}, \citenamefont {Cao},\ and\ \citenamefont
  {Wang}}]{Chen:2025sbz}%
  \BibitemOpen
  \bibfield  {author} {\bibinfo {author} {\bibfnamefont {C.}~\bibnamefont
  {Chen}}, \bibinfo {author} {\bibfnamefont {J.}~\bibnamefont {Jing}}, \bibinfo
  {author} {\bibfnamefont {Z.}~\bibnamefont {Cao}},\ and\ \bibinfo {author}
  {\bibfnamefont {M.}~\bibnamefont {Wang}},\ }\bibfield  {title} {\bibinfo
  {title} {{Complete quasinormal modes of type-D black holes}},\ }\href
  {https://doi.org/10.1103/f8m8-vr4l} {\bibfield  {journal} {\bibinfo
  {journal} {Phys. Rev. D}\ }\textbf {\bibinfo {volume} {112}},\ \bibinfo
  {pages} {103036} (\bibinfo {year} {2025})},\ \Eprint
  {https://arxiv.org/abs/2506.14635} {arXiv:2506.14635 [gr-qc]} \BibitemShut
  {NoStop}%
\bibitem [{\citenamefont {{Changkai Chen}}(2025)}]{ChenQNM}%
  \BibitemOpen
  \bibfield  {author} {\bibinfo {author} {\bibnamefont {{Changkai Chen}}},\
  }\href@noop {} {\bibinfo {title} {{Complete-QNMs}}},\ \bibinfo {howpublished}
  {\href{https://github.com/IronChen1/Complete-QNMs}{https://github.com/IronChen1/Complete-QNMs}}
  (\bibinfo {year} {{2025}})\BibitemShut {NoStop}%
\bibitem [{\citenamefont {Dreyer}\ \emph {et~al.}(2004)\citenamefont {Dreyer},
  \citenamefont {Kelly}, \citenamefont {Krishnan}, \citenamefont {Finn},
  \citenamefont {Garrison},\ and\ \citenamefont
  {Lopez-Aleman}}]{Dreyer:2003bv}%
  \BibitemOpen
  \bibfield  {author} {\bibinfo {author} {\bibfnamefont {O.}~\bibnamefont
  {Dreyer}}, \bibinfo {author} {\bibfnamefont {B.~J.}\ \bibnamefont {Kelly}},
  \bibinfo {author} {\bibfnamefont {B.}~\bibnamefont {Krishnan}}, \bibinfo
  {author} {\bibfnamefont {L.~S.}\ \bibnamefont {Finn}}, \bibinfo {author}
  {\bibfnamefont {D.}~\bibnamefont {Garrison}},\ and\ \bibinfo {author}
  {\bibfnamefont {R.}~\bibnamefont {Lopez-Aleman}},\ }\bibfield  {title}
  {\bibinfo {title} {{Black hole spectroscopy: Testing general relativity
  through gravitational wave observations}},\ }\href
  {https://doi.org/10.1088/0264-9381/21/4/003} {\bibfield  {journal} {\bibinfo
  {journal} {Class. Quant. Grav.}\ }\textbf {\bibinfo {volume} {21}},\ \bibinfo
  {pages} {787} (\bibinfo {year} {2004})},\ \Eprint
  {https://arxiv.org/abs/gr-qc/0309007} {arXiv:gr-qc/0309007} \BibitemShut
  {NoStop}%
\bibitem [{\citenamefont {Berti}\ \emph {et~al.}(2006)\citenamefont {Berti},
  \citenamefont {Cardoso},\ and\ \citenamefont {Will}}]{Berti:2005ys}%
  \BibitemOpen
  \bibfield  {author} {\bibinfo {author} {\bibfnamefont {E.}~\bibnamefont
  {Berti}}, \bibinfo {author} {\bibfnamefont {V.}~\bibnamefont {Cardoso}},\
  and\ \bibinfo {author} {\bibfnamefont {C.~M.}\ \bibnamefont {Will}},\
  }\bibfield  {title} {\bibinfo {title} {{On gravitational-wave spectroscopy of
  massive black holes with the space interferometer LISA}},\ }\href
  {https://doi.org/10.1103/PhysRevD.73.064030} {\bibfield  {journal} {\bibinfo
  {journal} {Phys. Rev. D}\ }\textbf {\bibinfo {volume} {73}},\ \bibinfo
  {pages} {064030} (\bibinfo {year} {2006})},\ \Eprint
  {https://arxiv.org/abs/gr-qc/0512160} {arXiv:gr-qc/0512160} \BibitemShut
  {NoStop}%
\bibitem [{\citenamefont {Cardoso}\ and\ \citenamefont
  {Gualtieri}(2016)}]{Cardoso:2016ryw}%
  \BibitemOpen
  \bibfield  {author} {\bibinfo {author} {\bibfnamefont {V.}~\bibnamefont
  {Cardoso}}\ and\ \bibinfo {author} {\bibfnamefont {L.}~\bibnamefont
  {Gualtieri}},\ }\bibfield  {title} {\bibinfo {title} {{Testing the black hole
  \textquoteleft{}no-hair\textquoteright{} hypothesis}},\ }\href
  {https://doi.org/10.1088/0264-9381/33/17/174001} {\bibfield  {journal}
  {\bibinfo  {journal} {Class. Quant. Grav.}\ }\textbf {\bibinfo {volume}
  {33}},\ \bibinfo {pages} {174001} (\bibinfo {year} {2016})},\ \Eprint
  {https://arxiv.org/abs/1607.03133} {arXiv:1607.03133 [gr-qc]} \BibitemShut
  {NoStop}%
\bibitem [{\citenamefont {Berti}\ \emph {et~al.}(2018)\citenamefont {Berti},
  \citenamefont {Yagi}, \citenamefont {Yang},\ and\ \citenamefont
  {Yunes}}]{Berti:2018vdi}%
  \BibitemOpen
  \bibfield  {author} {\bibinfo {author} {\bibfnamefont {E.}~\bibnamefont
  {Berti}}, \bibinfo {author} {\bibfnamefont {K.}~\bibnamefont {Yagi}},
  \bibinfo {author} {\bibfnamefont {H.}~\bibnamefont {Yang}},\ and\ \bibinfo
  {author} {\bibfnamefont {N.}~\bibnamefont {Yunes}},\ }\bibfield  {title}
  {\bibinfo {title} {{Extreme Gravity Tests with Gravitational Waves from
  Compact Binary Coalescences: (II) Ringdown}},\ }\href
  {https://doi.org/10.1007/s10714-018-2372-6} {\bibfield  {journal} {\bibinfo
  {journal} {Gen. Rel. Grav.}\ }\textbf {\bibinfo {volume} {50}},\ \bibinfo
  {pages} {49} (\bibinfo {year} {2018})},\ \Eprint
  {https://arxiv.org/abs/1801.03587} {arXiv:1801.03587 [gr-qc]} \BibitemShut
  {NoStop}%
\bibitem [{\citenamefont {Isi}\ \emph {et~al.}(2019)\citenamefont {Isi},
  \citenamefont {Giesler}, \citenamefont {Farr}, \citenamefont {Scheel},\ and\
  \citenamefont {Teukolsky}}]{Isi:2019aib}%
  \BibitemOpen
  \bibfield  {author} {\bibinfo {author} {\bibfnamefont {M.}~\bibnamefont
  {Isi}}, \bibinfo {author} {\bibfnamefont {M.}~\bibnamefont {Giesler}},
  \bibinfo {author} {\bibfnamefont {W.~M.}\ \bibnamefont {Farr}}, \bibinfo
  {author} {\bibfnamefont {M.~A.}\ \bibnamefont {Scheel}},\ and\ \bibinfo
  {author} {\bibfnamefont {S.~A.}\ \bibnamefont {Teukolsky}},\ }\bibfield
  {title} {\bibinfo {title} {{Testing the no-hair theorem with GW150914}},\
  }\href {https://doi.org/10.1103/PhysRevLett.123.111102} {\bibfield  {journal}
  {\bibinfo  {journal} {Phys. Rev. Lett.}\ }\textbf {\bibinfo {volume} {123}},\
  \bibinfo {pages} {111102} (\bibinfo {year} {2019})},\ \Eprint
  {https://arxiv.org/abs/1905.00869} {arXiv:1905.00869 [gr-qc]} \BibitemShut
  {NoStop}%
\bibitem [{\citenamefont {Isi}\ and\ \citenamefont {Farr}(2021)}]{Isi:2021iql}%
  \BibitemOpen
  \bibfield  {author} {\bibinfo {author} {\bibfnamefont {M.}~\bibnamefont
  {Isi}}\ and\ \bibinfo {author} {\bibfnamefont {W.~M.}\ \bibnamefont {Farr}},\
  }\bibfield  {title} {\bibinfo {title} {{Analyzing black-hole ringdowns}},\
  }\href@noop {} {\  (\bibinfo {year} {2021})},\ \Eprint
  {https://arxiv.org/abs/2107.05609} {arXiv:2107.05609 [gr-qc]} \BibitemShut
  {NoStop}%
\bibitem [{\citenamefont {Capano}\ \emph {et~al.}(2023)\citenamefont {Capano},
  \citenamefont {Cabero}, \citenamefont {Westerweck}, \citenamefont {Abedi},
  \citenamefont {Kastha}, \citenamefont {Nitz}, \citenamefont {Wang},
  \citenamefont {Nielsen},\ and\ \citenamefont {Krishnan}}]{Capano:2021etf}%
  \BibitemOpen
  \bibfield  {author} {\bibinfo {author} {\bibfnamefont {C.~D.}\ \bibnamefont
  {Capano}}, \bibinfo {author} {\bibfnamefont {M.}~\bibnamefont {Cabero}},
  \bibinfo {author} {\bibfnamefont {J.}~\bibnamefont {Westerweck}}, \bibinfo
  {author} {\bibfnamefont {J.}~\bibnamefont {Abedi}}, \bibinfo {author}
  {\bibfnamefont {S.}~\bibnamefont {Kastha}}, \bibinfo {author} {\bibfnamefont
  {A.~H.}\ \bibnamefont {Nitz}}, \bibinfo {author} {\bibfnamefont {Y.-F.}\
  \bibnamefont {Wang}}, \bibinfo {author} {\bibfnamefont {A.~B.}\ \bibnamefont
  {Nielsen}},\ and\ \bibinfo {author} {\bibfnamefont {B.}~\bibnamefont
  {Krishnan}},\ }\bibfield  {title} {\bibinfo {title} {{Multimode Quasinormal
  Spectrum from a Perturbed Black Hole}},\ }\href
  {https://doi.org/10.1103/PhysRevLett.131.221402} {\bibfield  {journal}
  {\bibinfo  {journal} {Phys. Rev. Lett.}\ }\textbf {\bibinfo {volume} {131}},\
  \bibinfo {pages} {221402} (\bibinfo {year} {2023})},\ \Eprint
  {https://arxiv.org/abs/2105.05238} {arXiv:2105.05238 [gr-qc]} \BibitemShut
  {NoStop}%
\bibitem [{\citenamefont {Liang}\ and\ \citenamefont {Wang}(2026)}]{wanghe}%
  \BibitemOpen
  \bibfield  {author} {\bibinfo {author} {\bibfnamefont {B.}~\bibnamefont
  {Liang}}\ and\ \bibinfo {author} {\bibfnamefont {H.}~\bibnamefont {Wang}},\
  }\bibfield  {title} {\bibinfo {title} {Recent advances in simulation-based
  inference for gravitational wave data analysis},\ }\href
  {https://doi.org/10.61977/ati2025020} {\bibfield  {journal} {\bibinfo
  {journal} {Astronomical Techniques and Instruments}\ }\textbf {\bibinfo
  {volume} {3}},\ \bibinfo {pages} {93} (\bibinfo {year} {2026})}\BibitemShut
  {NoStop}%
\bibitem [{\citenamefont {Wald}(1973)}]{Wald:1973wwa}%
  \BibitemOpen
  \bibfield  {author} {\bibinfo {author} {\bibfnamefont {R.~M.}\ \bibnamefont
  {Wald}},\ }\bibfield  {title} {\bibinfo {title} {{On perturbations of a Kerr
  black hole}},\ }\href {https://doi.org/10.1063/1.1666203} {\bibfield
  {journal} {\bibinfo  {journal} {J. Math. Phys.}\ }\textbf {\bibinfo {volume}
  {14}},\ \bibinfo {pages} {1453} (\bibinfo {year} {1973})}\BibitemShut
  {NoStop}%
\bibitem [{\citenamefont {Chandrasekhar}(1984)}]{Chandrasekhar:1984mgh}%
  \BibitemOpen
  \bibfield  {author} {\bibinfo {author} {\bibfnamefont {S.}~\bibnamefont
  {Chandrasekhar}},\ }\bibfield  {title} {\bibinfo {title} {{On algebraically
  special perturbations of black holes}},\ }\href
  {https://doi.org/10.1098/rspa.1984.0021} {\bibfield  {journal} {\bibinfo
  {journal} {Proceedings of the Royal Society of London. A. Mathematical and
  Physical Sciences}\ }\textbf {\bibinfo {volume} {392}},\ \bibinfo {pages} {1}
  (\bibinfo {year} {1984})}\BibitemShut {NoStop}%
\bibitem [{\citenamefont {Couch}\ and\ \citenamefont
  {Newman}(1973)}]{Couch:1973zc}%
  \BibitemOpen
  \bibfield  {author} {\bibinfo {author} {\bibfnamefont {W.~E.}\ \bibnamefont
  {Couch}}\ and\ \bibinfo {author} {\bibfnamefont {E.~T.}\ \bibnamefont
  {Newman}},\ }\bibfield  {title} {\bibinfo {title} {{Algebraically special
  perturbations of the Schwarzschild metric}},\ }\href
  {https://doi.org/10.1063/1.1666311} {\bibfield  {journal} {\bibinfo
  {journal} {J. Math. Phys.}\ }\textbf {\bibinfo {volume} {14}},\ \bibinfo
  {pages} {285} (\bibinfo {year} {1973})}\BibitemShut {NoStop}%
\bibitem [{\citenamefont {Cook}\ and\ \citenamefont
  {Zalutskiy}(2016)}]{Cook:2016fge}%
  \BibitemOpen
  \bibfield  {author} {\bibinfo {author} {\bibfnamefont {G.~B.}\ \bibnamefont
  {Cook}}\ and\ \bibinfo {author} {\bibfnamefont {M.}~\bibnamefont
  {Zalutskiy}},\ }\bibfield  {title} {\bibinfo {title} {{Purely imaginary
  quasinormal modes of the Kerr geometry}},\ }\href
  {https://doi.org/10.1088/0264-9381/33/24/245008} {\bibfield  {journal}
  {\bibinfo  {journal} {Class. Quant. Grav.}\ }\textbf {\bibinfo {volume}
  {33}},\ \bibinfo {pages} {245008} (\bibinfo {year} {2016})},\ \Eprint
  {https://arxiv.org/abs/1603.09710} {arXiv:1603.09710 [gr-qc]} \BibitemShut
  {NoStop}%
\bibitem [{\citenamefont {Cook}\ \emph {et~al.}(2019)\citenamefont {Cook},
  \citenamefont {Annichiarico},\ and\ \citenamefont {Vickers}}]{Cook:2018ses}%
  \BibitemOpen
  \bibfield  {author} {\bibinfo {author} {\bibfnamefont {G.~B.}\ \bibnamefont
  {Cook}}, \bibinfo {author} {\bibfnamefont {L.~S.}\ \bibnamefont
  {Annichiarico}},\ and\ \bibinfo {author} {\bibfnamefont {D.~J.}\ \bibnamefont
  {Vickers}},\ }\bibfield  {title} {\bibinfo {title} {{Unknown branch of the
  total-transmission modes for the Kerr geometry}},\ }\href
  {https://doi.org/10.1103/PhysRevD.99.024008} {\bibfield  {journal} {\bibinfo
  {journal} {Phys. Rev. D}\ }\textbf {\bibinfo {volume} {99}},\ \bibinfo
  {pages} {024008} (\bibinfo {year} {2019})},\ \Eprint
  {https://arxiv.org/abs/1808.00987} {arXiv:1808.00987 [gr-qc]} \BibitemShut
  {NoStop}%
\bibitem [{\citenamefont {Cook}\ and\ \citenamefont {Lu}(2023)}]{Cook:2022kbb}%
  \BibitemOpen
  \bibfield  {author} {\bibinfo {author} {\bibfnamefont {G.~B.}\ \bibnamefont
  {Cook}}\ and\ \bibinfo {author} {\bibfnamefont {S.}~\bibnamefont {Lu}},\
  }\bibfield  {title} {\bibinfo {title} {{New total transmission modes of the
  Kerr geometry with Schwarzschild limit frequencies at complex infinity}},\
  }\href {https://doi.org/10.1103/PhysRevD.107.044043} {\bibfield  {journal}
  {\bibinfo  {journal} {Phys. Rev. D}\ }\textbf {\bibinfo {volume} {107}},\
  \bibinfo {pages} {044043} (\bibinfo {year} {2023})},\ \Eprint
  {https://arxiv.org/abs/2211.14955} {arXiv:2211.14955 [gr-qc]} \BibitemShut
  {NoStop}%
\bibitem [{\citenamefont {Cook}(2025)}]{cook_2025_16801267}%
  \BibitemOpen
  \bibfield  {author} {\bibinfo {author} {\bibfnamefont {G.~B.}\ \bibnamefont
  {Cook}},\ }\bibfield  {title} {\bibinfo {title} {Kerr modes: Phase fixed
  gravitational qnms and ttms},\ }\href
  {https://doi.org/10.5281/zenodo.16801267} {10.5281/zenodo.16801267} (\bibinfo
  {year} {2025})\BibitemShut {NoStop}%
\bibitem [{\citenamefont {Cook}(2026)}]{greg_cook_2026_19070504}%
  \BibitemOpen
  \bibfield  {author} {\bibinfo {author} {\bibfnamefont {G.}~\bibnamefont
  {Cook}},\ }\href {https://doi.org/10.5281/zenodo.19070504} {\bibinfo {title}
  {cookgb/kerrmodes: Initial public release (changes to swspheroidal paclet)}}
  (\bibinfo {year} {2026})\BibitemShut {NoStop}%
\bibitem [{\citenamefont {Zhou}\ \emph {et~al.}(2026)\citenamefont {Zhou},
  \citenamefont {Ji}, \citenamefont {Wu},\ and\ \citenamefont
  {Cao}}]{Zhou:2025xdo}%
  \BibitemOpen
  \bibfield  {author} {\bibinfo {author} {\bibfnamefont {Y.-S.}\ \bibnamefont
  {Zhou}}, \bibinfo {author} {\bibfnamefont {M.-F.}\ \bibnamefont {Ji}},
  \bibinfo {author} {\bibfnamefont {L.-B.}\ \bibnamefont {Wu}},\ and\ \bibinfo
  {author} {\bibfnamefont {L.-M.}\ \bibnamefont {Cao}},\ }\bibfield  {title}
  {\bibinfo {title} {{Pseudospectrum and (in)stability of black hole total
  transmission modes}},\ }\href {https://doi.org/10.1103/plvy-f3mf} {\bibfield
  {journal} {\bibinfo  {journal} {Phys. Rev. D}\ }\textbf {\bibinfo {volume}
  {113}},\ \bibinfo {pages} {124042} (\bibinfo {year} {2026})},\ \Eprint
  {https://arxiv.org/abs/2512.16372} {arXiv:2512.16372 [gr-qc]} \BibitemShut
  {NoStop}%
\bibitem [{\citenamefont {Wu}\ \emph {et~al.}(2026)\citenamefont {Wu},
  \citenamefont {Zhou}, \citenamefont {Yu}, \citenamefont {Jia},\ and\
  \citenamefont {Cao}}]{Wu:2026hvf}%
  \BibitemOpen
  \bibfield  {author} {\bibinfo {author} {\bibfnamefont {L.-B.}\ \bibnamefont
  {Wu}}, \bibinfo {author} {\bibfnamefont {Y.-S.}\ \bibnamefont {Zhou}},
  \bibinfo {author} {\bibfnamefont {Z.}~\bibnamefont {Yu}}, \bibinfo {author}
  {\bibfnamefont {M.-F.}\ \bibnamefont {Jia}},\ and\ \bibinfo {author}
  {\bibfnamefont {L.-M.}\ \bibnamefont {Cao}},\ }\bibfield  {title} {\bibinfo
  {title} {{Virtual absorption modes of Schwarzschild-de Sitter spacetimes in
  semi-open systems}},\ }\href@noop {} {\  (\bibinfo {year} {2026})},\ \Eprint
  {https://arxiv.org/abs/2603.22897} {arXiv:2603.22897 [gr-qc]} \BibitemShut
  {NoStop}%
\bibitem [{\citenamefont {Dong}\ \emph {et~al.}(2026)\citenamefont {Dong},
  \citenamefont {Wu}, \citenamefont {Zhou},\ and\ \citenamefont
  {Guo}}]{Dong:2026zxy}%
  \BibitemOpen
  \bibfield  {author} {\bibinfo {author} {\bibfnamefont {X.-N.}\ \bibnamefont
  {Dong}}, \bibinfo {author} {\bibfnamefont {L.-B.}\ \bibnamefont {Wu}},
  \bibinfo {author} {\bibfnamefont {Y.-S.}\ \bibnamefont {Zhou}},\ and\
  \bibinfo {author} {\bibfnamefont {Z.-K.}\ \bibnamefont {Guo}},\ }\bibfield
  {title} {\bibinfo {title} {{The (in)stability on total transmission modes
  with small bumps}},\ }\href@noop {} {\  (\bibinfo {year} {2026})},\ \Eprint
  {https://arxiv.org/abs/2609.07167} {arXiv:2609.07167 [gr-qc]} \BibitemShut
  {NoStop}%
\bibitem [{\citenamefont {{Changkai Chen}}(2026)}]{ChenTTM}%
  \BibitemOpen
  \bibfield  {author} {\bibinfo {author} {\bibnamefont {{Changkai Chen}}},\
  }\href@noop {} {\bibinfo {title} {{Complete-TTMs}}},\ \bibinfo {howpublished}
  {\href{https://github.com/IronChen1/Complete-TTMs}{https://github.com/IronChen1/Complete-TTMs}}
  (\bibinfo {year} {{2026}})\BibitemShut {NoStop}%
\bibitem [{\citenamefont {Cook}\ and\ \citenamefont
  {Wang}(2026)}]{Cook:2026gpm}%
  \BibitemOpen
  \bibfield  {author} {\bibinfo {author} {\bibfnamefont {G.~B.}\ \bibnamefont
  {Cook}}\ and\ \bibinfo {author} {\bibfnamefont {X.}~\bibnamefont {Wang}},\
  }\bibfield  {title} {\bibinfo {title} {{Choosing the phase for the
  spin-weighted spheroidal functions}},\ }\href
  {https://doi.org/10.1103/ffpl-dyld} {\bibfield  {journal} {\bibinfo
  {journal} {Phys. Rev. D}\ }\textbf {\bibinfo {volume} {114}},\ \bibinfo
  {pages} {064052} (\bibinfo {year} {2026})},\ \Eprint
  {https://arxiv.org/abs/2603.22485} {arXiv:2603.22485 [gr-qc]} \BibitemShut
  {NoStop}%
\bibitem [{\citenamefont {Cavalcante}\ \emph {et~al.}(2024)\citenamefont
  {Cavalcante}, \citenamefont {Richartz},\ and\ \citenamefont
  {da~Cunha}}]{Cavalcante:2024swt}%
  \BibitemOpen
  \bibfield  {author} {\bibinfo {author} {\bibfnamefont {J.~P.}\ \bibnamefont
  {Cavalcante}}, \bibinfo {author} {\bibfnamefont {M.}~\bibnamefont
  {Richartz}},\ and\ \bibinfo {author} {\bibfnamefont {B.~C.}\ \bibnamefont
  {da~Cunha}},\ }\bibfield  {title} {\bibinfo {title} {{Exceptional Point and
  Hysteresis in Perturbations of Kerr Black Holes}},\ }\href
  {https://doi.org/10.1103/PhysRevLett.133.261401} {\bibfield  {journal}
  {\bibinfo  {journal} {Phys. Rev. Lett.}\ }\textbf {\bibinfo {volume} {133}},\
  \bibinfo {pages} {261401} (\bibinfo {year} {2024})},\ \Eprint
  {https://arxiv.org/abs/2407.20850} {arXiv:2407.20850 [gr-qc]} \BibitemShut
  {NoStop}%
\bibitem [{\citenamefont {Cavalcante}\ \emph {et~al.}(2026)\citenamefont
  {Cavalcante}, \citenamefont {Richartz},\ and\ \citenamefont
  {da~Cunha}}]{Cavalcante:2025abr}%
  \BibitemOpen
  \bibfield  {author} {\bibinfo {author} {\bibfnamefont {J.~P.}\ \bibnamefont
  {Cavalcante}}, \bibinfo {author} {\bibfnamefont {M.}~\bibnamefont
  {Richartz}},\ and\ \bibinfo {author} {\bibfnamefont {B.~C.}\ \bibnamefont
  {da~Cunha}},\ }\bibfield  {title} {\bibinfo {title} {{Ergodic hysteresis of
  the Kerr black hole spectrum}},\ }\href {https://doi.org/10.1103/mssm-ws7d}
  {\bibfield  {journal} {\bibinfo  {journal} {Phys. Rev. D}\ }\textbf {\bibinfo
  {volume} {113}},\ \bibinfo {pages} {L101504} (\bibinfo {year} {2026})},\
  \Eprint {https://arxiv.org/abs/2511.16640} {arXiv:2511.16640 [gr-qc]}
  \BibitemShut {NoStop}%
\bibitem [{\citenamefont {Cavalcante}\ and\ \citenamefont
  {Richartz}(2026)}]{Cavalcante:2026vgr}%
  \BibitemOpen
  \bibfield  {author} {\bibinfo {author} {\bibfnamefont {J.~P.}\ \bibnamefont
  {Cavalcante}}\ and\ \bibinfo {author} {\bibfnamefont {M.}~\bibnamefont
  {Richartz}},\ }\bibfield  {title} {\bibinfo {title} {{Exceptional lines in
  the Kerr-Newman black hole spectrum}},\ }\href@noop {} {\  (\bibinfo {year}
  {2026})},\ \Eprint {https://arxiv.org/abs/2607.09878} {arXiv:2607.09878
  [gr-qc]} \BibitemShut {NoStop}%
\bibitem [{\citenamefont {Panosso~Macedo}\ \emph {et~al.}(2026)\citenamefont
  {Panosso~Macedo}, \citenamefont {Katagiri}, \citenamefont {Kubota},\ and\
  \citenamefont {Motohashi}}]{PanossoMacedo:2025xnf}%
  \BibitemOpen
  \bibfield  {author} {\bibinfo {author} {\bibfnamefont {R.}~\bibnamefont
  {Panosso~Macedo}}, \bibinfo {author} {\bibfnamefont {T.}~\bibnamefont
  {Katagiri}}, \bibinfo {author} {\bibfnamefont {K.-i.}\ \bibnamefont
  {Kubota}},\ and\ \bibinfo {author} {\bibfnamefont {H.}~\bibnamefont
  {Motohashi}},\ }\bibfield  {title} {\bibinfo {title} {{Exceptional points and
  resonance in black hole ringdown}},\ }\href
  {https://doi.org/10.1103/5w5d-kr25} {\bibfield  {journal} {\bibinfo
  {journal} {Phys. Rev. D}\ }\textbf {\bibinfo {volume} {113}},\ \bibinfo
  {pages} {L121504} (\bibinfo {year} {2026})},\ \Eprint
  {https://arxiv.org/abs/2512.02110} {arXiv:2512.02110 [gr-qc]} \BibitemShut
  {NoStop}%
\bibitem [{\citenamefont {Rossi}\ \emph {et~al.}(2026)\citenamefont {Rossi},
  \citenamefont {Oshita}, \citenamefont {Gualtieri},\ and\ \citenamefont
  {Berti}}]{Rossi:2026als}%
  \BibitemOpen
  \bibfield  {author} {\bibinfo {author} {\bibfnamefont {D.}~\bibnamefont
  {Rossi}}, \bibinfo {author} {\bibfnamefont {N.}~\bibnamefont {Oshita}},
  \bibinfo {author} {\bibfnamefont {L.}~\bibnamefont {Gualtieri}},\ and\
  \bibinfo {author} {\bibfnamefont {E.}~\bibnamefont {Berti}},\ }\bibfield
  {title} {\bibinfo {title} {{Excitation factors and exceptional points of
  Kerr-de Sitter quasinormal modes}},\ }\href@noop {} {\  (\bibinfo {year}
  {2026})},\ \Eprint {https://arxiv.org/abs/2609.03009} {arXiv:2609.03009
  [gr-qc]} \BibitemShut {NoStop}%
\bibitem [{\citenamefont {Chen}\ and\ \citenamefont
  {Jing}(2023)}]{Chen:2023ese}%
  \BibitemOpen
  \bibfield  {author} {\bibinfo {author} {\bibfnamefont {C.}~\bibnamefont
  {Chen}}\ and\ \bibinfo {author} {\bibfnamefont {J.}~\bibnamefont {Jing}},\
  }\bibfield  {title} {\bibinfo {title} {{Radiation fluxes of gravitational,
  electromagnetic, and scalar perturbations in type-D black holes: an exact
  approach}},\ }\href {https://doi.org/10.1088/1475-7516/2023/11/070}
  {\bibfield  {journal} {\bibinfo  {journal} {JCAP}\ }\textbf {\bibinfo
  {volume} {11}},\ \bibinfo {pages} {070}},\ \Eprint
  {https://arxiv.org/abs/2307.14616} {arXiv:2307.14616 [gr-qc]} \BibitemShut
  {NoStop}%
\bibitem [{\citenamefont {Chen}\ and\ \citenamefont
  {Jing}(2024)}]{Chen:2023lsa}%
  \BibitemOpen
  \bibfield  {author} {\bibinfo {author} {\bibfnamefont {C.}~\bibnamefont
  {Chen}}\ and\ \bibinfo {author} {\bibfnamefont {J.}~\bibnamefont {Jing}},\
  }\bibfield  {title} {\bibinfo {title} {{Gravitational wave fluxes on generic
  orbits in near-extreme Kerr spacetime: Higher spin and large eccentricity}},\
  }\href {https://doi.org/10.1007/s11433-024-2431-0} {\bibfield  {journal}
  {\bibinfo  {journal} {Sci. China Phys. Mech. Astron.}\ }\textbf {\bibinfo
  {volume} {67}},\ \bibinfo {pages} {110411} (\bibinfo {year} {2024})},\
  \Eprint {https://arxiv.org/abs/2311.15295} {arXiv:2311.15295 [gr-qc]}
  \BibitemShut {NoStop}%
\bibitem [{\citenamefont {Chen}\ \emph {et~al.}(2026)\citenamefont {Chen},
  \citenamefont {Cao},\ and\ \citenamefont {Jing}}]{Chen:2026bym}%
  \BibitemOpen
  \bibfield  {author} {\bibinfo {author} {\bibfnamefont {C.}~\bibnamefont
  {Chen}}, \bibinfo {author} {\bibfnamefont {Z.}~\bibnamefont {Cao}},\ and\
  \bibinfo {author} {\bibfnamefont {J.}~\bibnamefont {Jing}},\ }\bibfield
  {title} {\bibinfo {title} {{Efficient and stable computation of
  gravitational-wave fluxes from generic Kerr orbits via a unified
  Heun-function framework}},\ }\href {https://doi.org/10.1103/nngm-dhpl}
  {\bibfield  {journal} {\bibinfo  {journal} {Phys. Rev. D}\ }\textbf {\bibinfo
  {volume} {114}},\ \bibinfo {pages} {063002} (\bibinfo {year} {2026})},\
  \Eprint {https://arxiv.org/abs/2605.09250} {arXiv:2605.09250 [gr-qc]}
  \BibitemShut {NoStop}%
\end{thebibliography}%

\end{document}